\documentclass[aps,twocolumn]{revtex4}

\RequirePackage[T2A]{fontenc}

\usepackage{amssymb}
\usepackage{amsmath}
\usepackage{bm}
\usepackage[dvips]{epsfig}
\usepackage{graphicx}

\begin{document}

\title{Features of the motion of ultracold atoms in quasiperiodic potentials} 

\author{Ivan Dynnikov$^{1}$ and Andrei Maltsev$^{2}$}

\affiliation{
\centerline{$^{1}$ \it{V.A. Steklov Mathematical Institute 
of Russian Academy of Sciences}}
\centerline{\it 119991 Moscow, Gubkina str. 8}
\centerline{$^{2}$ \it{L.D. Landau Institute for Theoretical Physics 
of Russian Academy of Sciences}}
\centerline{\it 142432 Chernogolovka, pr. Ak. Semenova 1A}
}

\begin{abstract}
 We consider here quasiperiodic potentials on the plane, which can serve 
as a ``transitional link'' between ordered (periodic) and chaotic (random)
potentials.  As can be shown, in almost any family of quasiperiodic 
potentials depending on a certain set of parameters, it is possible to
distinguish a set (in the parameter space) where, according to a certain
criterion, potentials with features of ordered potentials arise, and 
a set where we have potentials with features of random potentials. 
These sets complement each other in the complete parameter space, and 
each of them has its own specific structure.  The difference between 
``ordered'' and ``chaotic'' potentials will manifest itself, in particular, 
in the transport properties at different energies, which we consider 
here in relation to systems of ultracold atoms. It should be noted
here that the transport properties of particles in the considered potentials
can be accompanied by the phenomena of ``partial integrability'' inherent in
two-dimensional Hamiltonian systems.
\end{abstract}

\maketitle

\vspace{5mm}

\section{Introduction}

 In this paper, we consider applications of relatively recent 
results in the theory of quasiperiodic functions on the plane to 
the dynamics of ultracold atoms in quasiperiodic potentials. The 
general theory of quasiperiodic functions, the origins of which 
go back to the studies of H. Bohr and A.S. Besicovich (see, for example 
\cite{Bohr,Besicovitch}), is currently a classical area of mathematics 
and mathematical physics. In different papers, actually, one can find 
slightly different definitions of a quasiperiodic function. 
Here we will call a quasiperiodic function on the plane any function
$\, f (x, y) \, $, obtained by restricting a smooth $d$-periodic function
$\, F (z^{1}, \dots , z^{d}) \, $  to a generic affine embedding
$ \, \mathbb{R}^{2} \, \subset \, \mathbb{R}^{d} \, $. In this case, 
the dimension $\, d \, $ will be called the number of quasiperiods of 
the corresponding quasiperiodic function on the plane. 

The questions we consider here are mainly related 
to the geometry of the regions defined by the relation 
\begin{equation}
\label{Oblasti}
 f (x, y) \,\,\, \leqslant \,\,\, \epsilon_{0} 
\end{equation}
for different energy values $\, \epsilon_{0} \, $.
The function $\, f (x, y) \, $ will play the role of a potential 
in which the two-dimensional dynamics of a particle is considered. 
It is easy to see that if the dynamics is purely classical, then such 
regions play the role of ``accessibility regions'' in which a particle 
with energy $\, \epsilon_{0} \, $ may appear. As we will see below, 
the complexity of the motion of a particle, inherent in the general 
case of any Hamiltonian dynamics, can add up to the complexity of the 
geometry of the corresponding ``regions of accessibility'', which may 
have certain ``chaotic properties''. 

 In addition to the geometry of the ``accessibility regions'' 
for a particle with a given energy, we will also be interested in 
describing the changes in this geometry when changing the parameters 
that define the function $\, f (x, y) \, $ in real systems. 
In particular, it may be of interest to control the dynamics of a gas 
of particles in quasiperiodic potentials by changing such parameters.

  It is easy to see that the geometry of the domains (\ref{Oblasti}) 
is in fact closely related to the geometry of the level lines
\begin{equation}
\label{Urovni}
f (x, y) \,\,\, =  \,\,\, \epsilon_{0} 
\end{equation}
of the function $\, f (x, y) \, $, bounding these domains for the 
corresponding values of $\, \epsilon_{0} \, $. We can also say that 
each of the regions (\ref{Oblasti}) is the union of the level lines
(\ref{Urovni}) for all
$\, \epsilon_{0}^{\prime} \, \leqslant \, \epsilon_{0} \, $.

 The general problem of describing the geometry of the level lines 
of quasiperiodic functions on the plane was set by S.P. Novikov in the 
early 1980s (see \cite{MultValAnMorseTheory}) and then was actively 
studied in his topological school 
(\cite{zorich1,dynn1992,Tsarev,dynn1,zorich2,DynnBuDA,dynn2,dynn3}).
In its original setting, it was focused on the study of the geometry 
of the level lines of functions with three quasiperiods, which is actually 
equivalent to describing the geometry of the intersection lines 
of an arbitrary 3-periodic two-dimensional surface in 
$\, \mathbb{R}^{3} \, $ by planes of a given direction. 
In this setting, Novikov's  problem is most directly related 
to the problem of describing the dynamics of electrons in metals with 
complex Fermi surfaces, which, in turn, plays an important role in 
describing galvanomagnetic phenomena in metals (see for example
\cite{lifkag1,lifkag2,lifkag3,Kittel,etm,Abrikosov,KaganovPeschansky}).
We note here that the results obtained in the study of Novikov's 
problem in this case turn out to be very important for the description 
of galvanomagnetic phenomena in metals in the most general situation. 
In particular, it is possible to define nontrivial 
topological numbers observable in the conductivity of normal metals 
with complex Fermi surfaces in strong enough magnetic fields 
(\cite{PismaZhETF}). In addition, a complete classification of possible 
types of dynamics (trajectories) of electrons on the Fermi surface allowed 
to give also a complete description of various asymptotic 
behaviors of conductivity in metals with arbitrary Fermi surfaces 
in the limit of strong magnetic fields (see, for example
\cite{UFN,BullBrazMathSoc,JournStatPhys}).

 In general, Novikov's problem for an arbitrary number of quasiperiods 
is closely related to the theory of foliations and the theory of dynamical 
systems on manifolds. In particular, this connection plays an important 
role in the study of Novikov's problem in the case of four quasi-periods 
(see \cite{NovKvazFunc,DynNov}), which allows to obtain a number of 
fundamental results for this case.

 In this paper, we try to use, in the most complete way, the 
results obtained so far in the study of Novikov's problem to 
describe the main features of the particle dynamics in the 
potentials associated with this problem. In addition, the analysis 
of the results obtained for Novikov's problem allows, in fact, 
to propose a criterion dividing all quasiperiodic potentials 
(with three or more quasiperiods) into two types, namely, potentials 
with topologically regular level lines and potentials with chaotic 
level lines. If we consider quasiperiodic potentials as a model of the 
transition between periodic and random potentials, then the potentials 
of the first type can be classified as potentials in which the dynamics 
retains some topological integrability, and potentials of the second 
type as potentials in which the dynamics is close to the dynamics in 
random potentials. In this sense, only potentials of the second 
type can represent models of random potentials.

 In fact, quasiperiodic potentials arise quite often as members of a family 
(and this will be so in this work), i.e. we usually have a whole family 
of functions $\, V (x, y, {\bf U}) \, $, depending smoothly on a number 
of additional parameters $\, {\bf U} = (U^{1}, \dots, U^{N}) \, $.
In this situation, it is important that the subsets of the parameter space 
corresponding to potentials of different types 
have, as a rule, very specific structure. Namely, potentials 
of the first type, being stable with respect to small variations of 
parameters, form an open subset in the complete space of parameters. 
Moreover, this subset is a union of a (finite or countable) number 
of pairwise disjoint regions (stability zones) each of which is determined 
by its individual value of some topological invariant (a set of integers). 
On the contrary, potentials of the second type form a subset of fractal type 
complementary to the set described above.

 It can be seen, therefore, that in order to create quasiperiodic 
potentials that are most alike to truly random potentials, it is necessary 
to find a rather complicated set in the parameter space defining 
the family  of quasiperiodic potentials under consideration. 
It can also be noted here that the larger the number of 
quasiperiods the richer becomes the structure of this set.

 We believe that the most convenient systems for the experimental 
study of dynamics in quasiperiodic potentials are systems of ultracold 
atoms in optical traps, where such potentials can be easily constructed 
by superposition of several standing waves. It must be said that, despite 
the special technique for creating such potentials, they, in fact, have 
all the features of Novikov's problem in the general setting, so 
such systems allow us to study all essential aspects of the problem 
under consideration.

  As we have already said, here we study the dynamics of particles 
of different energies in the potentials described above. Thus, the presented 
results will be directly related to the description of transport phenomena 
in the limit of almost noninteracting atoms in traps with quasiperiodic 
potentials. This limit, in fact, arises quite often in the case of low 
concentration of (neutral) atoms trapped in a trap, as well as small 
radius of their interaction. In our case, we should naturally require 
that the mean free path of the atoms is greater than the typical length 
at which the global geometric features of their trajectories manifest 
themselves. What is also essential, as is known, two-dimensional 
Hamiltonian systems have, as a rule, very special dynamics 
(see \cite{Ott,LikhtLib}), being integrable at sufficiently low energy 
levels and passing to chaotic regimes with increasing energy. In our case, 
we will be able to observe how this circumstance agrees with the geometry 
of the level lines of our potentials.

\section{General analytical results. Regular and random potentials.}
\setcounter{equation}{0}

 As is well known, the most common method for creating external 
potentials for atoms in optical traps is a superposition of standing 
waves from additional laser sources (see, for example
\cite{Letokhov1968,LetokhovBook,Bloch,BlochDalibardZwerger,GreinerFolling}).
In the leading approximation, such potentials usually have a finite number 
of harmonics, i.e., they can be represented as the sum of a finite number 
of sinusoidal potentials. Despite this circumstance, as we will see below, 
in the situations we are considering, such potentials already have 
sufficient complexity, and to describe the atomic dynamics in them, 
it is necessary to describe the complete picture that arises in the 
study of the general Novikov's problem. 

 It is easy to see that the problem under consideration is rapidly 
becoming more complicated with an increase in the number of quasiperiods. 
As we have already said, rigorous analytical results exist at the moment 
only for the case of three and four quasiperiods. For comparison, it 
is convenient here to briefly consider also the situations of ``one'' and 
``two'' quasiperiods corresponding to periodic potentials depending on a 
single coordinate, and doubly periodic potentials on the plane, respectively.

 In our situation, the case of ``one'' quasiperiod will correspond in 
reality to the presence of a very simple potential, which can often be 
written approximately in the form 
\begin{equation}
\label{OnePeriod}
V (x,y) \,\,\, = \,\,\, V_{0} \,\, \sin k x \,\, , 
\end{equation}
where $\, T \, = \, 2 \pi / k \, $ is the ordinary period of the potential 
along the $x$ axis. The creation of potentials of this type is the simplest one 
from the technical point of view, and, certainly, such potentials are widely 
used in systems of cold atoms. 

 The level lines of potential (\ref{OnePeriod}), obviously, represent 
vertical straight lines, and the motion of atoms in such a potential 
occurs in straight vertical stripes at $\, E < V_{0} \, $. In fact, it is 
easy to see in this situation, that the motion of an atom is confined to a 
vertical strip provided
$$E \, - \, p_{y}^{2} / 2 m \,\,\, < \,\,\, V_{0}, $$
and in this case, the atom performs periodic oscillations along 
the $x$ axis and moves uniformly along the $y$ axis. If the above 
condition is violated, the atom obviously moves along a periodic 
trajectory having nonzero mean inclination with respect to the $y$ axis. 
It is easy to see that similar conditions can also be written for a 
uniformly moving coordinate system, which also allows to give 
a similar description of the motion of atoms in potentials of the form 
$$ V (x,y,t)= V_{0}\, \sin\bigl( k (x - u t)\bigr) .$$

 It can be seen here that at sufficiently low atomic energies and the 
value of the velocity $u$, the moving potential carries out a 
complete ``transportation'' of the atomic gas along the $x$ axis. 
In the general case, with a significant spread of energies, the moving 
potential allows only a partial ``transportation'' of the atomic gas. 

\vspace{1mm}

 The geometry of the level lines of a doubly periodic potential on the 
plane (with two independent periods
$\, {\bf e}_{1} \, $ and $\, {\bf e}_{2} $)
also has a relatively simple description. As in the case of ``one'' 
quasiperiod, the values of the potential $\, V(x,y) \, $ 
lie here in some closed segment $\, [ V_{\min}, V_{\max}] \, $.  
It is easy to see that, for generic potentials, the level lines of the 
potential are closed for values of $\, E \, $ sufficiently close to
$\, V_{\min} \, $ or $\, V_{\max} \, $. In the first case, however, 
closed level lines bound areas of lower potential values, while 
in the second case they bound areas of larger values. In the 
generic case, we have here two different values 
$\, V_{1} , V_{2} \, $:
$$ V_{\min} \,\, < \,\, V_{1} \,\, < \,\, V_{2}
\,\, < \,\, V_{\max}, $$
such that for all fixed values of $\, V (x,y)$
lying in the interval $\, (V_{1} , V_{2}) \, $
the corresponding levels contain open (non-closed) components. 
All open level components (lines) of the potential
$\, V(x,y) \, $ are, in this case, periodic curves having the same 
mean direction in the plane (for all levels in the interval 
$\, (V_{1} , V_{2}) $). The average direction of the open level 
lines can be any integer direction, i.e.\ a direction given by a
vector of the form
$${\bf l} \,\,\, = \,\,\, n_{1} \, {\bf e}_{1} 
\,\, + \,\, n_{2} \, {\bf e}_{2} $$
with some integers $\, n_{1}, n_{2} \, $. Note that it is
natural to demand the numbers $\, n_{1}, n_{2} \, $ to be relatively prime, 
given up to a common sign. 

 All open level lines for a given value of the potential can be 
divided into a finite number of families, such that all lines in
one family pass into each other when shifted by some period of
$\, V(x,y) \, $. The number of such different families is always 
even (although it can vary within the interval $\, (V_{1} , V_{2}) $ 
for sufficiently complex potentials). 

  Thus, it can be seen that the motion of atoms in generic periodic 
potentials occurs in bounded regions at 
$\, V_{\min} \, < \, \epsilon_{0} \, < \, V_{1} \, $,
in periodic stripes (as well as, possibly, isolated bounded regions) at 
$\, V_{1} \, < \, \epsilon_{0} \, < \, V_{2} \, $,
in the plane with excluded bounded domains (and possibly isolated 
bounded areas) at
$\, V_{2} \, < \, \epsilon_{0} \, < \, V_{\max} \, $ 
and in the whole plane at $\, \epsilon_{0} \, > \, V_{\max} \, $. 
In the case of adiabatic shifts of the potential in the plane, the atoms 
in the bounded regions shift with them, while the atoms in the periodic 
stripes move along with the stripes only when the shift is in the 
direction perpendicular to the stripes. Here, in contrast to the case 
of ``one'' quasiperiod, however, the shift of the potential 
along the direction of the stripes also affects the motion of atoms, 
since the stripes now have a nontrivial shape. 

\begin{figure}[t]
\begin{center}
\includegraphics[width=\linewidth]{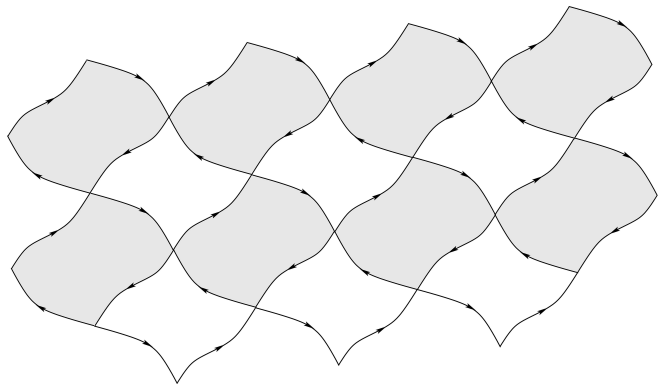}
\end{center}
\caption{Periodic net of singular level lines separating areas 
of the lower values from areas of the larger values of 
a non-generic periodic potential.}
\label{Fig1}
\end{figure}

 For non-generic periodic potentials (for example, having elements 
of rotational symmetry), the values $\, V_{1} \, $ and $\, V_{2} \, $ 
may coincide. Such potentials do not have open level lines,
but at the level $\, V_{1} = V_{2} \, $ they have a singular net 
(Fig.~\ref{Fig1}), separating the areas of smaller values of 
$\, V (x, y) \, $ from the areas of its larger values. Atoms move 
in bounded regions at
$\, V_{\min} \, < \, \epsilon_{0} \, < \, V_{1} = V_{2} \, $ 
and in the entire plane with excluded bounded regions (and, 
possibly, isolated bounded domains) at
$\, V_{1} = V_{2} \, < \, \epsilon_{0} \, < \, V_{\max} \, $.

 Methods for creating periodic potentials in a plane naturally 
imply the presence of a certain finite number of parameters 
describing such potentials. For example, when making a potential 
by using a superposition of two sinusoidal standing waves (with 
or without generation of higher harmonics), such parameters 
can be the orientations of both sinusoids, their amplitudes, 
periods, positions of the maxima, possibly the angle between 
their polarizations and the relative phase shift. It is easy 
to see that in this case the shift of the maxima of any of the 
sinusoids actually leads to a shift of the resultant potential 
as a whole and does not qualitatively change the picture we are 
considering.

 As for more general (continuous) variations of the parameters 
described above, one important common feature can be noted here.  
Namely, generic potentials form an open set, i.e. the relation 
$\, V_{2} > V_{1} \, $ is stable with respect to an arbitrary 
small change in parameters. On the contrary, the condition 
$\, V_{1} = V_{2} \, $ is unstable and can break down under an 
arbitrarily small variation of the parameters of the general 
form. In addition, the numbers $\, (n_{1} , n_{2}) \, $, 
relating the average directions of the open level lines 
to the periods of the potential, are also locally stable 
(although $\, {\bf e}_{1} \, $ and $\, {\bf e}_{2} \, $ 
can change with a change in parameters), and can change only 
when the potential passes through a non-generic situation 
($V_{1} = V_{2}$). Thus, it can be seen that the complete space 
of parameters can generally be divided into regions corresponding 
to different integer pairs $\, (n_{1} , n_{2}) \, $ and separated 
by boundaries on which the relation $\, V_{1} = V_{2} \, $ holds.

 Here we would also like to note that despite the relatively 
simple description of the ``accessibility regions'' in periodic 
potentials at any energy values, the conservative dynamics of 
cold atoms in such potentials can have very nontrivial properties 
(see, for example 
\cite{HemmSchroppHansch,HennequinVerkerk1,HennequinVerkerk2}).

\vspace{1mm}

  We now turn to considering potentials with larger numbers of 
quasiperiods, which are the main subject of our work. 

 The creation of quasiperiodic potentials in systems of ultracold 
atoms using superposition of standing waves has also attracted 
interest from both the theoretical and the experimental points of view.  
In particular, such potentials were considered both in the case of 
three-dimensional (see, for example, \cite{Guidoni1}) and in the 
case of two-dimensional 
(see \cite{Guidoni2,SanchezPalenciaSantos,ViebSbrosCartYuSchneid,
GautierYaoSanchezPalencia}) 
optical lattices for atoms trapped in magneto-optical traps 
(see also a review on the creation, confinement and monitoring of 
the behavior of gases of ultracold atoms, including that in potentials of 
various shapes, in the work \cite{Guidoni3}). It can also be noted 
that quasiperiodic (quasicrystalline) structures in two-dimensional 
systems of interacting ultracold atoms can arise even in the absence of 
special external modulation (see, for example, \cite{GopMartinDemler}).

 As we have said above, here we are interested in quasiperiodic 
potentials for two-dimensional systems of ultracold atoms.  
As we have also already noted, we will especially focus here on the 
cases of 3 and 4 quasi-periods, for which profound analytical 
results are known to date. In this section, we will 
simply formulate the rigorous analytical results obtained 
so far for such potentials. In the next section, we will consider 
in more detail the features of the geometry arising here, with 
specific examples. 

\begin{figure}[t]
\begin{center}
\includegraphics[width=\linewidth]{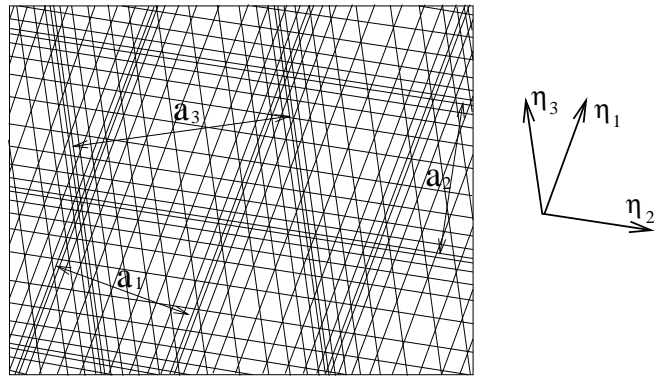}
\end{center}
\caption{Superposition of three standing waves in a plane 
with the formation of a potential with three quasi-periods 
(schematically). (The vectors $\, \bm{\eta}_{i} \, $ indicate 
the directions of the wave fronts, and the vectors 
$\, {\bf a}_{i} \, $ are the shifts between the maxima 
of their amplitudes).}
\label{Fig2}
\end{figure}

 As we have already said, Novikov's problem for the case of three 
quasiperiods, which we consider here in the most detail, 
is currently most thoroughly studied. 
In the considered method of creating potentials in the plane, 
this situation corresponds to the potentials obtained by 
superimposing three sinusoidal standing waves oriented at 
different angles (with or without the presence of higher 
harmonics) (Fig.~\ref{Fig2}).

 It should be noted here, of course, that a superposition of 
three (or more) standing waves in lattices of cold atoms can be used 
to create not only quasiperiodic, but also interesting periodic 
potentials in two-dimensional systems (in particular, such a scheme 
was proposed in the work \cite{DuanDemlerLukin} to create hexagonal 
(honeycomb) lattices and in \cite{KagomeLattice} to create 
tri-hexagonal (Kagome) lattices). In this case, the wave numbers of 
the corresponding waves must, generally speaking, satisfy a number of 
special additional conditions. In our situation, we will assume that 
the wave numbers are independent parameters in the space of the 
considered potentials. In the general case, the potentials we are 
considering will not have exact periods in the plane, which can arise 
only for special values of the parameters.

 As we have already said, the results presented below will be based 
only on quasiperiodic properties of the potentials arising in our 
case, therefore, many additional details of their origin will not, 
in fact, play any essential role.

 It is easy to see that a potential generated by three 
sinusoidal waves (and higher harmonics) 
$$V ({\bf r}) \,\,\, = \,\,\,  \sum_{i=1}^{3} \,
V_{i} \cos ({\bf k}_{(i)} {\bf r} + \delta_{i}) 
\,\,\, + \,\,\, \dots \,\, , $$
represents a restriction of the periodic in $\, {\mathbb{R}^{3}} \, $
function
$$V (X^{1}, X^{2}, X^{3}) \,\,\, = \,\,\, \sum_{i=1}^{3} \, 
V_{i} \cos X^{i} \,\,\, + \,\,\, \dots \,\, $$
to an affine embedding 
$\, \mathbb{R}^{2} \rightarrow \mathbb{R}^{3}$,
given by the formulas 
\begin{equation}
\label{AffEmb}
\left(
\begin{array}{c} 
x \cr 
y 
\end{array}
\right)
\quad \rightarrow \quad
\left( 
\begin{array}{c}
k_{(1)}^{1} x + k_{(1)}^{2} y + \delta_{1}   \cr
k_{(2)}^{1} x + k_{(2)}^{2} y + \delta_{2}   \cr
k_{(3)}^{1} x + k_{(3)}^{2} y + \delta_{3} 
\end{array}
\right) 
\, = \,
\left( 
\begin{array}{c}
{\bf k}_{(1)} {\bf r} + \delta_{1}   \cr
{\bf k}_{(2)} {\bf r} + \delta_{2}   \cr
{\bf k}_{(3)} {\bf r} + \delta_{3} 
\end{array}
\right) 
\end{equation}

 It can be noted here that the potentials differing only in
shifts of the maxima of the standing waves ($\delta_{i}$) correspond 
to the same function $\, V (X^{1}, X^{2}, X^{3}) \, $ in
$\, {\mathbb{R}^{3}}$, and their difference is due only to a change 
in the affine embedding $\, \mathbb{R}^{2} \rightarrow \mathbb{R}^{3}$, 
under which the plane $\, \mathbb{R}^{2} \, $ is shifted in
$\, {\mathbb{R}^{3}} \, $ keeping its direction fixed.
 
 As we have already said, even quasiperiodic potentials having 
a small number of harmonics are already sufficiently complex to 
observe all aspects of the general Novikov's problem for three 
quasiperiods; for this reason, we will consider the problem described 
above on the grounds of the general results for functions with three 
quasiperiods. Below we will formulate, in the most convenient form for us, 
a number of fundamental results for the general Novikov's problem following 
from the works  
\cite{zorich1,dynn1992,Tsarev,dynn1,zorich2,DynnBuDA,dynn2,dynn3}. 
Similar statements for artificially created potentials in the plane, 
in fact, were given in the work \cite{JMathPhys}, where electron 
transport phenomena in such potentials in the presence of a strong 
magnetic field were studied. Note, however, that in the work 
\cite{JMathPhys}, the main role was played by the geometry of the 
level lines of quasiperiodic potentials, rather than by that of the 
regions of lower values, which we consider here.

 It is convenient for us to start with the remark that, 
although this does not happen in the general case, the potentials 
obtained by the method considered here can also be periodic.  
This situation arises whenever the corresponding plane 
$\, \mathbb{R}^{2} \subset \mathbb{R}^{3} \, $ is integral (rational), 
that is, contains two independent integer vectors in 
$\, {\mathbb{R}^{3}} \, $. It is easy to see that the corresponding 
potentials are everywhere dense among all potentials of our interest, 
while their periodic properties are determined by the values 
of the parameters $\, ({\bf k}_{1}, {\bf k}_{2}, {\bf k}_{3}) \, $.
To describe the level lines of such potentials, all previously made 
statements about periodic potentials can be used. However, it 
should be noted here right away that the overwhelming majority of these 
potentials will have very large (in absolute value) periods 
$\, {\bf e}_{1} \, $ and $\, {\bf e}_{2} \, $. As a consequence, 
the above-described properties of the level lines of such potentials 
will be observed only on very large scales, while on smaller scales 
their level lines may have completely different nontrivial properties 
that are more important for describing the observed experimental data. 
As a result, for potentials with large periods, as a rule, more 
informative will be more general statements about the level lines of 
quasiperiodic potentials, which we cite below.

\begin{figure}[t]
\begin{center}
\includegraphics[width=\linewidth]{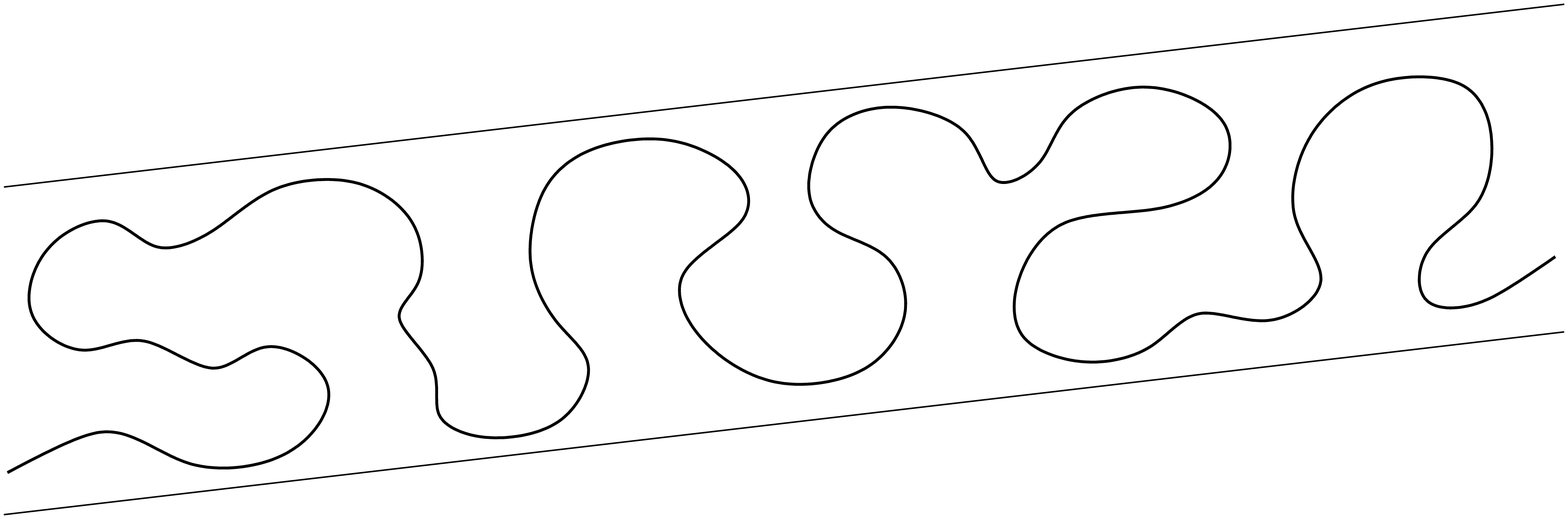}
\end{center}
\caption{An open level line of a quasiperiodic potential lying 
in a straight strip of finite width and passing through it 
(schematically).}
\label{Fig3}
\end{figure}

 At the same time, it turns out that the role of the everywhere dense 
set of periodic potentials in the parameter space is, in fact, extremely 
important in the study of Novikov's problem for the quasiperiodic 
potentials we are considering. Below we will formulate an extremely 
important statement concerning small deformations of periodic 
potentials and following from the results of works 
\cite{zorich1,dynn1992}. Note here that in works 
\cite{zorich1,dynn1992} it is actually assumed that certain regularity 
conditions (of general position) are fulfilled, which we do not formulate
here in detail, assuming that they are always satisfied for arising 
in reality physical potentials (\cite{Conditions}). In this case, in the 
simplest form, the corollaries we need from \cite{zorich1, dynn1992} 
can be formulated as follows:

\vspace{1cm}

 Let some complete set of parameters 
$${\bf U}_{0} \, = \, ({\bf k}_{(1)}^{0}, {\bf k}_{(2)}^{0},
{\bf k}_{(3)}^{0}, V_{1}^{0}, V_{2}^{0}, V_{3}^{0}, \dots ) $$
correspond to some periodic potential in the plane. 
Then there is an open neighborhood 
$\, \Omega \, $ of the point  $\, {\bf U}_{0} \, $
(``stability zone'') in the parameter space, such that for any 
$\, {\bf U} \in \, \Omega \, $ the following holds:

\vspace{1mm}

A1. All open level lines of the corresponding two-dimensional 
potential $\, V (x, y, {\bf U}) \, $ lie in straight strips of 
finite width, passing through them (Fig.~\ref{Fig3});

\vspace{1mm}

A2. The mean direction $\, {\bf l} ({\bf U}) \, $ of strips, 
containing open level lines of the potential 
$\, V (x, y, {\bf U}) \, $, is defined in the entire region 
$\, \Omega \, $, for some (irreducible) integer triple
$\, (m^{1}, m^{2}, m^{3}) \, $, by the relation
\begin{equation}
\label{TripleDir}
\left( m^{1} {\bf k}_{(1)} \, + \, m^{2} {\bf k}_{(2)} \, + \,
m^{3} {\bf k}_{(3)} \, , \, {\bf l} ({\bf U}) \right) 
\,\,\, = \,\,\, 0 
\end{equation}

\vspace{1mm}

 Note here that, for embeddings (\ref{AffEmb}) of maximal 
irrationality degree, relation (\ref{TripleDir}) uniquely determines
(up to the sign) the triple $\, (m^{1}, m^{2}, m^{3}) \, $.

\vspace{1mm}

 The triples $\, (m^{1}, m^{2}, m^{3}) \, $ have, actually, a
topological origin and can also be introduced in another way.
Namely, we recall that among the parameters of our potentials, 
there are the positions of the maxima of the standing 
waves used to create the potential. In the picture we are now 
considering, the shift of an individual standing wave is no longer 
equivalent to a simple shift of the resulting potential and 
represents a somewhat more complex transformation. It is also 
easy to see that the shift of the front of a standing wave 
(perpendicular to itself) by the period of this wave is 
equivalent to the identical transformation. In general, the full 
set of all such transformations forms a three-parameter group 
($\mathbb{T}^{3}$) containing simple shifts as an algebraic 
subgroup. At the same time, simple shifts form an everywhere 
dense set in the considered group of transformations for 
generic potentials; therefore, all such potentials associated 
with the described transformations are, in a sense, mutually related. 
In particular, such potentials have similar level lines for 
any energy value $\, \epsilon_{0} \, $.

\begin{figure}[t]
\begin{center}
\includegraphics[width=\linewidth]{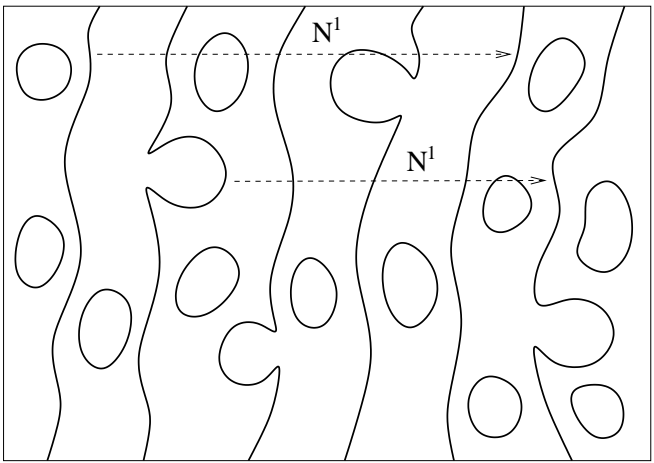}
\end{center}
\caption{The shift of regular open level lines of a quasiperiodic 
potential under a shift of the maxima of one of the standing waves 
by a full period in the direction of the phase growth.}
\label{Fig4}
\end{figure}

 If we now consider a generic potential $\, V (x, y, {\bf U}) \, $
corresponding to some values $\, {\bf U} \in \Omega \, $, and fix a
level $\, \epsilon_{0} \, $, corresponding to the appearance of 
open level lines, then we can follow the change of any of these 
lines for each of the three continuous shifts of the standing wave 
fronts (in the direction of the phase growth) for the corresponding 
periods. In this case, each of the continuous shifts will correspond 
to the movement of an open level line in the plane (possibly with 
merging with it and separation from it of closed level lines in the 
process of movement), as a result of which it will finally transform 
into some other level line of the same potential (Fig.~\ref{Fig4}). 
It can be shown in this case that the number of positions $\, N \, $, 
by which the line is shifted, is the same for all open lines on this 
level, and is determined only by the selected transformation.

 The triplet of numbers $\, (N^{1}, N^{2}, N^{3}) \, $, determined 
at three successive shifts of the first, second and third waves, 
respectively, can be represented as 
$$(N^{1}, N^{2}, N^{3}) \,\,\, = \,\,\, M \, 
(m^{1}, m^{2}, m^{3}) \,\,\, , $$
where $\, M \in \mathbb{Z} \, $ and 
$\, (m^{1}, m^{2}, m^{3}) \, $ is an irreducible integer triple. 
The number $\, M \, $ is always even and also has a topological 
origin (the number of ``equivalence classes'' of open level lines 
in the plane), and the triple $\, (m^{1}, m^{2}, m^{3}) \, $ 
coincides with the one introduced earlier up to a sign.

\vspace{1mm}

 Thus, we can see that in the space of our parameters we can 
distinguish a set of ``stability zones'' $\, \Omega_{{\bf m}, M}$, 
parameterized by integer triples 
$\, {\bf m} \, = \, (m^{1}, m^{2}, m^{3}) \, $ 
(and also by numbers $\, M$). The numbers 
$\, (m^{1}, m^{2}, m^{3}) \, $, generally speaking, do not run 
through the entire set of integer (relatively prime) triples, but their 
number is generally infinite, and they can take arbitrarily large 
values. Each ``stability zone'' represents some open region in the 
parameter space with a piecewise smooth boundary (the boundaries of 
different zones can be adjacent to each other). The set of 
``stability zones'' is a rather rich structure in the parameter 
space, in particular, the ``stability zones'' contain all the 
values of parameters corresponding to the emergence of periodic 
potentials $\, V (x, y, {\bf U}) \, $ (see \cite{Conditions}).

 It can be seen that the above description of open level lines 
of potentials arising in the zones $\, \Omega_{{\bf m}, M}$ is 
rather simple and very informative (especially in the case of 
small values of $\, (m^{1}, m^{2}, m^{3}) $). In particular, 
it gives much more information about the behavior of level lines 
of arising in $\, \Omega_{{\bf m}, M} \, $ periodic potentials 
with large periods than can be obtained from the fact of their 
periodicity. As the values of $\, (m^{1}, m^{2}, m^{3}) \, $ grow, 
the sizes of the stability zones $\, \Omega_{{\bf m}, M} \, $
decrease, and this also makes the strips containing 
open level lines (and the corresponding areas of lower values) wider. 
Along with this, the above description also refers to larger 
and larger scales in the plane, giving less detailed information 
about the geometry of the level lines at smaller scales 
(Fig.~\ref{Fig5}). As we will see below, in this situation, 
the behavior of the level lines (on small scales) can no longer 
be described in such a simple way and has much more complex 
(chaotic) properties. As we will also see, in limiting cases 
such behavior can lead to completely chaotic behavior of the 
level lines of quasiperiodic potentials, which has complex 
chaotic properties at all scales.

\begin{figure}[t]
\begin{center}
\includegraphics[width=\linewidth]{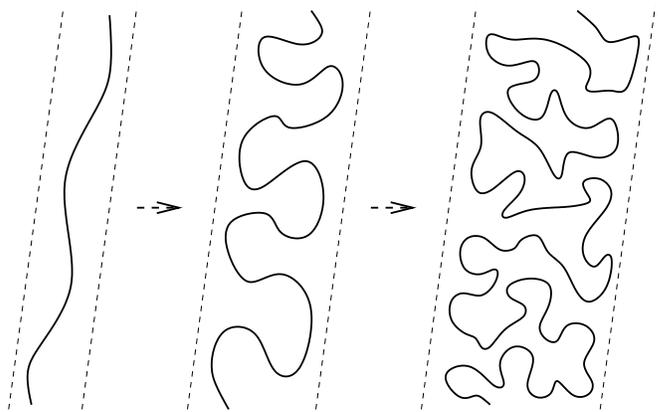}
\end{center}
\caption{Complication of the geometry of ``topologically regular'' 
open level lines of a quasiperiodic potential with increasing 
values of $\, (m^{1}, m^{2}, m^{3}) \, $ (schematically). }
\label{Fig5}
\end{figure}

\vspace{1mm}

\begin{figure}[t]
\begin{center}
\includegraphics[width=\linewidth]{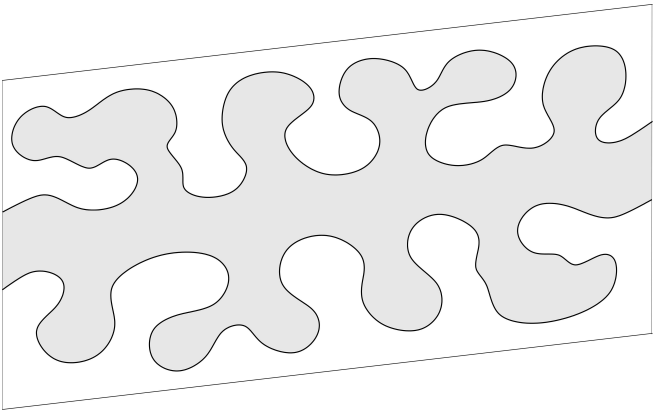}
\end{center}
\caption{The region of smaller values of a quasiperiodic potential, 
lying in a straight strip of finite width and passing through it 
(schematically).}
\label{Fig6}
\end{figure}

 Based on the properties of the level lines of quasiperiodic 
potentials in stability zones, it can be seen here
that the corresponding regions 
$\,  V (x, y, {\bf U}) \, < \, \epsilon_{0} \, $ 
should also possess the same properties if the level 
lines $\,  V (x, y, {\bf U}) \, = \, \epsilon_{0} \, $ are open 
(Fig.~\ref{Fig6}).

\vspace{1mm}

 Namely, if the level $\, \epsilon_{0} \, $ contains open
level lines, then any open connected domain
$\,  V (x, y, {\bf U}) \, < \, \epsilon_{0} \, $
also lies in a straight strip of finite width and passes through it. 
The mean direction of such a strip, obviously, coincides with the 
mean direction of the open level lines and is given by equation 
(\ref{TripleDir}). We note here that now both the level lines and 
the regions of lower values of the potential are no longer periodic 
for the general values of the parameters. As can be seen, however, 
properties (A1) - (A2) give here a certain analogy with the case 
of periodic potentials. 

 It is easy to see that transport phenomena in a potential with 
parameters lying in one of the stability zones can have an explicit  
anisotropy. This property should, as a rule, be observed if the 
ensemble of particles placed in such a potential contains 
particles with energies corresponding to the emergence of open 
level lines of the potential. It can be seen, in fact, that such 
anisotropy can also be observed under more general assumptions, 
in particular, in systems of strongly interacting particles or in 
the hydrodynamic approximation.

 Here, however, we must immediately make an important remark 
about transport phenomena in the case we are considering. 
Namely, in stability zones with large values of the numbers
$\, (m^{1}, m^{2}, m^{3}) \, $ the width of the strips containing 
the domains $\,  V (x, y, {\bf U}) \, < \, \epsilon_{0} \, $
(in the presence of open level lines at
$\,  V (x, y, {\bf U}) \, = \, \epsilon_{0} $)
becomes rather large, and the shape of such regions becomes 
more and more complex, which is (very schematically) shown at 
Fig.~\ref{Fig6}. As a consequence of this, the motion of particles 
in such regions becomes more and more complicated, gradually 
acquiring the features of wandering in a random potential.

\vspace{1mm}

 We also note here that the above-described movement of the 
open level lines (and the ``regions of accessibility'' restricted 
by them) under the shifts of the maxima of each of the standing 
waves depends in the most significant way on the values 
$\, (m^{1}, m^{2}, m^{3}) \, $. In particular, the speed of
movement of the ``accessibility regions'' at large values of 
$\, (m^{1}, m^{2}, m^{3}) \, $ can significantly exceed the 
speed of the maxima of standing waves. The latter circumstance 
can, in fact, play an essential role for many questions of the 
transport of atoms in optical lattices (see, for example, 
\cite{LPGARMNMCA} and the references therein). Moreover, 
such a movement, generally speaking, produces an incomplete 
transportation of atoms in the corresponding direction due to 
separation of some closed areas from the ``accessibility areas'' 
and joining of others to them in the process of their 
movement.

\vspace{1mm}

 We also point out here one more important circumstance. Namely, as we 
have already said, the presence of a stability zone in the space of 
parameters means, naturally, the preservation of the picture 
determined by it under sufficiently small variations of the
parameters. In fact, as follows from the topological considerations, 
this picture is also stable under much more general 
variations of the potential, in particular, those arising 
on finite scales provided that they are small enough. Thus, it can be seen 
that the above description of the geometry of the regions
$\,  V (x, y, {\bf U}) \, < \, \epsilon_{0} \, $, as well as 
the features of transport phenomena, is stable under
disturbances or ``defects'' of a sufficiently small magnitude. 
This circumstance is also important in the presence of additional 
(non-quasiperiodic) slowly varying potentials, which are also 
often present in experimental techniques. The above description 
(with the same numbers $\, (m^{1}, m^{2}, m^{3}) $) is preserved 
in general in regions where the maximum change of such potentials 
does not exceed the energy interval of the existence of open level lines 
of the potential $\, V (x, y, {\bf U}) \, $. It should be said here that 
the estimate of the admissible variations of the potential 
$\, V (x, y, {\bf U}) \, $ also rapidly decreases with 
increase in numbers $\, (m^{1}, m^{2}, m^{3}) \, $.

\vspace{1mm}

 Despite the fact that the set of all stability zones 
 $\, \Omega_{{\bf m}, M} \, $ forms an open cover of an everywhere 
dense set in the parameter space, it, generally speaking, does not 
cover it entirely, and quasiperiodic potentials with three quasi-periods 
can have, as we said, level lines that are more complex than those 
described above (\cite{Tsarev,dynn2}). It can be said, nevertheless, 
that the ``regular'' situation described above is, in a sense, basic 
for the case of three quasi-periods, while the more complex behavior 
of the level lines requires a special construction of the corresponding 
potential. For a more complete description of the picture arising in 
the general case, we present here, based on the results of works 
\cite{dynn1,dynn3}, a number of important statements about the 
structure of the level lines of potentials with three quasiperiods 
of the most general form. In fact, together with the above statements, 
the results presented below represent, in a certain sense, a complete 
theory of the level lines of potentials with three quasiperiods in 
the plane.

\vspace{1mm} 
 
 We note here at once that all the potentials obtained from 
3-periodic functions by means of an embedding 
$\, \mathbb{R}^{2} \subset \mathbb{R}^{3} \, $ can in fact 
be divided into 3 types. Namely, first of all, as we have 
already seen, for certain values of
$\, ({\bf k}_{(1)}, {\bf k}_{(2)}, {\bf k}_{(3)}) \, $
the corresponding potential $\, V (x, y, {\bf U}) \, $
may actually turn out to be doubly periodic. 
Here we call such potentials type I potentials. 
The second possibility is that the potential 
$\, V (x, y, {\bf U}) \, $, while not being doubly periodic, 
still has one (up to a factor) period in the plane
$\, \mathbb{R}^{2} \, $. Such potentials will be called 
type II potentials. Like potentials of type I, they arise 
on an everywhere dense set of zero measure in the space of parameters
$\, ({\bf k}_{(1)}, {\bf k}_{(2)}, {\bf k}_{(3)}) \, $. Finally, into 
the family of potentials of type III 
we include potentials that do not have exact periods 
in the plane $\, \mathbb{R}^{2} \, $. In our case, only 
potentials of type III are in general position and correspond 
to a set of the full measure in the parameter space. 
In the statements about the level lines of quasiperiodic 
potentials formulated below, we will assume that the 
corresponding potentials are of type II or III, 
since potentials of type I have already been actually 
considered above. Under the assumptions made, the following 
statements can be formulated.

\vspace{1mm}

 Let $\, V (x, y, {\bf U}) \, $ be a potential with 
three quasi-periods taking values in the interval
$\, [ V_{\min} ({\bf U}), \, V_{\max} ({\bf U})] \, $.
Then:

\vspace{1mm}

B1. Open level lines of $\, V (x, y, {\bf U}) \, $
exist in a connected interval 
$\, [ V_{1} ({\bf U}), \, V_{2} ({\bf U})] \, $,
$$V_{\min} ({\bf U}) \,\,\, < \,\,\, 
V_{1} ({\bf U}) \,\,\, \leqslant \,\,\, V_{2} ({\bf U})
\,\,\, < \,\,\, V_{\max} ({\bf U}) \,\,\, , $$
which can degenerate to a single point
$\, V_{0} ({\bf U}) = V_{1} ({\bf U}) = 
V_{2} ({\bf U}) \, $.

\vspace{1mm}

B2. Whenever open level lines appear in a finite interval
$\, [ V_{1} ({\bf U}), \, V_{2} ({\bf U})] \, $,
they have the properties (A1)-(A2), presented above. 

\vspace{1mm}

B3. In the case when the interval
$\, [ V_{1} ({\bf U}), \, V_{2} ({\bf U})] \, $ 
shrinks to a single point
$\, V_{0} ({\bf U}) = V_{1} ({\bf U}) = 
V_{2} ({\bf U}) \, $, the open level lines arising at the 
corresponding level can either satisfy conditions (A1)-(A2) 
(this occurs at the boundaries of the stability zones
$\, \Omega_{{\bf m}, M} $), or have more complex chaotic 
behavior (this occurs at the accumulation points of an 
infinite number of zones $\, \Omega_{{\bf m}, M} \, $ 
with unboundedly increasing values of
$\, (m^{1}, m^{2}, m^{3}) $).

\vspace{1mm}

 It can be seen, therefore, that the appearance of open level 
lines in ``stability zones'' does not at all resemble the similar 
phenomenon for truly random potentials, where, 
as a rule, open level lines arise at a single energy level 
(if we consider random potentials from the point of view of 
the percolation theory, see e.g. \cite{Stauffer,Essam}).
Also, the potentials arising at the boundaries of the 
``stability zones'', although they have open level lines 
only for one value of $\, \epsilon_{0} \, $, are not very 
suitable for the role of random potentials due to the 
``too regular behavior'' of their open level lines. 
Therefore, it can be seen that it is natural to consider 
only potentials with chaotic level lines as models of 
a random potential. As we said above, in the case of three 
quasi-periods, such potentials always arise at accumulation 
points of ``stability zones'' with an increasingly complex 
geometry of open level lines, so that there is 
always a passage from the ``regular'' behaviour to the ``chaotic''
one in this case.

\vspace{1mm}

 It should also be noted here that chaotic level lines 
arising in the case of type II potentials are very different from 
those of type III potentials. Namely, chaotic level lines of potentials 
of type II always have the form of curves with an asymptotic direction 
in the plane $\, \mathbb{R}^{2} \, $ (see \cite{dynn2}). Thus, such level 
lines resemble to some extent ``regular'' level lines described above, 
and, in general, pass through the plane along some fixed direction. 
The difference between the two cases lies in the fact that the deviations 
of a chaotic level line of a type~II potential in the direction 
perpendicular to the asymptotic one are not necessarily bounded, 
so the line may not be enclosed in any straight strip of finite width. 
In this case, the same can be said also about the corresponding regions 
of lower values of the corresponding potential
$$V (x, y, {\bf U}) \,\, \leqslant \,\, V_{0} \,\,\, , $$
namely, they have the form of ``strips'' passing through plane 
in some fixed direction. The width of these strips, however, 
can vary indefinitely in their different parts, and they cannot be 
enclosed in straight strips of a fixed width. When the boundary 
value $\, \epsilon_{0} \, $ is shifted downward by an arbitrarily 
small amount, the region of lower values of the corresponding 
potential consists of strongly elongated bounded regions, and 
when $\, \epsilon_{0} \, $ is shifted upward, this region 
becomes the entire plane with strongly elongated 
bounded regions removed (and, possibly, with bounded 
regions not associated with the main component and lying inside 
the excluded regions added).

 One may wonder, therefore, to what extent type II potentials 
with chaotic level lines can be considered as a model of a random 
potential. In a sense, they can be considered as an intermediate 
case between ``regular'' and ``chaotic'' potentials.

\vspace{1mm}

\begin{figure}[t]
\begin{center}
\includegraphics[width=\linewidth]{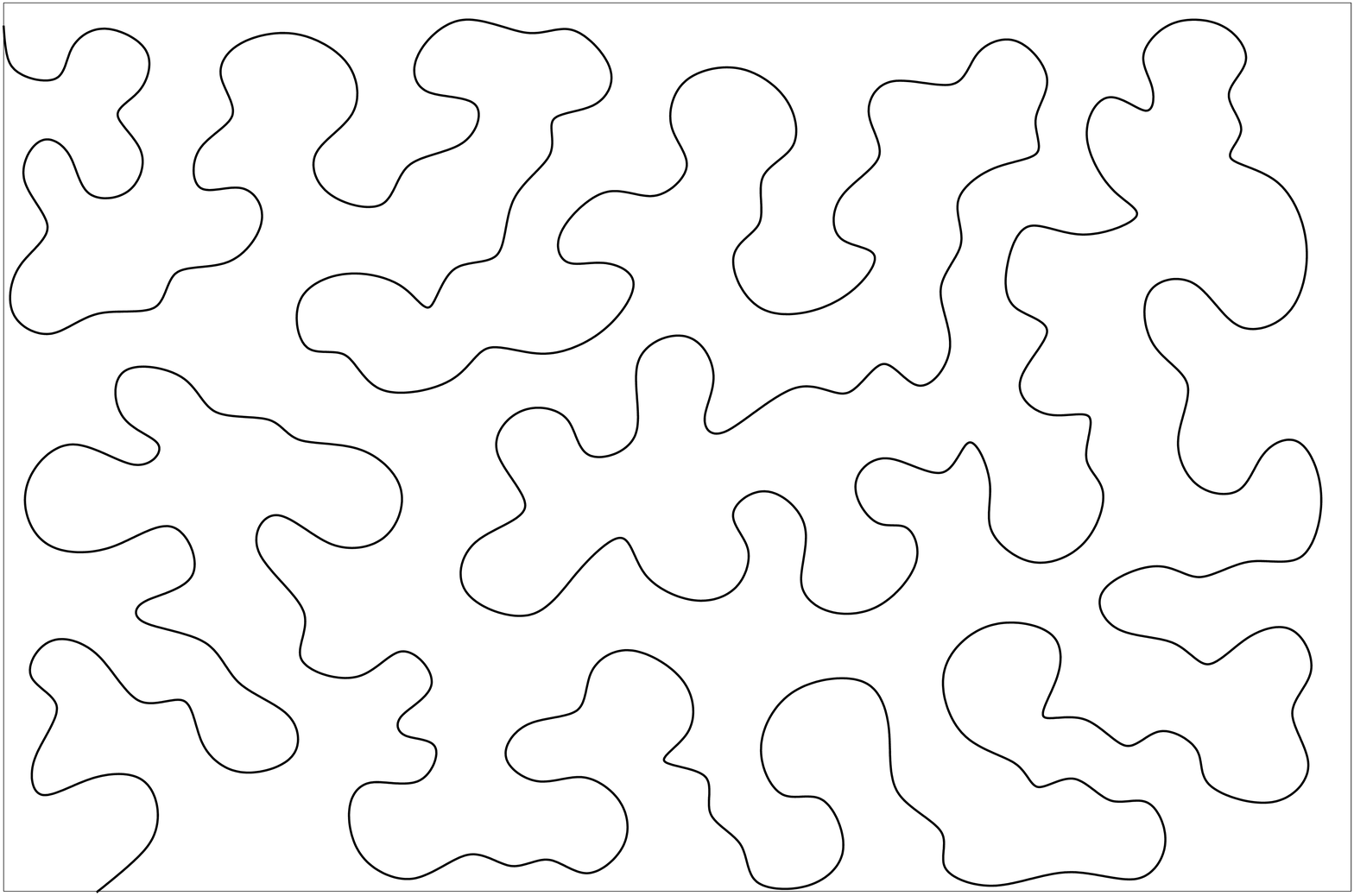}
\end{center}
\caption{A ``chaotic'' open level line of a potential with three 
quasi-periods (schematically).}
\label{Fig7}
\end{figure}

 Chaotic level lines arising in the case of type III potentials are 
much more complex and ``sweep'' the entire plane 
$\, \mathbb{R}^{2} \, $ in a chaotic manner (Fig.~\ref{Fig7}).
A similar behavior is exhibited in this case also by the 
regions of the smaller values
$\, V (x, y, {\bf U}) \,\, \leqslant \,\, V_{0} \, $.
When the boundary value $\, \epsilon_{0} \, $ is shifted 
downward by an arbitrarily small amount, the region of lower 
values of the corresponding potential splits into rather 
complicated bounded regions, and when $\, \epsilon_{0} \, $ 
is shifted upward, this region becomes the entire plane 
from which bounded areas of complex shape are excluded (and, 
possibly, additional bounded areas that are not associated 
with the main component and lying inside the excluded areas
added). An important circumstance here is that the 
linear dimensions of such regions grow according to a power 
law when approaching the value of $\, \epsilon_{0} \, $ 
with a fractional exponent
($ \sim |\epsilon - \epsilon_{0}|^{-\alpha}$),
which makes such potentials alike random potentials 
(see, for example, \cite{Riedel,Trugman}). It should be noted, 
however, that here, in the general case, a certain 
anisotropy can be observed, namely, the presence of two 
different growth degrees $\, \alpha \, $ and $\, \beta \, $, 
in a certain direction in the plane and in the one perpendicular to it
($0 < \alpha , \beta < 1$). In general, the stochastic properties 
of such level lines are quite complex and are currently the subject 
of intensive research (see, for example
\cite{dynn2,dynn3,ZhETF2,zorich3,DeLeo1,DeLeo2,
DeLeoPhysLettA,DeLeoPhysB,DeLeo3,DeLeoDynnikov1,DeLeoDynnikov2,
Skripchenko1,Skripchenko2,DynnSkrip1,DynnSkrip2,AvilaHubSkrip1,
AvilaHubSkrip2,TrMIAN}).

 The randomness of the level lines of a potential formed by 
three standing waves is invariant under shifts
of the phases $\, \delta_{i} \, $ (i.e., the positions of the 
standing wave maxima) with the remaining parameters fixed. 
In this case, the value of $\, V_{0} \, $, as well as the 
geometric features of the chaotic level lines (in particular, 
the degrees $\, \alpha \, $ and $\, \beta $), are preserved 
in a natural way under such shifts. The change in the 
``accessibility areas'' $\, V (x, y) \leqslant V_{0} \, $ during 
the shifts of the maxima of the standing waves is accompanied 
here by their rather complex movement, as well as numerous 
rearrangements at their boundaries. Generally speaking, the 
transportation of atomic gas with an adiabatic change in the 
positions of the maxima of standing waves in this situation 
must be computed separately for each such potential.

 In this work, we will try to represent the described level 
lines and the corresponding areas of lower potential values in 
the most visual way. In addition, as we said above, we will be 
interested in the dynamics of ultracold atoms in the regions we 
have described. Especially interesting in this case, in our opinion, 
is the superposition of the properties of chaotic dynamics itself 
and the chaotic properties of the ``accessibility regions'' for 
this dynamics.

\vspace{1mm}

\begin{figure}[t]
\begin{center}
\includegraphics[width=\linewidth]{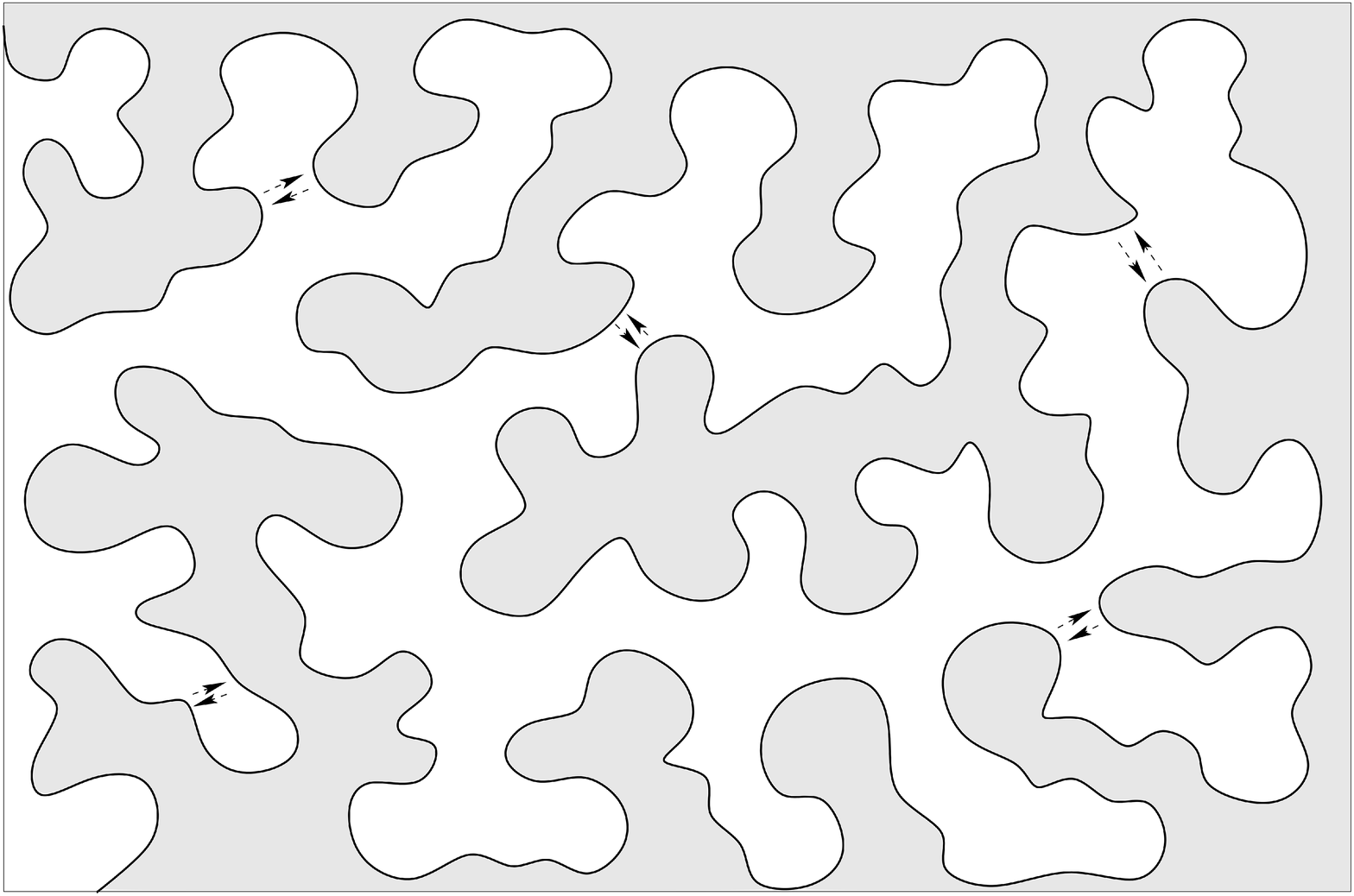}
\end{center}
\caption{Tunneling effects between different parts of the 
classical ``accessibility regions'' as they approach each 
other, for potentials with three quasi-periods. }
\label{Fig8}
\end{figure}

 Here we would like to note one more important property of 
the ``chaotic'' level lines of potentials  
$\, V (x, y, {\bf U}) \, $ with three quasi-periods, namely, 
the presence of their segments where the level line 
(as well as the corresponding region of smaller values
$\, V (x, y, {\bf U}) \,\, \leqslant \,\, V_{0} $)
passes ``very close'' to itself (Fig.~\ref{Fig8}). 
More precisely, by considering larger and larger areas in the 
plane, we can find segments of such a level line that are 
arbitrarily close to each other. As a consequence, when 
considering the semiclassical dynamics of atoms with energies 
close to the corresponding level $\, V_{0} \, $, it is always 
necessary to consider also the effects of tunneling from one 
part of the ``accessibility region'' to another near such places.

 Another consequence of the above circumstance is that, 
in contrast to the situation in stability zones, here the 
global geometry of the domains
$\, V (x, y, {\bf U}) \,\, \leqslant \,\, V_{0} \, $
is unstable with respect to arbitrarily small local variations 
of the potential $\, V (x, y) \, $ and can vary greatly 
(on large scales) in the presence of arbitrarily small 
perturbations or defects. As a consequence, the transport 
properties of particles in such potentials can also strongly 
depend on the presence of such defects in the plane of the 
potential. In the presence of additional slowly varying 
potentials, for example, of the type
$$V (x, y) \,\,\, = \,\,\, a \, x^{2} \,\,\, , \quad 
a \, \rightarrow \, 0 \,\,\, , $$
the global geometry of open level lines and the corresponding 
regions of lower values of the resulting potential will most often 
resemble the corresponding geometry for a smooth potential on large 
scales and behave in a typical ``chaotic'' manner on small scales.

\vspace{1mm}

\begin{figure}[t]
\begin{center}
\includegraphics[width=\linewidth]{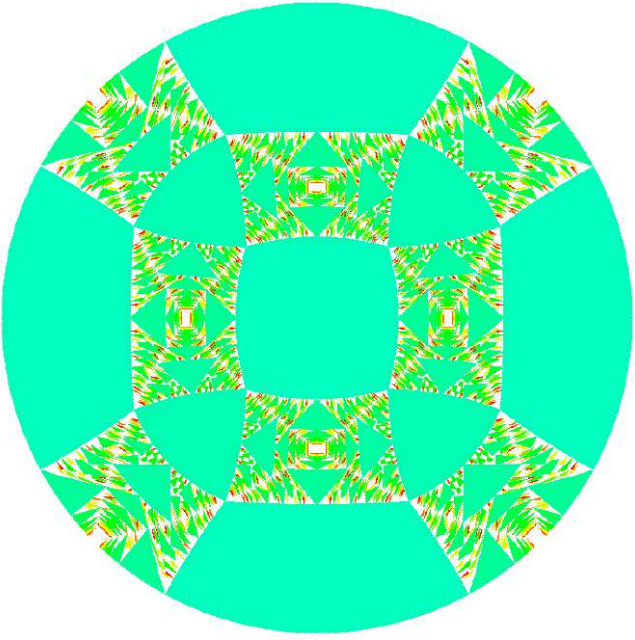}
\end{center}
\caption{Stability zones for quasiperiodic potentials in the 
plane obtained by restricting the potential 
$\, \cos X^{1} \, + \, \cos X^{2} \, + \, \cos X^{3} \, $
to all possible affine embeddings
$\, \mathbb{R}^{2} \rightarrow \mathbb{R}^{3} \, $
(\cite{DeLeoPhysLettA,DeLeoPhysB}). The zones are domains on the 
unit sphere formed by the endpoints of the unit vectors orthogonal to 
the direction of the embedding.}
\label{Fig9}
\end{figure}

 According to the general conjecture of S.P. Novikov, in the 
case of three quasi-periods, the emergence of potentials with 
chaotic level lines can occur only at a set of measure zero and, 
moreover, of the fractal codimension strictly greater than one, 
in the full space of parameters. It can be seen, therefore, that 
experimental construction of a potential with three 
quasi-periods which is close in properties to a random one in the 
sense described above requires a very special choice of parameters 
specifying such a potential. Fig.~\ref{Fig9} gives an 
example of the location of stability zones in the space of 
essential parameters under the restriction of the potential 
$$ \cos \, X^{1} \,\, + \,\, \cos \, X^{2} \,\, + \,\, 
\cos \, X^{3} $$
to two-dimensional planes for all possible affine embeddings 
$\, \mathbb{R}^{2} \rightarrow \mathbb{R}^{3} \, $.
In this case, only the direction of the embedding is essential, 
which can be specified by a unit vector in
$\, \mathbb{R}^{3} \, $ orthogonal to the corresponding plane
$\, \mathbb{R}^{2} \, $. The endpoints of such vectors lie on the unit 
sphere $\, \mathbb{S}^{2} \, $, and thus all stability zones can be 
viewed as regions on the unit sphere. It can be seen that the 
union of (an infinite number of) such regions defines a rather 
complex set on the sphere, and the complement to it has the
properties of a fractal. We also note here that Novikov's conjecture 
has not yet been rigorously proven, although it has been confirmed in 
a number of serious numerical experiments.

\vspace{1mm}

  Let us now formulate the analytical results known to date 
for potentials with four quasiperiods. It must be said at once 
that Novikov's problem for the case of four quasi-periods is 
more complicated in comparison with the case of three quasi-periods. 
At the same time, the case of four quasi-periods may turn out to be 
very important in the formulation we are considering in connection 
with the problem of modulation of two-dimensional quasicrystals in 
systems of cold atoms.

\begin{figure}[t]
\begin{center}
\includegraphics[width=\linewidth]{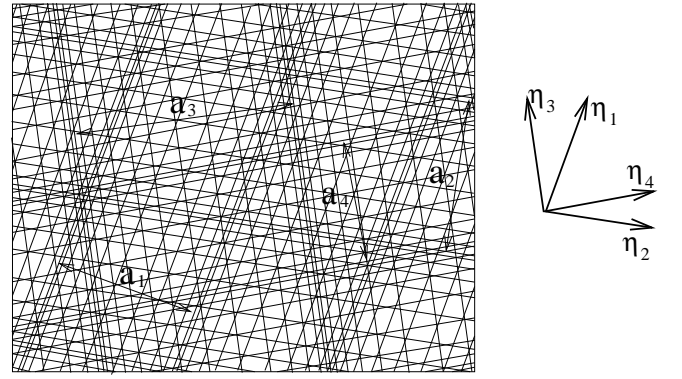}
\end{center}
\caption{Superposition of four standing waves in the plane 
with the formation of a potential with four quasi-periods 
(schematically). (The vectors $\, \bm{\eta}_{i} \, $ indicate 
the directions of the wave fronts, and the vectors 
$\, {\bf a}_{i} \, $ are the shifts between the maxima of 
their amplitudes).}
\label{Fig10}
\end{figure}

 In the setting we are considering, potentials with four 
quasi-periods are obtained as a result of a superposition of four 
independent sinusoidal standing waves (possibly with the generation 
of higher harmonics, Fig.~\ref{Fig10}) and can be written in the 
following general form 
$$V ({\bf r}) \,\,\, = \,\,\,  \sum_{i=1}^{4} \,
V_{i} \cos ({\bf k}_{(i)} {\bf r} + \delta_{i}) 
\,\,\, + \,\,\, \dots  $$

 The quasiperiodic properties of the potential 
are determined by the parameters
$\, ( {\bf k}_{(1)}, {\bf k}_{(2)}, {\bf k}_{(3)},
{\bf k}_{(4)} ) \, $ specifying the affine embedding
$\, \mathbb{R}^{2} \rightarrow \mathbb{R}^{4} \, $,
according to the formulas 
$$ \left(
\begin{array}{c} 
x \cr 
y 
\end{array}
\right)
\quad \rightarrow \quad
\left( 
\begin{array}{c}
k_{(1)}^{1} x + k_{(1)}^{2} y + \delta_{1}   \cr
k_{(2)}^{1} x + k_{(2)}^{2} y + \delta_{2}   \cr
k_{(3)}^{1} x + k_{(3)}^{2} y + \delta_{3}   \cr
k_{(4)}^{1} x + k_{(4)}^{2} y + \delta_{4}
\end{array}
\right) 
\, = \,
\left( 
\begin{array}{c}
{\bf k}_{(1)} {\bf r} + \delta_{1}   \cr
{\bf k}_{(2)} {\bf r} + \delta_{2}   \cr
{\bf k}_{(3)} {\bf r} + \delta_{3}   \cr
{\bf k}_{(4)} {\bf r} + \delta_{4}
\end{array}
\right) .
$$

 We present here some simplified consequences from the 
results of \cite{NovKvazFunc,DynNov}, which are analogous 
to the first of the statements formulated above for the case 
of three quasi-periods. As before, we will not consider here 
in detail all the regularity conditions (of general position) 
imposed in the works \cite{NovKvazFunc,DynNov},  
and we will assume that they are always satisfied for real 
potentials. Then, from the results of \cite{NovKvazFunc,DynNov}, 
the following statement follows:

\vspace{1cm}

 The complete space of parameters
$${\bf U}_{0} \, = \, ({\bf k}_{(1)}^{0}, {\bf k}_{(2)}^{0},
{\bf k}_{(3)}^{0}, {\bf k}_{(4)}^{0},
V_{1}^{0}, V_{2}^{0}, V_{3}^{0}, V_{4}^{0}, \dots ) \,\,\, , $$
defining potentials with four quasi-periods contains an everywhere 
dense set of open domains (``stability zones'' $\, \Omega $) such 
that in each of the domains $\, \Omega \, $ the following holds:

\vspace{1mm}

C1. open level lines of $\, V (x, y, {\bf U}) \, $
exist in a finite connected energy interval
$\, [V_{1} ({\bf U}) , V_{2} ({\bf U})] $:
$$V_{\min} ({\bf U}) \,\,\, < \,\,\, 
V_{1} ({\bf U}) \,\,\, \leqslant \,\,\, V_{2} ({\bf U})
\,\,\, < \,\,\, V_{\max} ({\bf U}) \,\,\, ; $$

\vspace{1mm}

C2. all open level lines of $\, V (x, y, {\bf U}) \, $
lie in straight strips of finite width and pass through 
them (Fig.~\ref{Fig3}); 

\vspace{1mm}

C3. the mean direction $\, {\bf l} ({\bf U}) \, $  
of strips containing open level lines of the potential
$\, V (x, y, {\bf U}) \, $ is defined in the entire domain
$\, \Omega \, $ by some (irreducible) integer quadruple
$\, (m^{1}, m^{2}, m^{3}, m^{4}) \, $ by the relation
$$\left( m^{1} {\bf k}_{(1)} \, + \, m^{2} {\bf k}_{(2)} \, + \,
m^{3} {\bf k}_{(3)} \, + \, m^{4} {\bf k}_{(4)}
\, , \, {\bf l} ({\bf U}) \right) \,\,\, = \,\,\, 0 .$$

\vspace{1mm}

 As in the case of three quasiperiods, the quadruples
$\, (m^{1}, m^{2}, m^{3}, m^{4}) \, $ are actually of topological 
origin and can be defined similarly to the earlier definition
of the triples~$(m^1,m^2,m^3)$ given in terms of the shift 
transformations of the parameters $\, \delta_{i} \, $. 

\vspace{1mm}

 Also, as in the case of three quasi-periods, the union of 
the stability zones (in the general case) does not cover here the entire 
parameter space. The complement to this union forms a complex 
set parametrizing potentials with chaotic level 
lines. It must be said that both the features of the chaotic 
behavior of the level lines and the structure of the corresponding 
set in the space of parameters almost have not been studied to date. 
In particular, it can be expected that the set of parameters 
corresponding to potentials with chaotic level lines has a nonzero 
measure here. Thus, in families of potentials with four quasi-periods, 
the construction of potentials with the properties of truly random 
potentials might be simpler from the experimental point of view. 
Note that, as in the case of three quasi-periods, each ``chaotic'' 
potential here is the limit of getting more and
more complicated ``regular'' 
potentials due to the accumulation of an infinite number of 
``stability zones'' near the point $\, {\bf U}_{0} \, $ 
defining this potential.

\vspace{1mm}

When passing to potentials with larger numbers of quasi-periods 
given by the embeddings 
$$ \left(
\begin{array}{c} 
x \cr 
y 
\end{array}
\right)
\,\,\, \rightarrow \,\,\,
\left( 
\begin{array}{c}
k_{(1)}^{1} x + k_{(1)}^{2} y + \delta_{1}   \cr
\vdots   \cr
k_{(d)}^{1} x + k_{(d)}^{2} y + \delta_{d}
\end{array}
\right) 
\, = \,
\left( 
\begin{array}{c}
{\bf k}_{(1)} {\bf r} + \delta_{1}   \cr
\vdots  \cr
{\bf k}_{(d)} {\bf r} + \delta_{d}
\end{array}
\right) \,\,\, ,
$$
it should be noted right away that there are currently 
no rigorous analytic results for the case $\, d > 4 \, $. 
It can be seen, however, that in this case too, quasiperiodic 
potentials can have a ``regular'' behavior of open level lines. 
In general, we will say that a potential with $\, d \, $ 
quasiperiods has regular open level lines if the following 
conditions are satisfied:

\vspace{1mm}

D1. Open level lines of $\, V (x, y) \, $
exist in a finite connected energy interval
$$V_{\min} \,\,\, < \,\,\, V_{1} \,\,\, \leqslant \,\,\, 
V (x, y) \,\,\, \leqslant \,\,\, V_{2} \,\,\, < \,\,\, V_{\max} 
\,\,\, ; $$

\vspace{1mm}

D2. All open level lines of $\, V (x, y, {\bf U}) \, $
lie in straight strips of finite width and pass through 
them (Fig.~\ref{Fig3});

\vspace{1mm}

D3. The mean direction $\, {\bf l} \, $ of strips containing 
open level lines of the potential $\, V (x, y) \, $ 
is determined, for some (irreducible) integer vector 
$\, (m^{1}, \dots , m^{d}) \, $, by the relation
$$\left( m^{1} {\bf k}_{(1)} \, + \dots \, + \,
m^{d} {\bf k}_{(d)} \, , \, {\bf l} \right) 
\,\,\, = \,\,\, 0 .$$

\vspace{1mm}

 Potentials with $ \, d \, $ quasi-periods with regular behavior 
of open level lines arise, in particular, whenever they are formed 
by means of sufficiently small (quasiperiodic) additions to 
potentials with a smaller number of quasi-periods (and regular 
behavior of open level lines). But in reality, such situations 
are not limited to this, and it is possible to construct a huge 
number of examples of very complex potentials with a large number 
of quasi-periods and regular behavior of open level lines in the 
sense described above.

\vspace{1mm}

 It can be shown that in the most general case there is a huge 
set of situations when for a generic potential from some family of 
quasiperiodic potentials $\, V (x, y, {\bf U}) \, $ properties 
(D1)-(D3) take place and, moreover, such a situation is locally 
stable. In this case we have a stability zone
$\, \Omega_{(m^{1}, \dots , m^{d})} \, $ in the space of parameters,
such that conditions (D1)-(D3) are satisfied at all its points 
for some values $\, V_{1} ({\bf U}) \, $ and 
$\, V_{2} ({\bf U}) \, $ and the same
$\, (m^{1}, \dots  , m^{d}) \, $. The boundaries of the zone 
$\, \Omega_{(m^{1}, \dots , m^{d})} \, $ are actually determined 
then by the condition
$\, V_{1} ({\bf U}) \, = \, V_{2} ({\bf U}) \, $.

 Returning to the considered method of creating quasiperiodic 
potentials in systems of cold atoms (a superposition of standing 
waves), we can show, as before, that the numbers 
$\, (m^{1}, \dots , m^{d}) \, $ can be determined in a purely 
topological way (by successive shifts of the maxima of the standing 
waves by a period and observation of the corresponding shifts of 
open level lines). 

\vspace{1mm}

 In the general case, it can be stated that when creating 
quasiperiodic potentials it is natural to divide them, according 
to the behavior of their level lines, into potentials that 
retain certain properties of ordered potentials, and potentials 
approaching random potentials. The potentials of the first type 
are stable and arise in some open regions in the space 
of their parameters, while each of these regions is determined 
by the respective value of the topological invariant---an integer vector 
$\, (m^{1}, \dots , m^{d}) \, $. Potentials of the second type 
are unstable and arise in rather complex sets of fractal type 
(the complements to the union of domains 
$\, \Omega_{(m^{1}, \dots , m^{d})} $). To construct potentials 
with the properties of random potentials, it is therefore necessary 
to fix a set of domains
$\, \Omega_{(m^{1}, \dots , m^{d})} \, $ 
in the parameter space. Considering the potentials in the 
complement, it can be expected that the corresponding potentials 
with a large number of quasi-periods can, in fact, serve as one of 
the models of random potentials, since the complexity of the behavior 
of their open level lines increases very rapidly with an increase in 
the number of quasi-periods. In conclusion of this section, we note 
that the emergence of random potentials has also been considered in systems 
of optical lattices for ultracold atoms 
(see, for example \cite{HorCourtGryn,Boiron}).

\section{Particle dynamics in potentials of different types}
\setcounter{equation}{0}

  The main subject of this section is the dynamics 
of ultracold atoms in a two-dimensional plane in the presence of 
a quasiperiodic potential $\, V (x, y) \, $. In the leading 
approximation, such dynamics can be considered in a classical way 
provided it is assumed that the atoms have fairly well-defined trajectories 
in $\, \mathbb{R}^{2} \, $. An exception in this case may be the 
motion of atoms at special segments of trajectories, where quantum 
tunneling from one segment of the trajectory to another can play 
an important role. As we have already said, we will be primarily 
interested in the dynamics of atoms with energies corresponding 
to the presence of open level lines of the considered potentials. 
To apply the semiclassical description, we must therefore assume 
the relation $\, h / \sqrt{2MV} \ll a \, $, where $\, a \, $ is
is the typical value of the periods of standing waves used in
experiment. In our situation we assume also that a sufficiently large 
number of atoms in the ensemble have energies corresponding to the 
presence of open level lines, which also implies the relation 
$\, T \simeq V \, $. Thus, in our case, we must also put 
$\, h / \sqrt{2MT} \ll a \, $. In the general case, the forces acting 
on atoms in optical traps can contain both conservative and dissipative 
parts (see, for example
\cite{GordonAshkin,DalibardCohen-Tannoudji,MinoginLetokhov,MechanicalAction}).
We will consider here the classical dynamics of noninteracting atoms 
in the nondissipative limit, which gives a good approximation to the 
real dynamics of heavy atoms in many important situations.

 As can be seen in this case, the features of the geometry of 
quasiperiodic potentials should manifest themselves most of all in 
the dynamics of particles at energies lying in the intervals of the 
presence of open level lines of the potential. Here we will focus 
specifically on the study of such dynamics. It is easy to see that 
the features of this dynamics should naturally manifest themselves 
in the transport properties of an ultracold gas in the presence of 
particles of the corresponding energies in the ensemble.

  As we saw in the previous section, quasiperiodic potentials 
can in fact be divided into two types according to the behavior 
of their open level lines. At the same time, this division is 
most directly related to the dynamics of particles in such 
potentials, since it determines the geometry of the 
``regions of accessibility'' for particles with certain energies. 
As a consequence of this, one can expect a difference in the dynamics 
of particles in the potentials of these two types, which is observable 
in the study of systems of ultracold atoms.

 As we mentioned above, dynamics in two-dimensional Hamiltonian 
systems has one more feature, namely, it is integrable at low energy 
levels and becomes chaotic at high energies (see \cite{Ott,LikhtLib}). 
At intermediate energy levels, the phase space of a system is divided 
into regions where the integrable case takes place, and domains where 
chaotic dynamics occurs. In particular, one can observe the effects 
of gradual chaotization of dynamics, when a particle can ``stick'' 
to invariant two-dimensional tori for a long time, performing rare 
``jumps'' (Levy flights) between different tori, as well as other 
similar effects. This feature of the dynamics has been studied in systems 
of cold atoms both for free atoms moving in periodic potentials 
(see \cite{HennequinVerkerk1,HennequinVerkerk2}), and in the presence 
of additional interaction of the atom motion with internal degrees of 
freedom (see \cite{ArgonovPrants,Prants2020}).

 In this work, we do not consider the internal degrees of freedom 
of atoms and focus only on the motion of the atom as a whole. As we 
have already said, we will be interested here in the dynamics of atoms 
with energies corresponding to the appearance of open level lines at 
the corresponding potentials. This is the area in which it is
especially interesting for us to observe the combination of different 
modes (integrability, its complication, and transition to chaotic 
dynamics) in different parts of the phase space. It can be immediately 
noted that, since open level lines of a quasiperiodic potential 
also appear at some ``intermediate'' (between the minimum and maximum) 
values, the ``gradual chaotization'' of dynamics described above 
will often arise precisely near such level lines. It can also be said that 
such chaotization must, of course, correlate with the nontrivial 
geometry of the ``accessible regions'' of the motion of atoms in the 
potentials we are considering. In particular, for developed chaotization, 
the motion of atoms should be close to diffusion, so that the transport 
properties of an atomic gas are determined by the diffusion of atoms in 
regions of a given geometry. Under conditions of ``intermediate'' 
chaotization, one can expect that the geometry of the accessibility 
regions significantly affects the Levy flights between two-dimensional 
tori, which, of course, is also determinant when considering the transport 
properties of an atomic gas. In addition, in the situation under 
consideration, two-dimensional tori corresponding to integrable 
dynamics can cut (three-dimensional) manifolds of constant 
energy in a nontrivial way, so that we can also observe non-integrable 
dynamics localized in different parts of the ``accessibility regions''. 
The geometry of such areas and their location is, of course, also 
related to the general geometry of the ``regions of accessibility'' 
for a given particle energy. On the whole, as it is easy to see, 
in the model we are considering, it is the diffusion or Levy flight 
regimes that determine the transport properties of an atomic gas 
that can make it possible to observe the differences between 
the ``regular'' and ``chaotic'' quasiperiodic potentials introduced 
above.

 Below we give the results of a numerical study of the classical 
dynamics of particles in the potentials consedered here. 
As a model, we consider here potentials with three 
quasiperiods obtained by the restrictions of the potential
$$V \left( \cos X^{1} \, + \, \cos X^{2} \, + \, \cos X^{3} \right) $$
to the planes defined by different embeddings (\ref{AffEmb}). 
It must be said that, from the point of view of Novikov's problem, this 
family of potentials contains absolutely all situations that can occur 
for potentials with three quasi-periods; namely, we will find here 
both stable ``regular'' potentials with very different features of the 
geometry of open level lines, and a rich set of ``chaotic'' potentials.
It is worth noting just one feature of the potentials from this 
set, which is as follows. For any ``regular'' potential arising 
in the described family, the interval of existence of open level lines 
is symmetric with respect to zero, i.e. we always have the relation 
$\, V_{1} ({\bf U}) \, = \, - \, V_{2} ({\bf U})\, $ for the values 
$\, V_{1} ({\bf U})\, $ and $\, V_{2} ({\bf U})\,$ introduced above.
Similarly, for all ``chaotic'' potentials of this family, open level 
lines arise exactly at the zero energy level ($V_{0} = 0$). This 
feature is specific for this family of potentials and, generally 
speaking, does not take place in the most general case. In all other 
respects, the geometric properties of the level lines of potentials 
from the above family reflect the most general situation. Since, as we 
have already said, we are interested in the dynamics of particles in 
the areas of the appearance of open potential level lines, we will 
often study such a dynamics here at the value $\, \epsilon = 0 \, $.

 Thus, all potentials will have the form
\begin{multline}
\label{GeneralFormula}
V(x, y) \,\,\, = \,\,\, 
\cos \left( a_{1} x + b_{1} y + c_{1} \right) \,\, +  \\
+ \,\, \cos  \left( a_{2} x + b_{2} y + c_{2} \right) 
\,\, + \,\, \cos \left( a_{3} x + b_{3} y + c_{3} \right) 
\end{multline}
with some coefficients $\, a_{i}$, $\, b_{i}$ and $\, c_{i}$.
In fact, since the choice of the direction of the coordinate 
axes in the plane $\, \mathbb{R}^{2} \, $ will not matter much 
to us, we will assume that the axis $\, x \, $ coincides with the 
intersection line of $\, \mathbb{R}^{2} \, $ with the plane 
$\, (X^{1}, X^{2})\, $ in $\, \mathbb{R}^{3} \,$. 
So we can always put here $\, a_{3} = 0 \, $.

 As we have already mentioned above, the type of the potential is 
determined in our case only by the direction of the embedding
$\, \mathbb{R}^{2} \rightarrow \mathbb{R}^{3} \, $
and can be obtained from the diagram shown in Fig.~\ref{Fig9}.
It can be seen that by selecting the embedding parameters,
we can easily implement any of the situations interesting 
for us.

\vspace{1mm} 
 
 To compare the dynamics in potentials of different types, 
we present here the results for three potentials, the first 
two of which have ``regular'' open level lines and belong 
to rather large stability zones in Fig.~\ref{Fig9}, and the 
third has ``chaotic'' open level lines, which appear only at the
zero energy value. 

\vspace{1mm}

\begin{figure}[t]
\begin{center}
\includegraphics[width=0.9\linewidth]{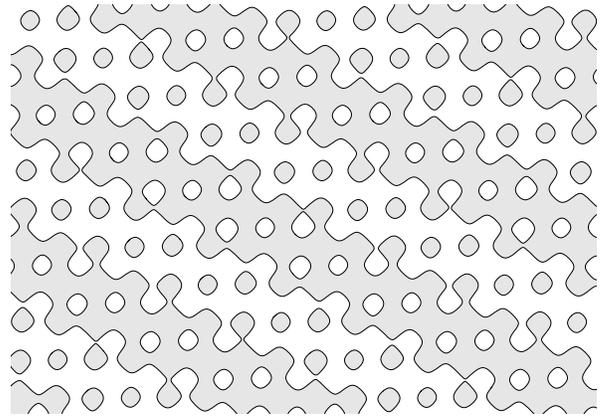}  
\end{center}
\caption{The potential (\ref{Potentsial1}) with ``regular'' 
open level lines from the largest stability zone in Fig.~\ref{Fig9}. 
(The filled areas correspond to the values $\, V (x, y) \leqslant 0 $).}
\label{Fig11}
\end{figure}

 Our first series of computations refers to the potential
 with coefficients
\begin{multline}
\label{Potentsial1}
a_{1} \, = \, - 0.12251993420338196 ,  \\
b_{1} \, = \, -0.2250221718850486 ,  \\
c_{1} \, = \, 1.5505542426422338 ,   \\
a_{2} \, = \, 0.9924660526802913 ,  \\
b_{2} \, = \, - 0.02777898711920925 ,  \\
c_{3} \, = \, 0.12374024573075965 ,   \\
a_{3} \, = \, 0 ,  \\
b_{3} \, = \, 0.9739575709622912 ,  \\
c_{3} \, = \, 3.1548761694687415
\end{multline} 
(Fig.~\ref{Fig11}), which lies inside the largest zone in
Fig.~\ref{Fig9} with topological numbers 
$\, (m^{1}, m^{2}, m^{3}) \, = \, (1, 0, 0) \, $.
It must be said that we are intentionally considering 
separately a potential from this zone here, since the latter
actually differs somewhat from the other zones. Namely, in 
addition to the above-mentioned ``partial'' integrability 
inherent in two-dimensional Hamiltonian systems, here there 
is an additional closeness to the integrable situation, due 
to the fact that the center of this zone corresponds 
to a potential integrable at all energies, which is
$$V (x, y) \,\,\, = \,\,\, \cos x \,\, + \,\, \cos y $$ 
 
 As a consequence of this, integrable dynamics can arise here 
for richer sets of initial conditions than it can for potentials from 
other zones. As we will actually see, this assumption is confirmed; 
in particular, we failed to find here a well-pronounced diffusion 
dynamics at the level $\, \epsilon = 0 \, $. This dynamics arises only 
with a noticeable increase in the particle energy. At the same 
time, the emergence of potentials from this particular zone is 
the most expected in the experiment in comparison with others
due to the significant size of this zone.

\begin{figure}[t]
\begin{center}
\includegraphics[width=0.9\linewidth]{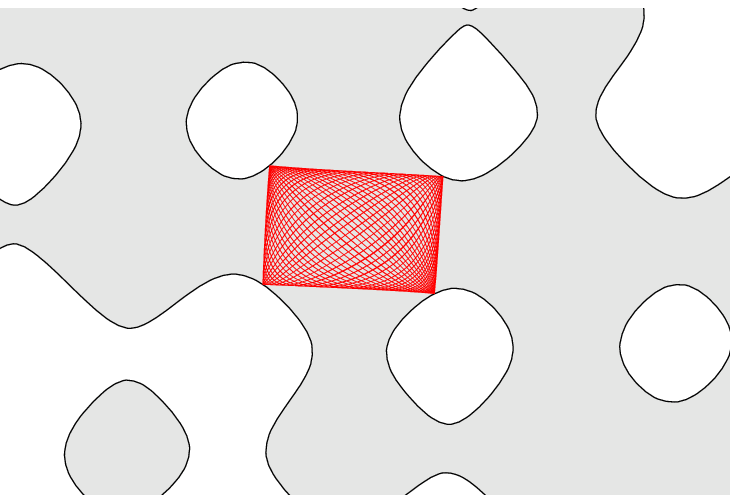}  
\end{center}
\vspace{5mm}
\begin{center}
\includegraphics[width=0.9\linewidth]{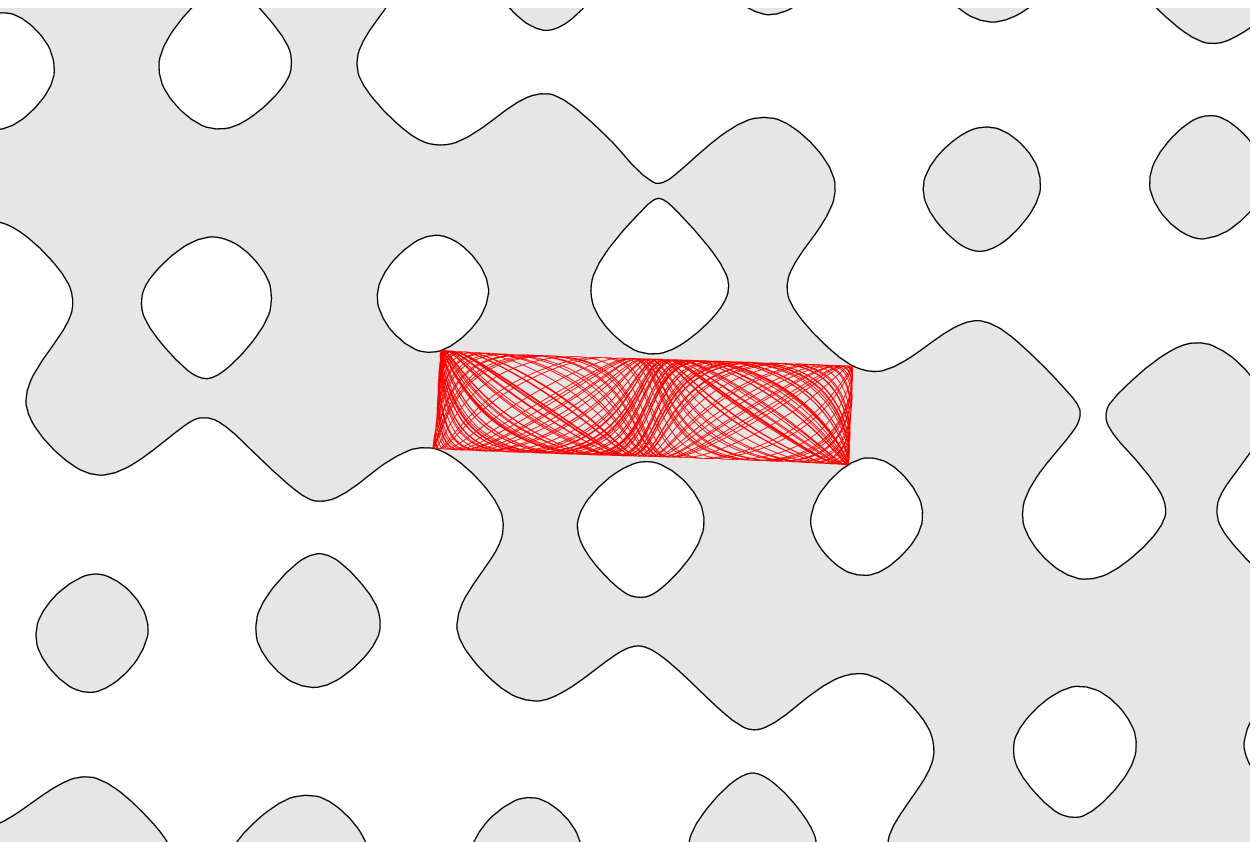}
\end{center}
\vspace{5mm}
\begin{center}
\includegraphics[width=0.9\linewidth]{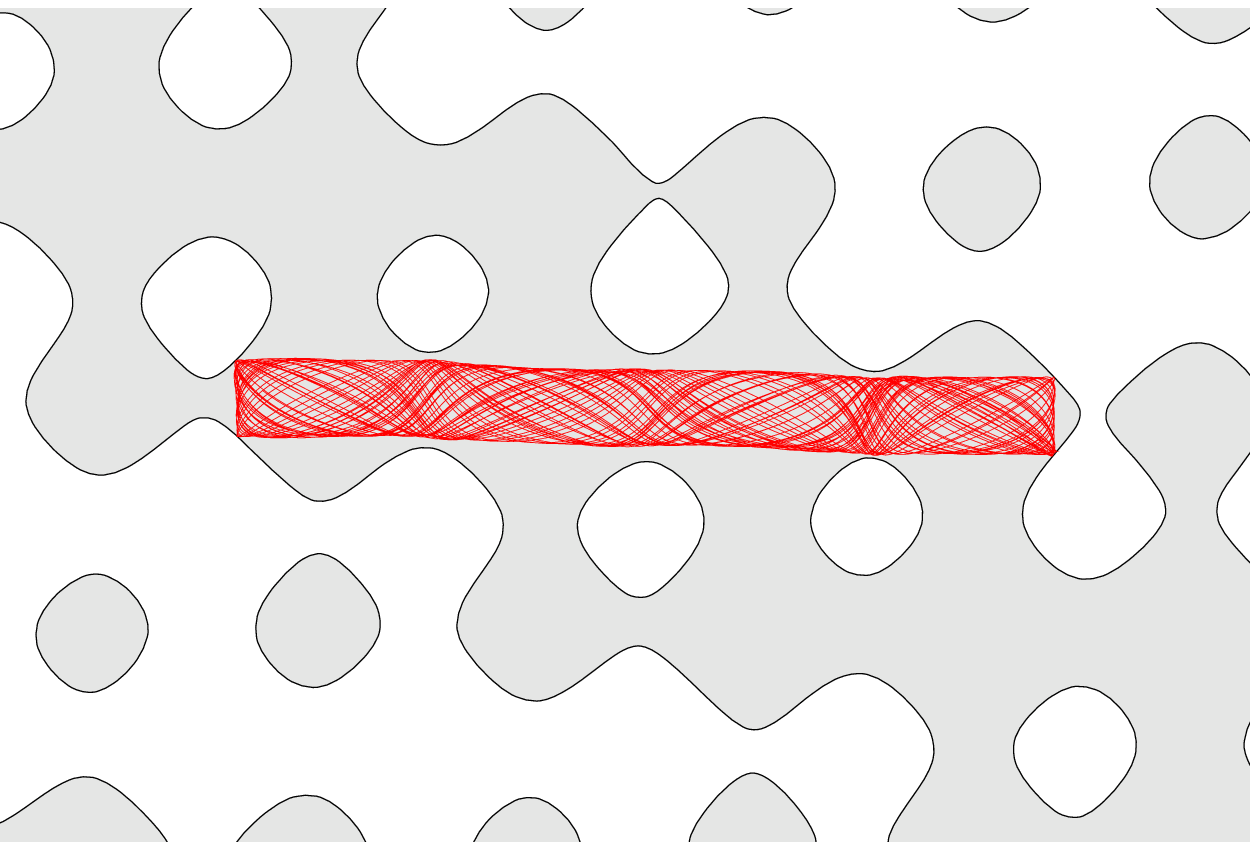}
\end{center}
\caption{Examples of invariant tori corresponding to 
a relatively simple dynamics of atoms in the ``regular'' 
potential (\ref{Potentsial1}) at zero total energy.}
\label{Fig12}
\end{figure}

 As we have already said, we restrict ourselves here to 
considering three-dimensional manifolds in the phase space by fixing 
the total energy of a particle. The potential we are considering 
has open level lines in rather wide energy interval
$$- 0.7493 \,\,\, \leqslant \,\,\, V  \,\,\, \leqslant \,\,\, 0.7493$$
(approximately). We first consider the dynamics of particles 
with energy $\, \epsilon = 0 \, $. As we said above, we can expect 
here the appearance of large regions where the dynamics is in fact 
integrable and occurs on two-dimensional tori embedded in the phase 
space. This is exactly what happens at the level 
$\, \epsilon = 0 \, $, moreover, by carefully choosing the 
initial data, one can find the regions of their values, where 
the dynamics on the tori corresponds to a rather simple motion 
in the coordinate space (Fig.~\ref{Fig12}).

\begin{figure}[t]
\begin{center}
\includegraphics[width=0.9\linewidth]{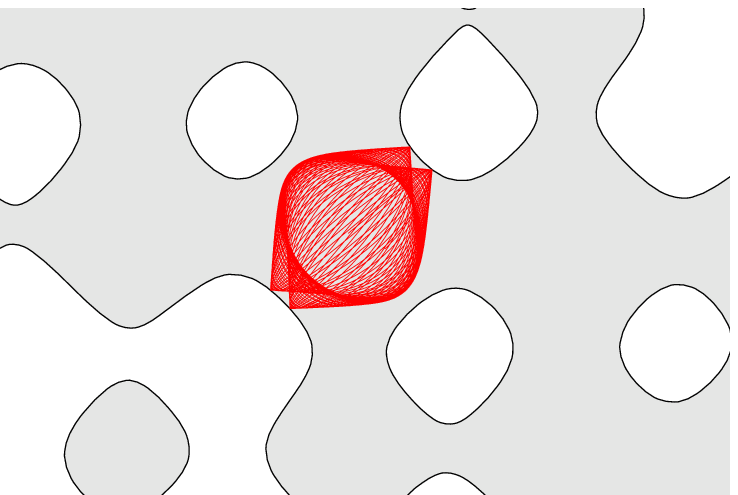}  
\end{center}
\vspace{5mm}
\begin{center}
\includegraphics[width=0.9\linewidth]{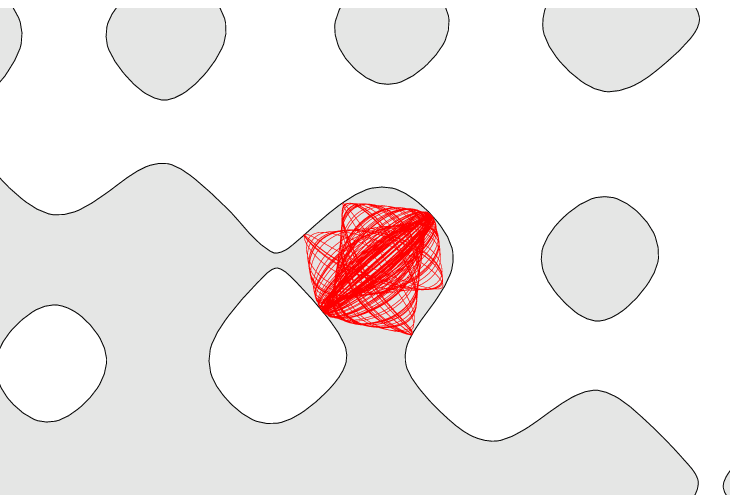}
\end{center}
\vspace{5mm}
\begin{center}
\includegraphics[width=0.9\linewidth]{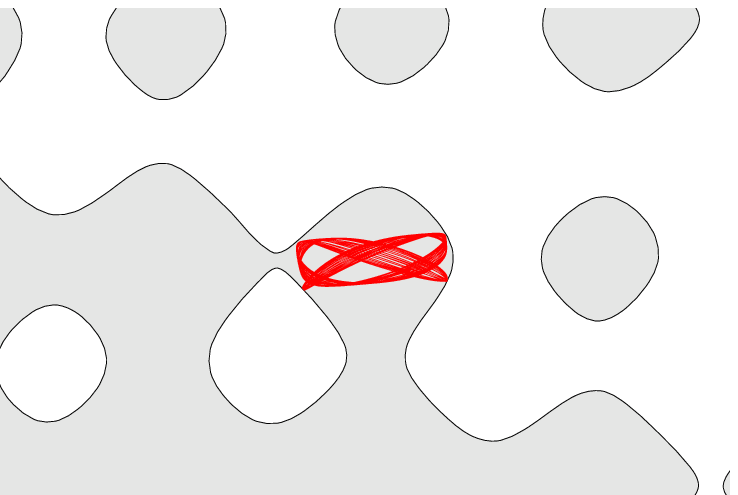}
\end{center}
\caption{Examples of invariant tori defining more complex 
dynamics of atoms in the ``regular'' potential 
(\ref{Potentsial1}) at zero total energy.}
\label{Fig13}
\end{figure}

By varying the initial data at the level $\, \epsilon = 0 \, $ 
(for the same potential), one can, however, discover (smaller) 
regions in which the geometry of invariant tori becomes more 
and more complicated, which also leads to complication of the 
motion of particles in the coordinate space (Fig.~\ref{Fig13}).

\begin{figure}[t]
\begin{center}
\includegraphics[width=0.8\linewidth]{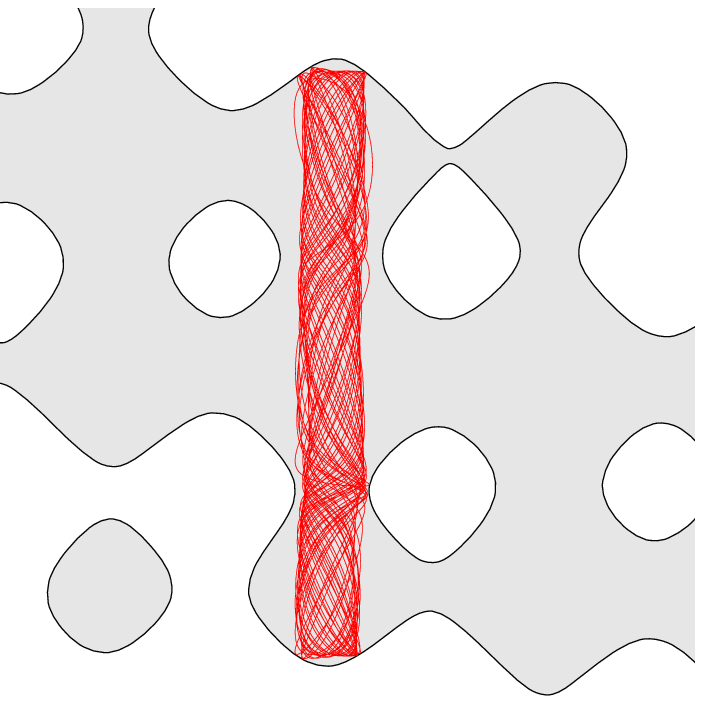}
\end{center}
\vspace{1mm}
\begin{center}
\includegraphics[width=0.8\linewidth]{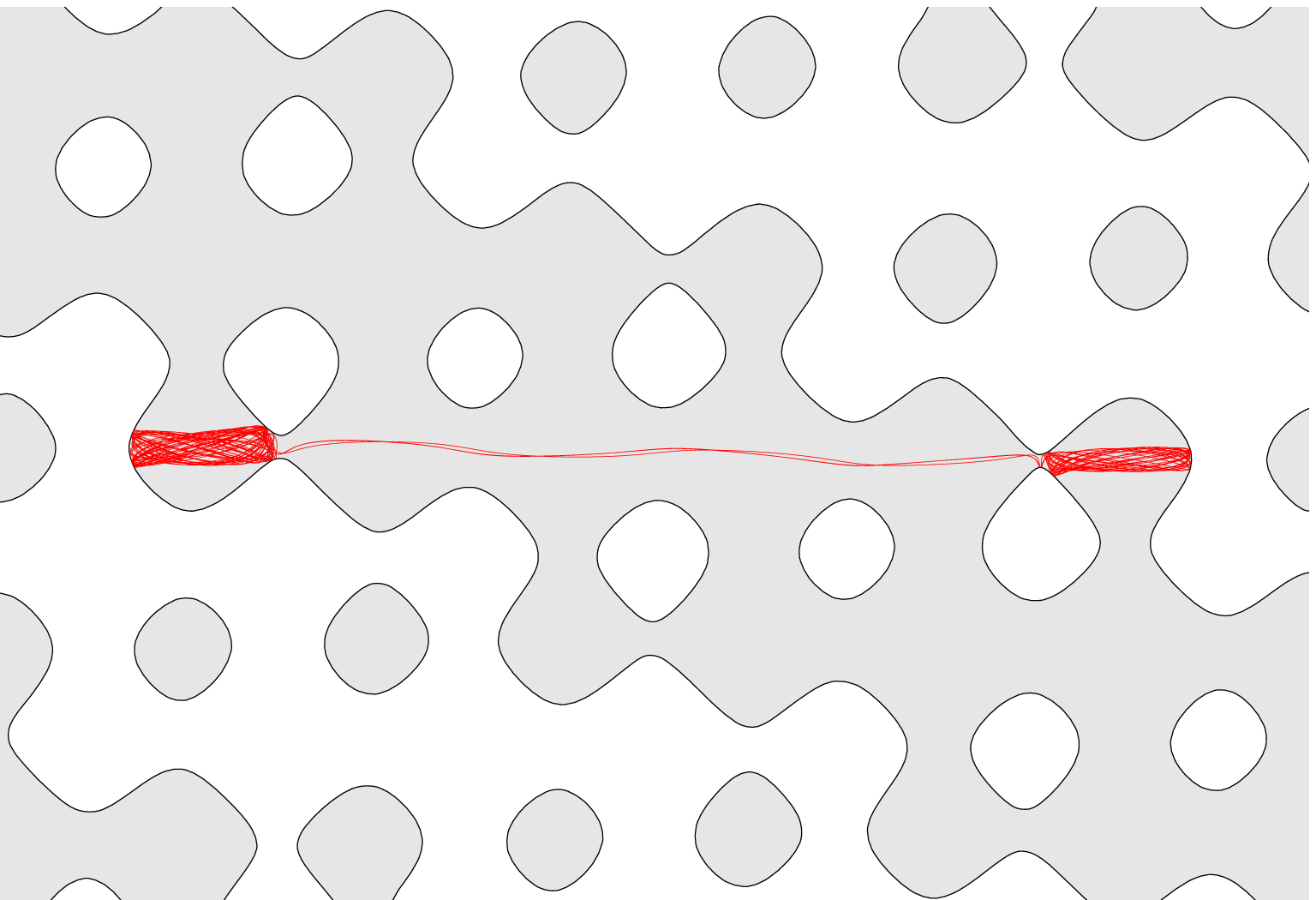}
\end{center}
\vspace{2mm}
\begin{center}
\includegraphics[width=0.8\linewidth]{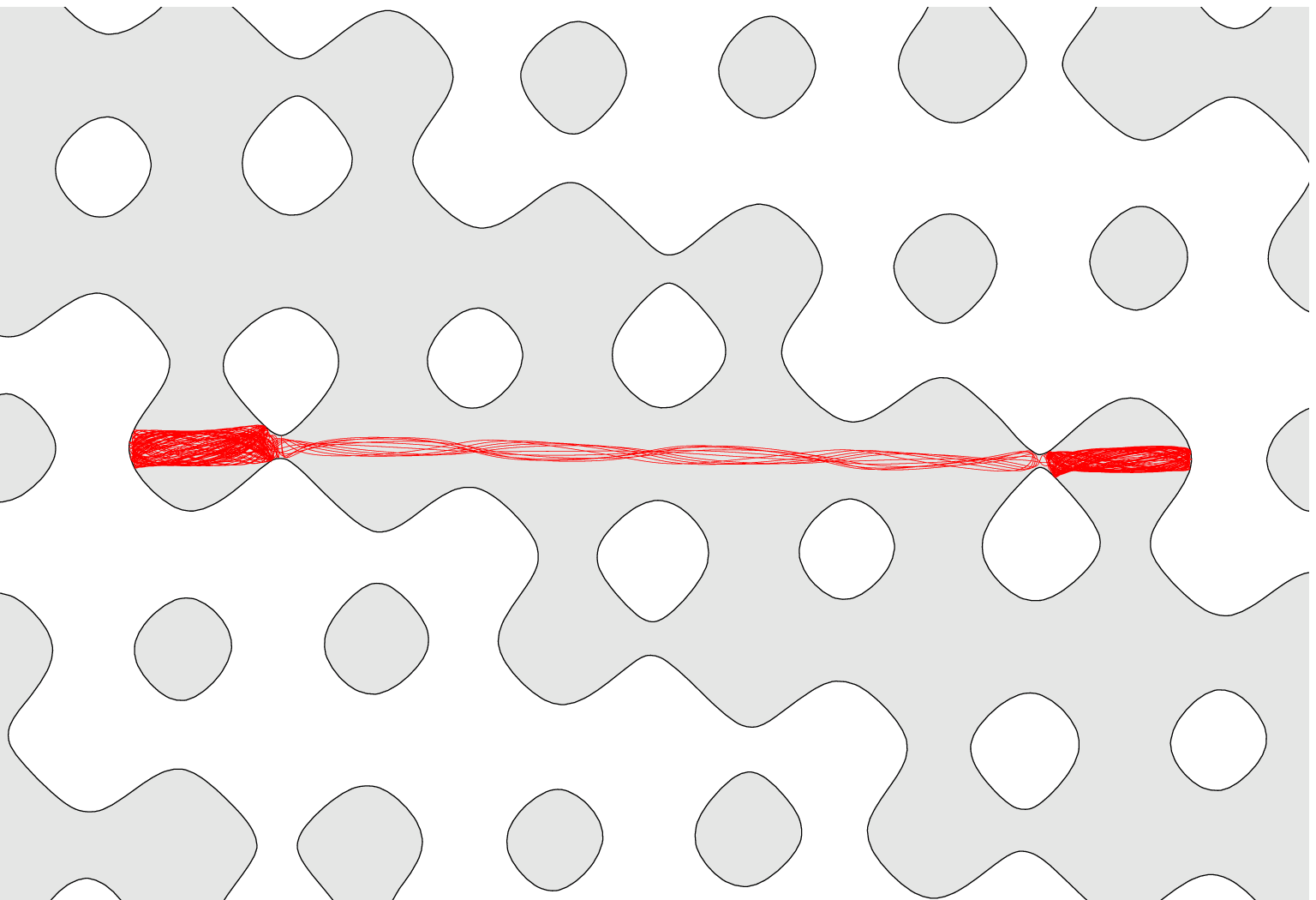}
\end{center}
\caption{Levy flights between ``close'' and ``distant'' 
tori and the complication of invariant tori when the 
initial conditions change in the case of the ``regular'' potential 
(\ref{Potentsial1}) at zero total particle energy.}
\label{Fig14}
\end{figure}

 Choosing the initial data in an even more special way 
(at the level $\, \epsilon = 0 $), one can also see even 
more complicated surgery of invariant tori, where 
particles ``stick'' for a rather long time to simpler 
invariant tori and ``jump'' from one of these tori to 
another (Levy flights) at certain times (Fig.~\ref{Fig14}).

\begin{figure}[t]
\begin{center}
\includegraphics[width=0.9\linewidth]{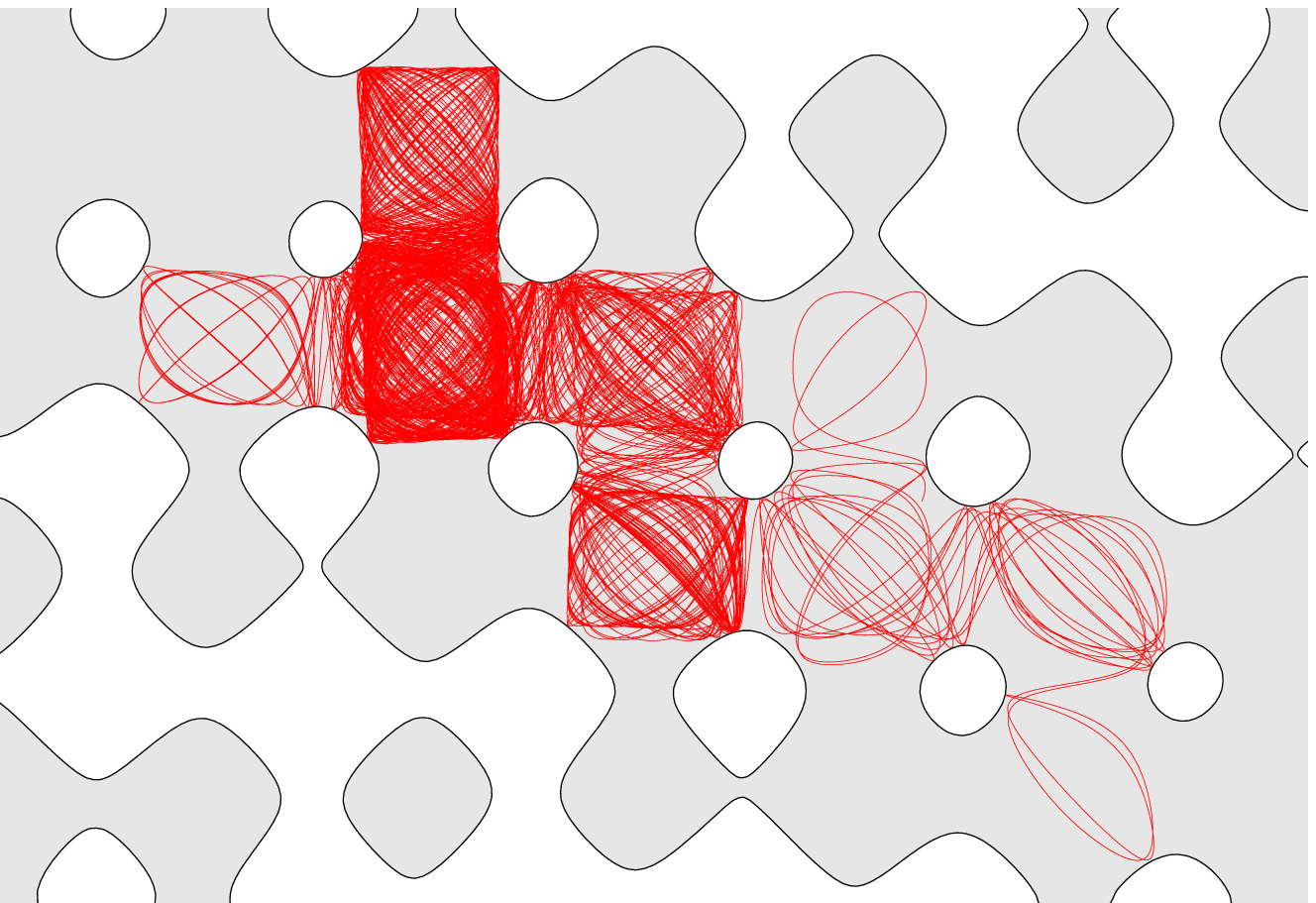}
\end{center}
\vspace{5mm}
\begin{center}
\includegraphics[width=0.9\linewidth]{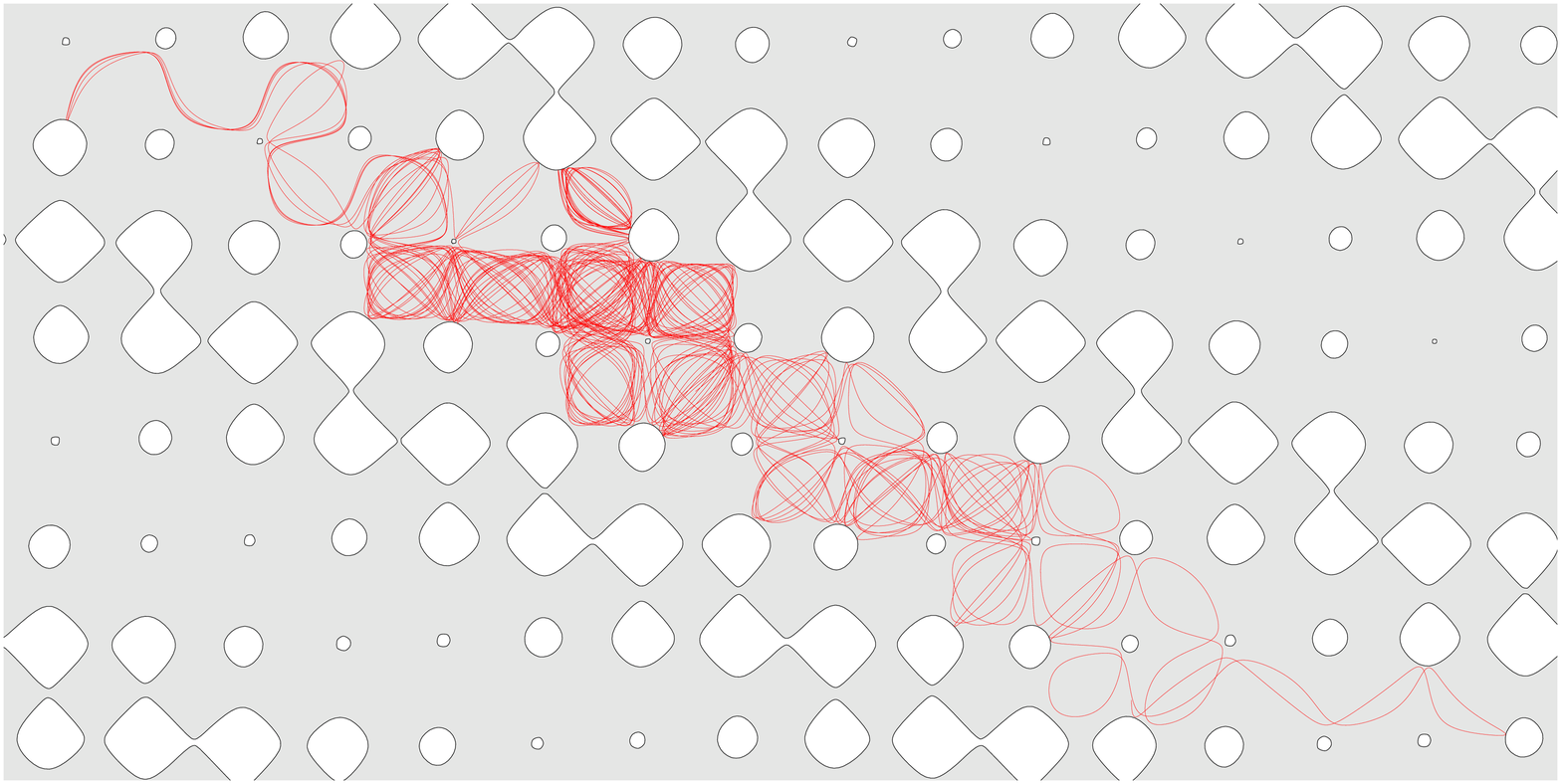}
\end{center}
\caption{Accessibility regions and transition to the 
diffusion regime in the case of the ``regular'' potential (\ref{Potentsial1}) 
with increasing particle energy 
($\epsilon = 0.5$ and $\epsilon = 0.98$).}
\label{Fig15}
\end{figure}

 The regimes shown in Fig.~\ref {Fig12}--\ref{Fig14}, 
correspond to the dynamics of particles with zero total energy. 
As we have already said, for this potential we failed to find 
diffusion regimes at $\, \epsilon = 0 \, $; however, they appear 
with increasing particle energy (Fig.~\ref{Fig15}). In fact, 
a clearly pronounced diffusion behavior arises here at energies 
for which the open level lines of the potential have already 
disappeared, and the region of accessibility extends infinitely 
in two dimensions. It is interesting that the diffusion dynamics 
here retains, nevertheless, a pronounced anisotropy, preserving 
the memory of the mean direction of the open level lines of the 
potential. It can also be noted here that even at these levels 
there remain quite a lot of invariant tori, and the diffusion 
dynamics has, at the same time, the form of Levy flights with 
``sticking'' to invariant tori. As we have already said, this 
behavior is apparently characteristic only of potentials from 
the zone with $\, (m^{1}, m^{2}, m^{3}) \, = \, (1, 0, 0) \, $
(and the zones identical with it) due to the circumstances mentioned 
above. In particular, we will present below a description of the 
dynamics in a potential from another large stability zone, which, 
apparently, is characteristic for most potentials with ``regular'' 
level lines.

\begin{figure}[t]
\begin{center}
\includegraphics[width=0.9\linewidth]{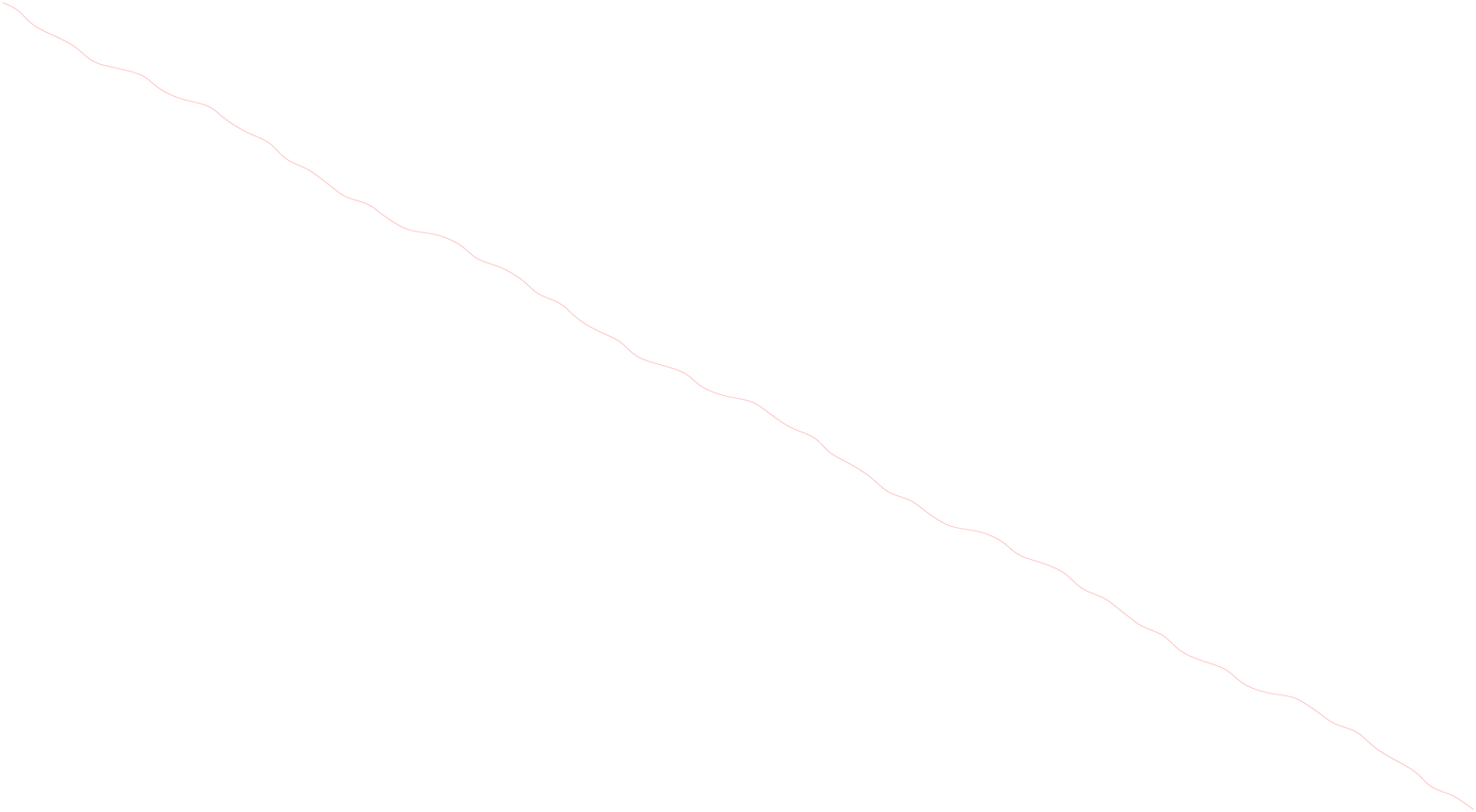}
\end{center}
\vspace{5mm}
\begin{center}
\includegraphics[width=0.1\linewidth]{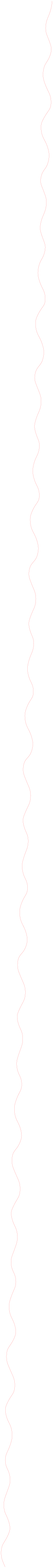}
\end{center}
\vspace{5mm}
\begin{center}
\includegraphics[width=0.9\linewidth]{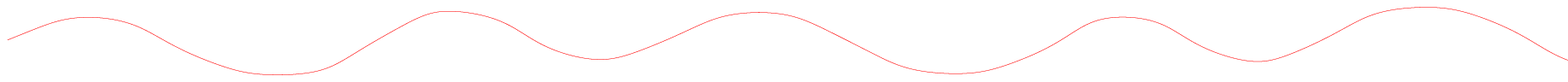}
\end{center}
\caption{The appearance of ballistic trajectories in the 
``regular'' potential (\ref{Potentsial1}) with increasing 
particle energy ($\epsilon = 4$).}
\label{Fig16}
\end{figure}

 With a further increase in energy, the motion of particles 
in the potential gradually changes from diffusion motion to ballistic 
motion. We should note here, however, that ballistic motion in 
quasiperiodic potentials also has, apparently, noticeable
features. In particular, for potentials from the family under 
consideration, even at rather high energies, a fairly large 
part of the phase volume is occupied by ballistic trajectories 
of certain directions. For the potential we are considering, 
three main directions can be distinguished at once, namely the 
directions of the level lines of the three cosines in the 
formula (\ref{GeneralFormula}), along which ballistic motion 
occurs already at sufficiently low energies (Fig.~\ref{Fig16}). 
With increasing energy, the number of such directions increases 
and ballistic trajectories with directions other than the three 
indicated ones can also be observed.

\begin{figure}[t]
\begin{center}
\includegraphics[width=0.9\linewidth]{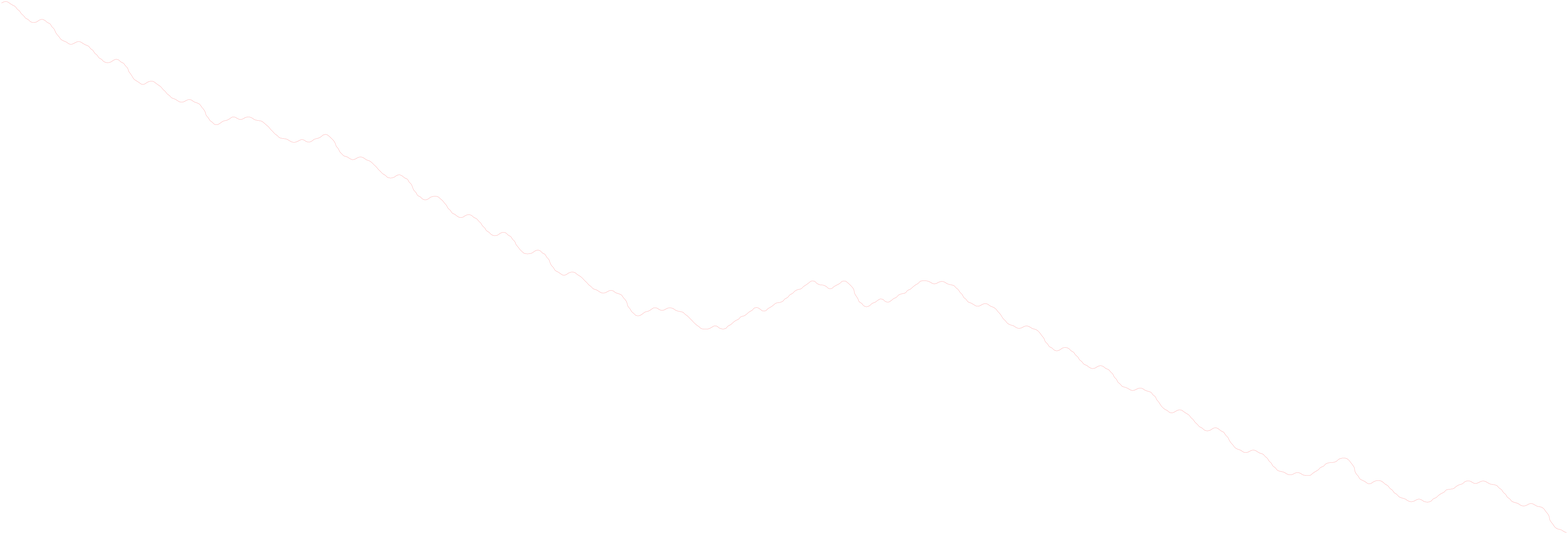}
\end{center}
\vspace{5mm}
\begin{center}
\includegraphics[width=0.9\linewidth]{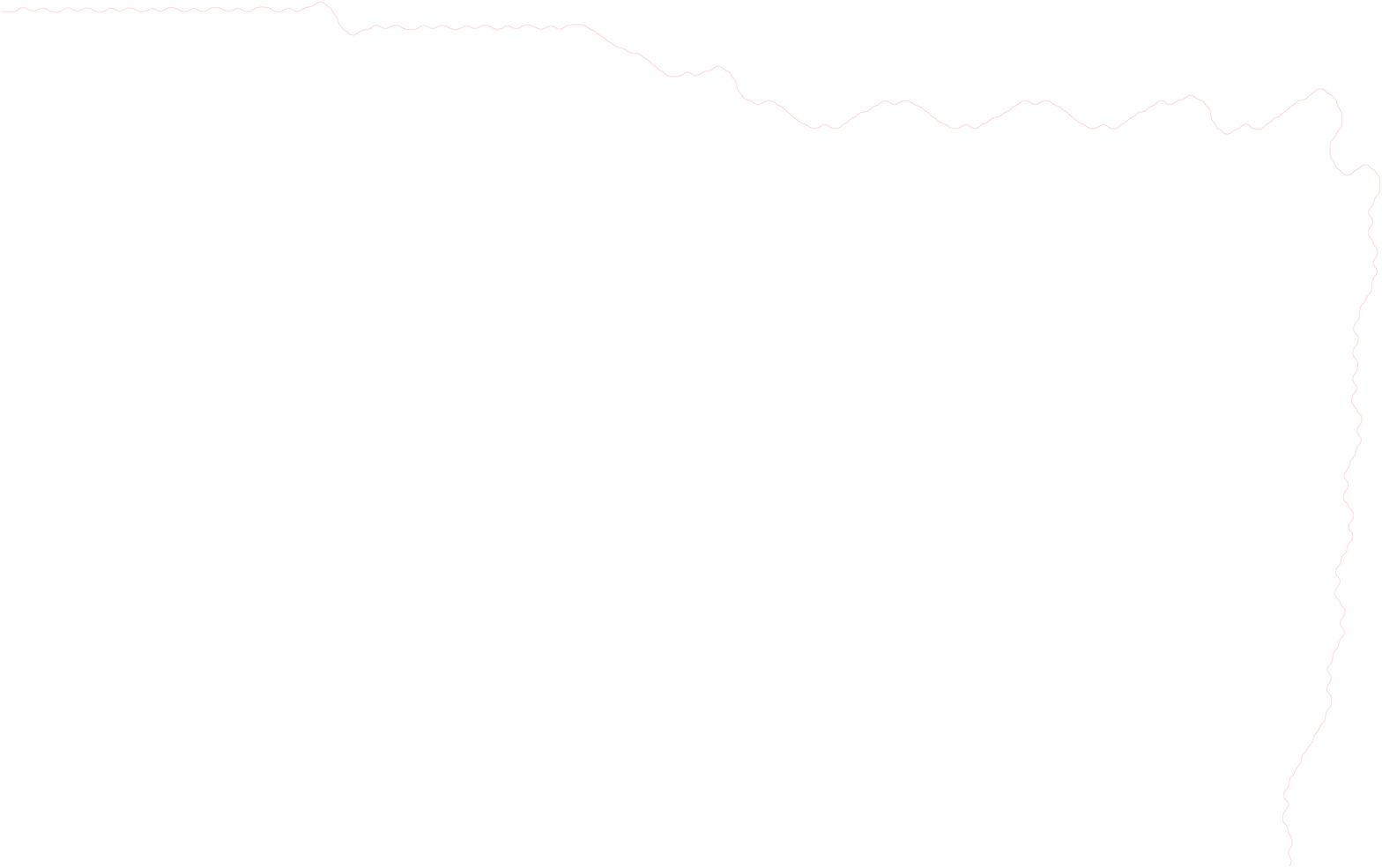}
\end{center}
\vspace{5mm}
\begin{center}
\includegraphics[width=0.9\linewidth]{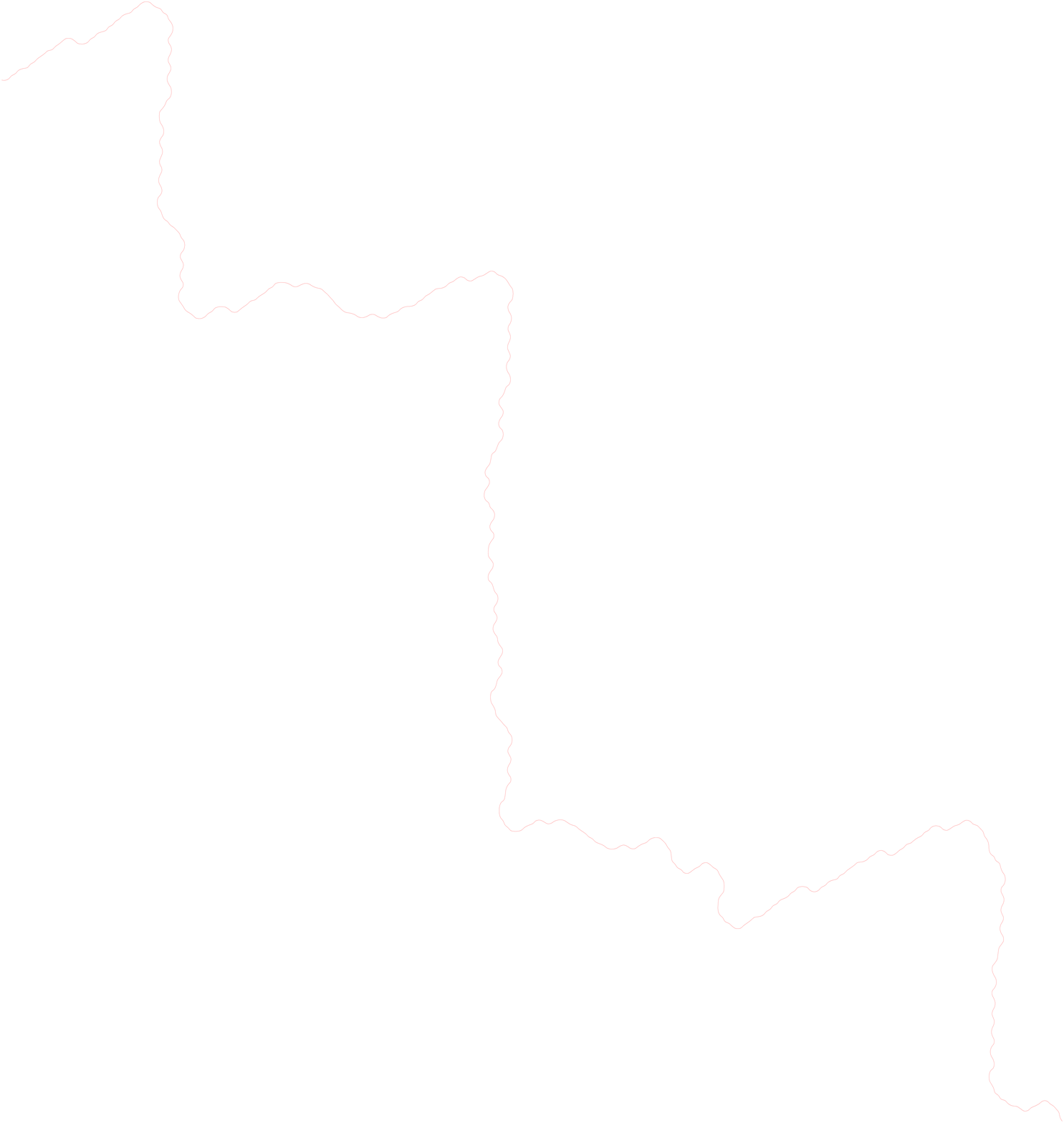}
\end{center}
\caption{``Almost'' ballistic trajectories in the case of a
``regular'' potential (\ref{Potentsial1}) ($\epsilon = 4$).}
\label{Fig17}
\end{figure}

 In addition to the ``purely'' ballistic trajectories described 
above, one can also observe trajectories consisting of long 
ballistic sections of the indicated directions, connected by 
short sections ``switching'' between two directions 
(Fig.~\ref{Fig17}). Like the ``purely'' ballistic trajectories, 
such trajectories should also introduce specific features into 
transport phenomena at the corresponding particle energies. 
It can be seen, however, that with an increase in the number 
of the corresponding ballistic directions, as well as the 
complication of the geometry of the ``quasi-ballistic'' 
trajectories, it will be more and more difficult to detect 
such properties.

 Thus, it can be seen that transport phenomena caused by the 
ballistic motion of atoms also reveal the geometric structure 
of quasiperiodic potentials. In comparison with the geometry
of open level lines, however, this structure is simpler and 
is directly related to the harmonics generating the potential. 
As we will see below, this property of ballistic motion manifests 
itself in fact for potentials of all types, and, in this sense, 
the corresponding transport phenomena almost do not distinguish 
between ``regular'' quasiperiodic potentials from ``chaotic'' 
ones. Also, in contrast to the case of diffusion motion, which 
gives a well-observed contribution to the transport processes 
associated with the geometry of open level lines, the experimental 
observation of transport contributions from ballistic trajectories 
of stable directions can be more complicated due to the addition 
of a large number of such contributions at high energies. 
On the other hand, ballistic motion in quasiperiodic potentials 
is also, apparently, a fundamental property of such potentials, 
in particular, ballistic directions also play an important role 
in quantum dynamics in potentials of this type
(see \cite{ViebSbrosCartYuSchneid}).
 
 As we have already noted above, diffusion dynamics, as well as 
dynamics including distant jumps between different types of 
localized dynamics, which give us the most information about 
the type and topological parameters of a potential, in this 
example arise mainly at energies lying above the interval of 
existence of open level lines. In this case, however, such 
dynamics keeps the ``memory'' of the geometry of open level 
lines of the potential and gives a strongly anisotropic 
contribution of the corresponding direction to transport 
phenomena at these energies. This feature, as we have already 
said above, is apparently associated with ``additional reasons'' 
for the appearance of integrable dynamics in this stability 
zone, leading to an increase in the phase volume filled with 
such dynamics.

\clearpage

\begin{figure}[t]
\begin{center}
\includegraphics[width=0.9\linewidth]{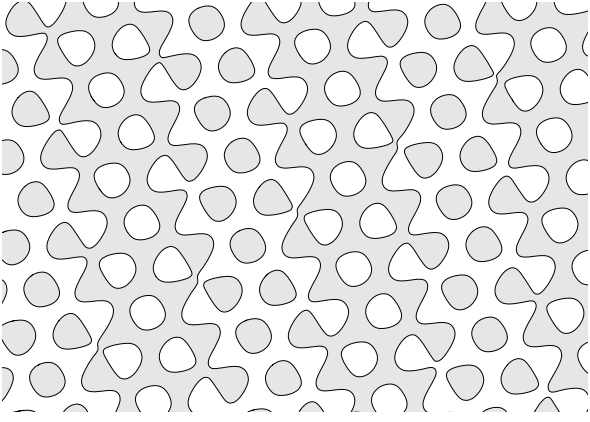}  
\end{center}
\caption{Potential (\ref{Potentsial2}) with ``regular'' open 
level lines from the stability zone in Fig.~\ref{Fig9} 
corresponding to the topological numbers 
$\, (m^{1}, m^{2}, m^{3}) \, = \, (1, 1, 1) \, $.
(The filled areas correspond to the values $\, V (x, y) \leqslant 0$).}
\label{Fig18}
\end{figure}

 The second series of our computations refers to the potential
with coefficients
\begin{multline}
\label{Potentsial2}
a_{1} \, = \, - 0.6194151736623348 ,  \\
b_{1} \, = \, - 0.44502823229775823 ,  \\
c_{1} \, = \, 1.4421279589366298 ,   \\
a_{2} \, = \, 0.7850635914605004 ,  \\
b_{2} \, = \, - 0.3511272752829312 ,  \\
c_{3} \, = \, 0.8986352554278761 ,   \\
a_{3} \, = \, 0 ,  \\
b_{3} \, = \, 0.8238079321117983 ,  \\
c_{3} \, = \, 2.3379002628621635 ,
\end{multline} 
lying inside the zone in Fig.~\ref{Fig9} with topological numbers 
$\, (m^{1}, m^{2}, m^{3}) \, = \, (1, 1, 1) \, $
(Fig.~\ref{Fig18}). Like the previous one, potential 
(\ref{Potentsial2}) has a fairly large energy interval 
$$- 0.7548 \,\,\, \leqslant \,\,\, V \,\,\, \leqslant \,\,\, 0.7548 \, , $$
containing open level lines of the potential.

\begin{figure}[t]
\begin{center}
\includegraphics[width=0.9\linewidth]{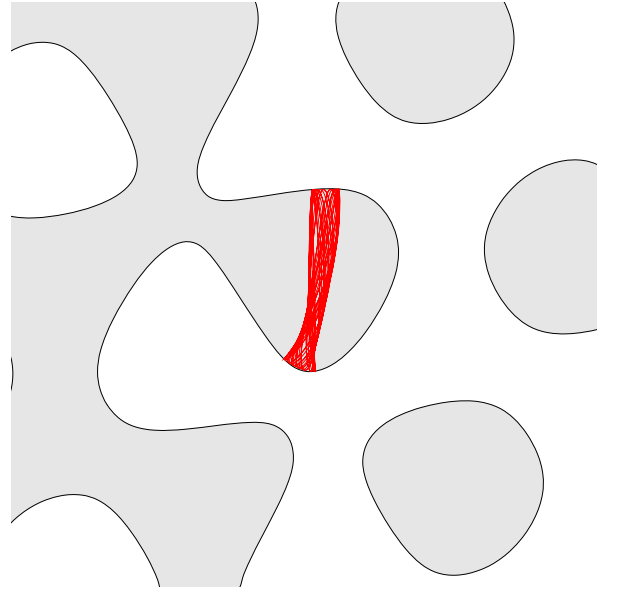}  
\end{center}
\vspace{5mm}
\begin{center}
\includegraphics[width=0.9\linewidth]{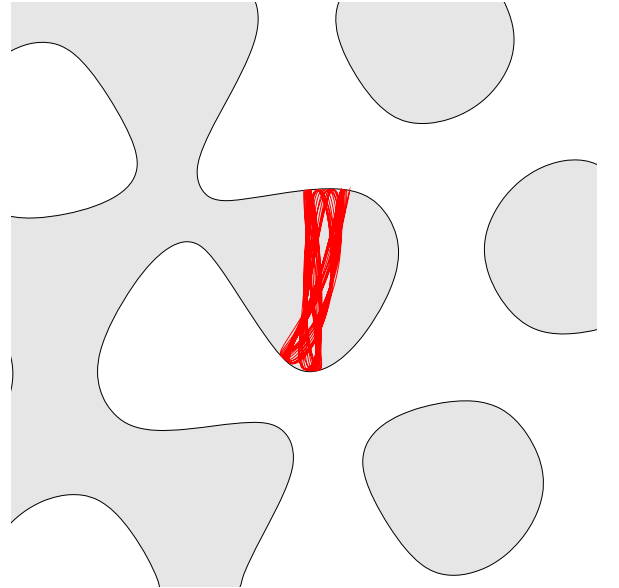}
\end{center}
\caption{Examples of invariant tori corresponding to the 
dynamics of atoms in the case of the ``regular'' potential 
(\ref{Potentsial2}) at zero total energy.}
\label{Fig19}
\end{figure}

\begin{figure}[t]
\begin{center}
\includegraphics[width=0.9\linewidth]{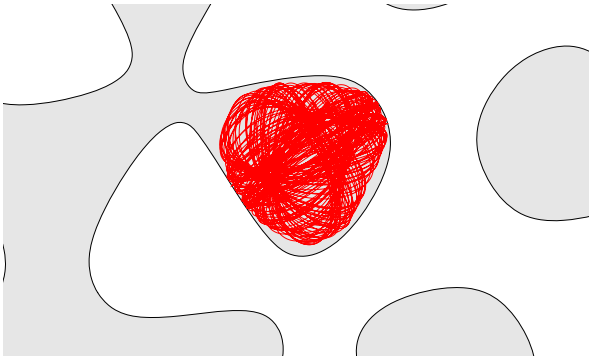}  
\end{center}
\caption{Non-integrable dynamics in a region separated from 
the rest of the phase space in the case of the ``regular'' potential 
(\ref{Potentsial2}) at zero total energy.}
\label{Fig20}
\end{figure}

 Here, in fact, to observe most of the described regimes, it is 
sufficient to study the dynamics of particles at the energy 
$\, \epsilon = 0 \, $. In particular, we can also observe here 
the presence of invariant tori of varying complexity 
(Fig.~\ref{Fig19}), as well as regions in the phase space, 
separated by such tori (Fig.~\ref{Fig20}).

\begin{figure}[t]
\vspace{5mm}
\begin{center}
\includegraphics[width=0.9\linewidth]{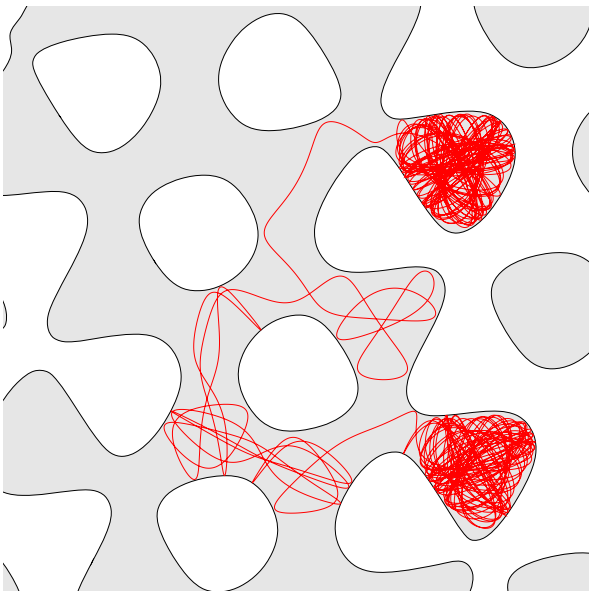}
\end{center}
\caption{Rare Levy flights in the case of the ``regular'' potential 
(\ref{Potentsial2}) at zero total particle energy.}
\label{Fig21}
\end{figure}

 Here one can also observe such a phenomenon as incomplete 
separation of the energy level by invariant tori, when there 
are ``gaps'' between the tori, allowing the trajectory to 
leave an ``almost isolated'' region at certain times. This 
situation is presented in the coordinate space by long 
wandering paths of particles in certain areas with very rare 
transitions (Levy flights) between them (Fig.~\ref{Fig21}).

\begin{figure}[t]
\begin{center}
\includegraphics[width=0.9\linewidth]{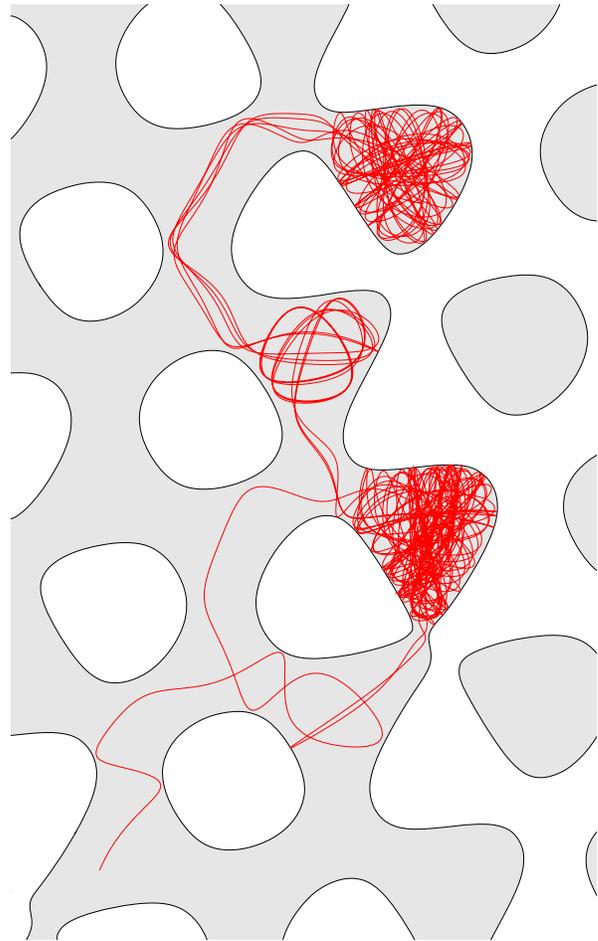}
\end{center}
\caption{Frequent Levy flights in the case of the ``regular'' 
potential (\ref{Potentsial2}) at zero total particle 
energy.}
\label{Fig22}
\end{figure}

\begin{figure}[t]
\begin{center}
\includegraphics[width=0.78\linewidth]{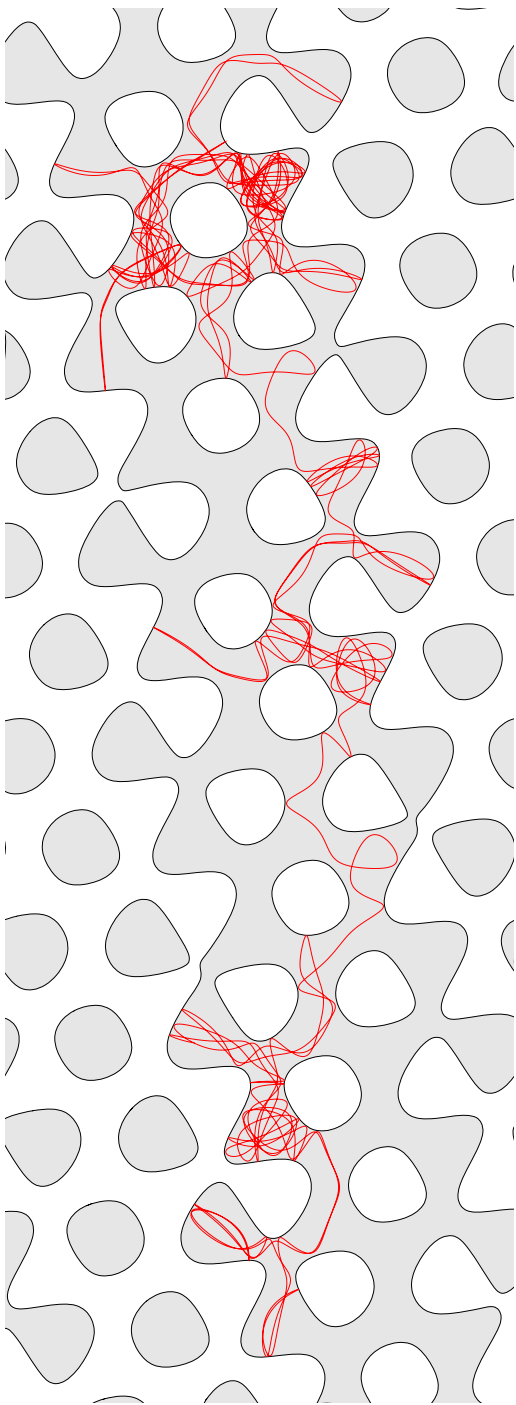}
\end{center}
\caption{Diffusion dynamics in the case of the ``regular'' 
potential (\ref{Potentsial2}) at zero total particle 
energy.}
\label{Fig23}
\end{figure}

  Finally, in certain areas of the initial data at zero 
energy level, one can also observe here much more complex 
Levy flights (Fig.~\ref{Fig22}), turning into diffusion 
modes (Fig.~\ref{Fig23}). As we have already said, we expect 
that the presence of clearly pronounced diffusion dynamics 
among other regimes in the interval of the existence of open 
level lines is in fact a general phenomenon for ``regular'' 
quasiperiodic potentials, if there are no special reasons 
suppressing such dynamics (as in the previous case). 
As it is easy to see, the geometry of the accessible 
regions for particles of fixed energy has here the most 
direct influence on the geometry of Levy flights and 
diffusion dynamics.

\begin{figure}[t]
\begin{tabular}{lr}
\includegraphics[width=0.55\linewidth]{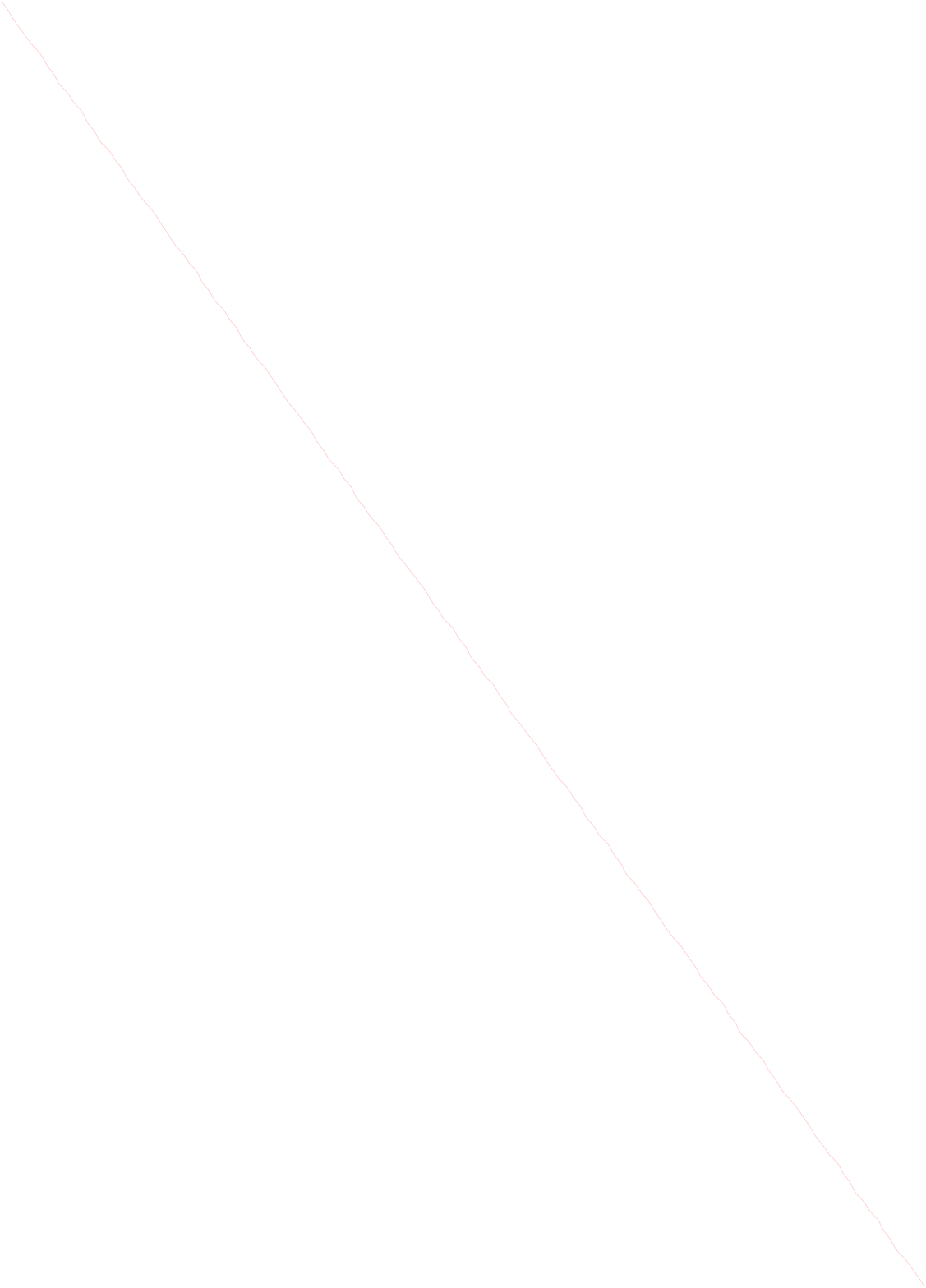}  &
\includegraphics[width=0.4\linewidth]{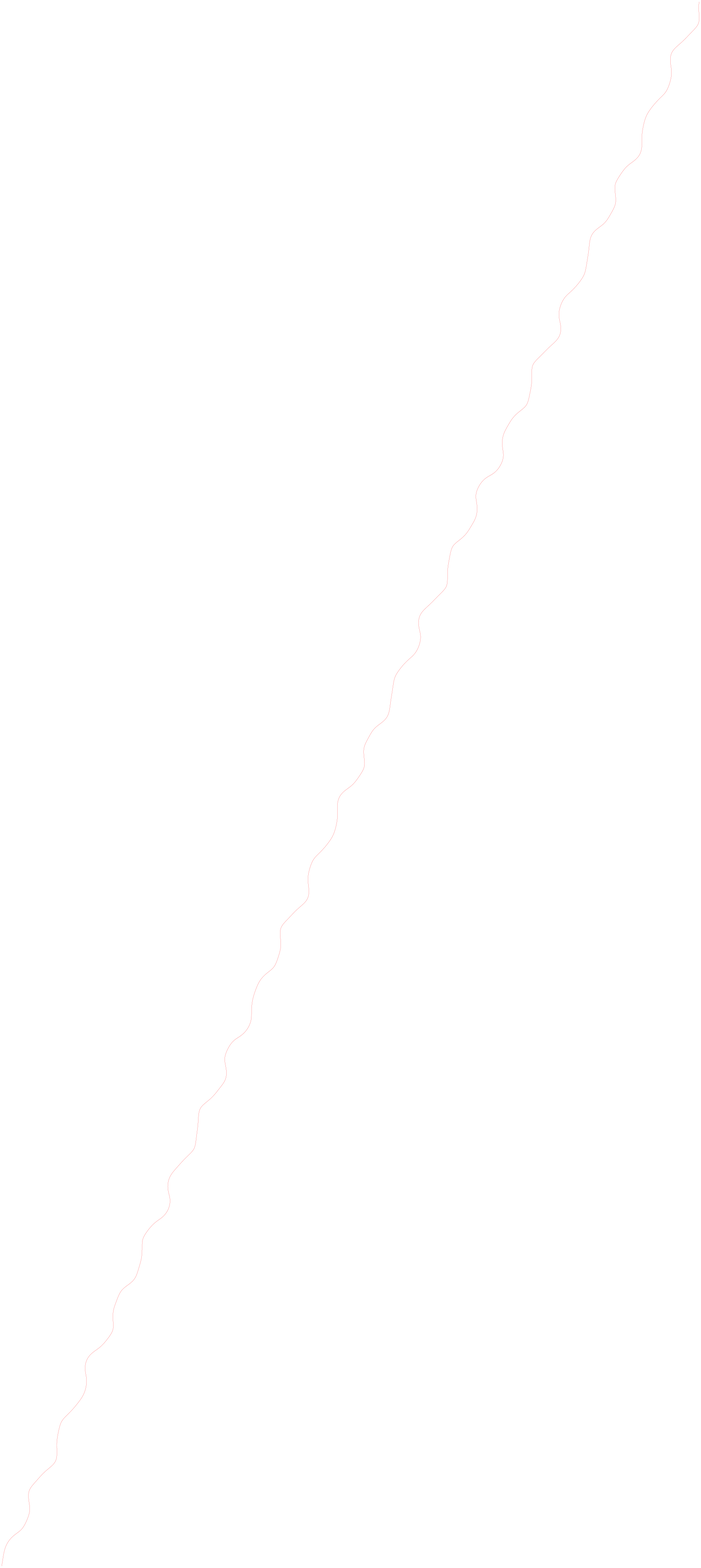}
\end{tabular}
\vspace{5mm}
\begin{center}
\includegraphics[width=0.95\linewidth]{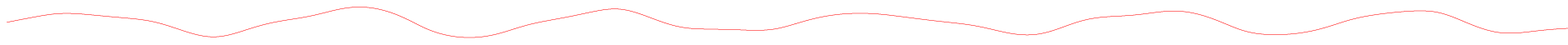}
\end{center}
\caption{Ballistic trajectories of the ``main'' directions 
in the case of the ``regular'' potential (\ref{Potentsial2}) ($\epsilon = 4$).}
\label{Fig24}
\end{figure}

 As in the previous case, a further increase in energy 
leads to the appearance of ballistic trajectories in the 
considered potential. The main stable directions of such 
trajectories (Fig.~\ref{Fig24}) are also determined here 
simply by the directions of the level lines of cosines 
present in (\ref{GeneralFormula}), and are 
not related, in fact, with the type of arising potential. 
Ballistic trajectories of stable directions, as we have 
already noted above, occupy a finite phase volume at 
a fixed energy level.

\begin{figure}[t]
\begin{center}
\includegraphics[width=0.9\linewidth]{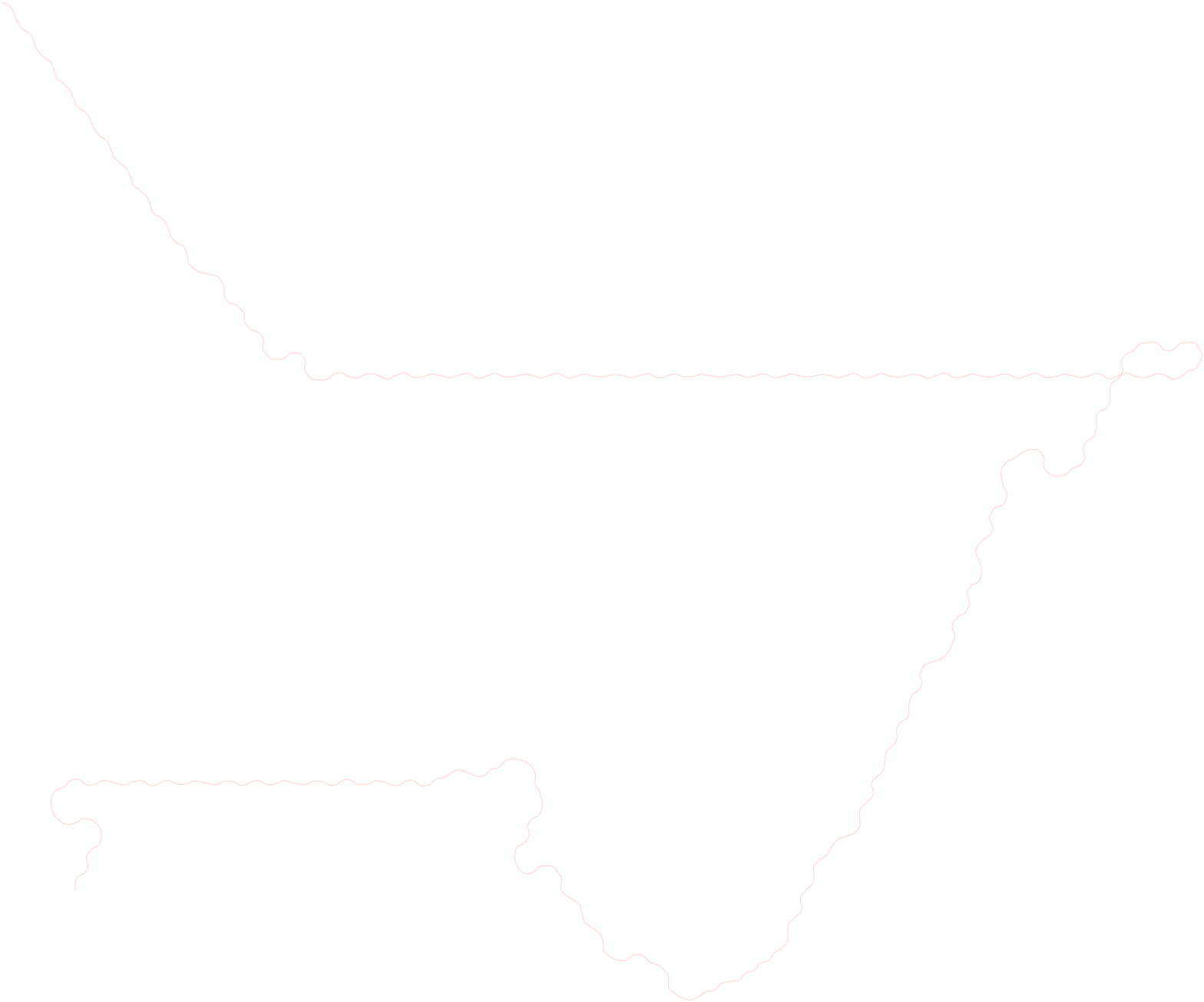}
\end{center}
\vspace{5mm}
\begin{center}
\includegraphics[width=0.9\linewidth]{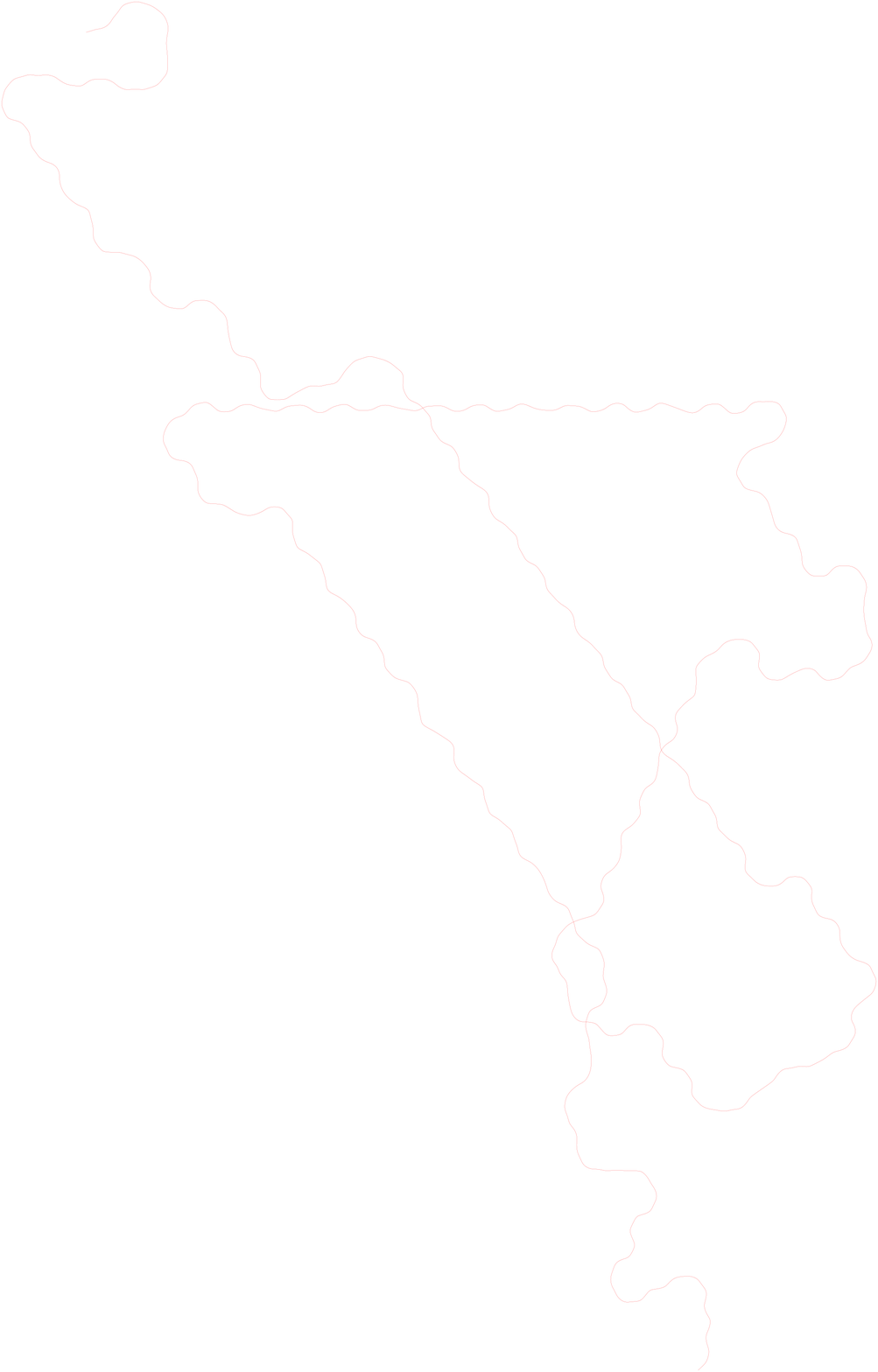}
\end{center}
\vspace{5mm}
\caption{``Almost'' ballistic trajectories in the case of the ``regular'' 
potential (\ref{Potentsial2}) ($\epsilon = 4$).}
\label{Fig25}
\end{figure}

 With a further increase in the particle energy, the number 
of stable directions of ballistic trajectories increases. 
In addition, as in the previous case, a lot of 
``quasi-ballistic'' trajectories also arise here, with 
rather long ballistic segments joined by short transitions 
between them (Fig.~\ref{Fig25}). As in the previous case, 
it can be noted that the geometric features of the contribution 
of ballistic trajectories to transport phenomena become more 
and more ``blurred'' with an increase in the number of stable 
directions of such trajectories, as well as with a complication 
of the geometry of ``quasi-ballistic'' trajectories. As for 
determining the type of potential, as well as its stable 
topological parameters (numbers $\, (m^{1}, m^{2}, m^{3}) $), 
they, as in the previous case, are best determined by the 
contribution of diffusion trajectories, as well as trajectories 
containing long ``hops'' (Levy flights) between segments of 
almost integrable or localized dynamics. It can be noted that 
in this example (in contrast to the previous one), the 
corresponding dynamics arises mostly in the 
interval of the existence of open level lines of the potential.
We expect that this property should actually manifest itself 
for most types of ``regular'' potentials (in the absence of 
additional reasons for increasing the phase volume occupied 
by the integrable dynamics) created by the method under 
consideration.

\clearpage

\begin{figure}[t]
\begin{center}
\includegraphics[width=0.9\linewidth]{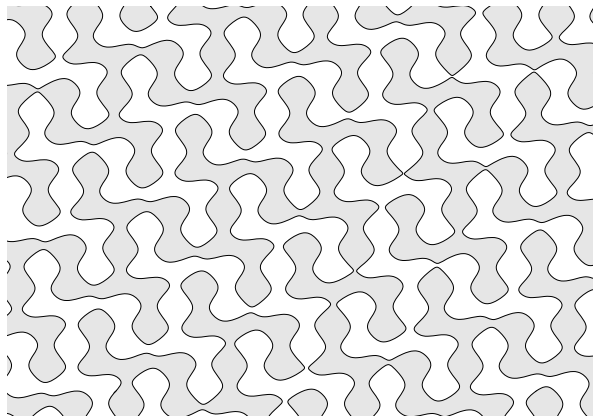}  
\end{center}
\caption{Potential (\ref{Potentsial3}) having ``chaotic'' 
open level lines. (The filled areas correspond to the values 
$\, V (x, y) \leqslant 0$).}
\label{Fig26}
\end{figure}

Our last series of computations refers to the potential
with coefficients
 \begin{multline}
\label{Potentsial3}
a_{1} \, = \, - 0.6190763027420052 ,  \\
b_{1} \, = \, - 0.2572674789786692 ,  \\
c_{1} \, = \, 1.311209111211166 ,   \\
a_{2} \, = \, 0.7853308419916342 ,  \\
b_{2} \, = \, - 0.20280395367888493 ,  \\
c_{3} \, = \, 0.8662242771884692 ,   \\
a_{3} \, = \, 0 ,  \\
b_{3} \, = \, 0.9448195598272575 ,  \\
c_{3} \, = \, 2.950743051151684 
\end{multline}
having ``chaotic'' level lines (Fig.~\ref{Fig26}).
In this case, open level lines, as we have already said, 
exist only at the value $\, V_{0} = 0 \, $.

\begin{figure}[t]
\begin{center}
\includegraphics[width=0.9\linewidth]{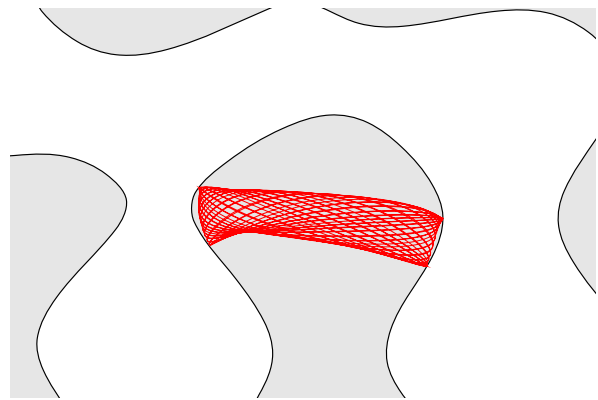}  
\end{center}
\vspace{5mm}
\begin{center}
\includegraphics[width=0.9\linewidth]{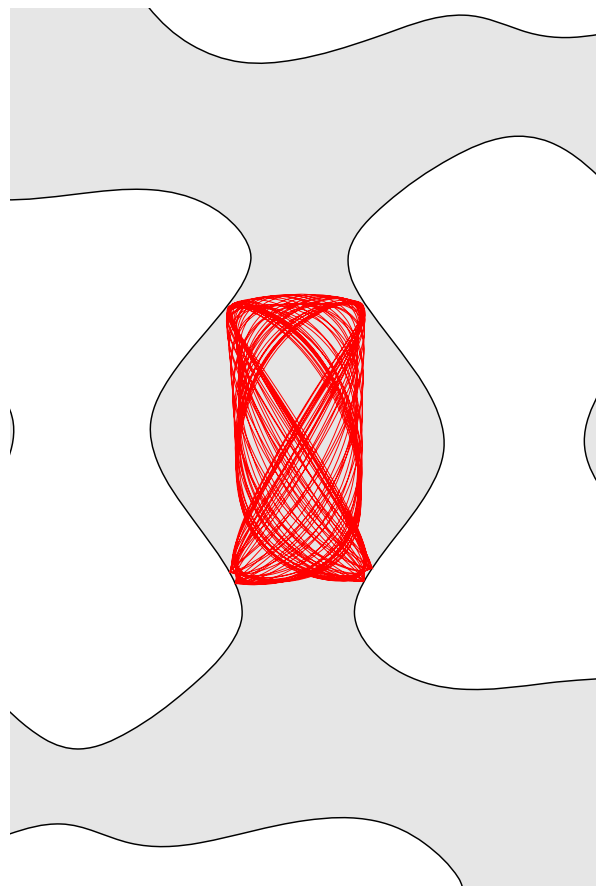}
\end{center}
\caption{Examples of invariant tori corresponding to a relatively 
simple dynamics of atoms in the case of the ``chaotic'' potential 
(\ref{Potentsial3}) at zero total energy.}
\label{Fig27}
\end{figure}

\begin{figure}[t]
\begin{center}
\includegraphics[width=0.9\linewidth]{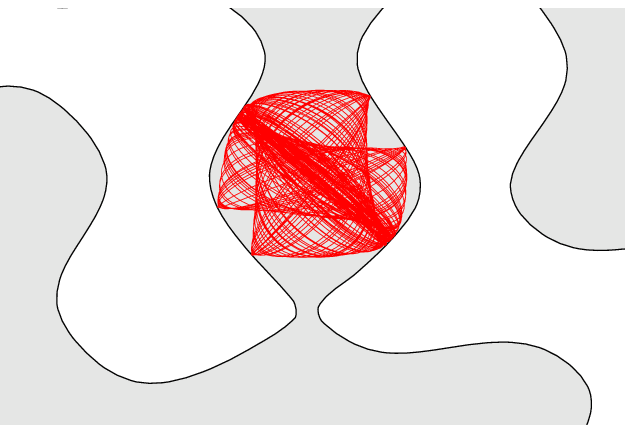}  
\end{center}
\vspace{5mm}
\begin{center}
\includegraphics[width=0.9\linewidth]{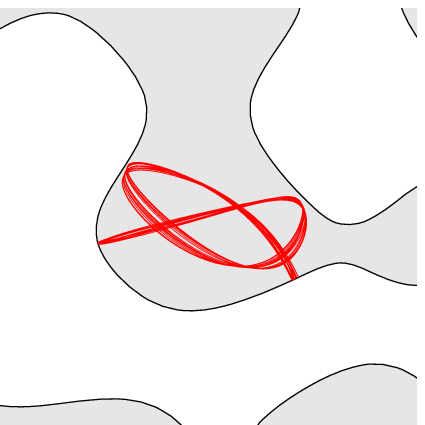}
\end{center}
\caption{Examples of invariant tori defining more complex 
dynamics of atoms in the case of the ``chaotic'' potential 
(\ref{Potentsial3}) at zero total energy.}
\label{Fig28}
\end{figure}

 As in the previous two cases, in this case, at the energy 
level $\, \epsilon = 0 \, $ we can also see regions corresponding 
to the motion along tori of  relatively simple 
(Fig.~\ref{Fig27}) as well as more complex geometry (Fig.~\ref{Fig28}).

\begin{figure}[t]
\begin{center}
\includegraphics[width=0.9\linewidth]{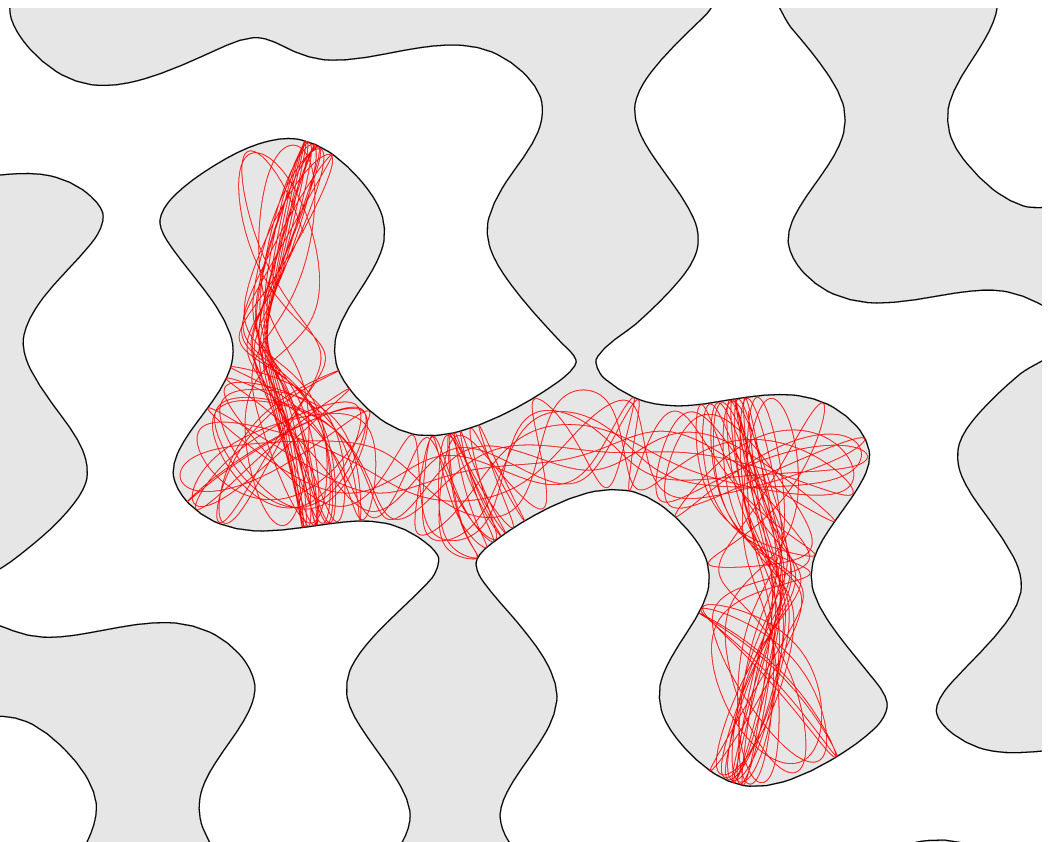}  
\end{center}
\vspace{5mm}
\begin{center}
\includegraphics[width=0.9\linewidth]{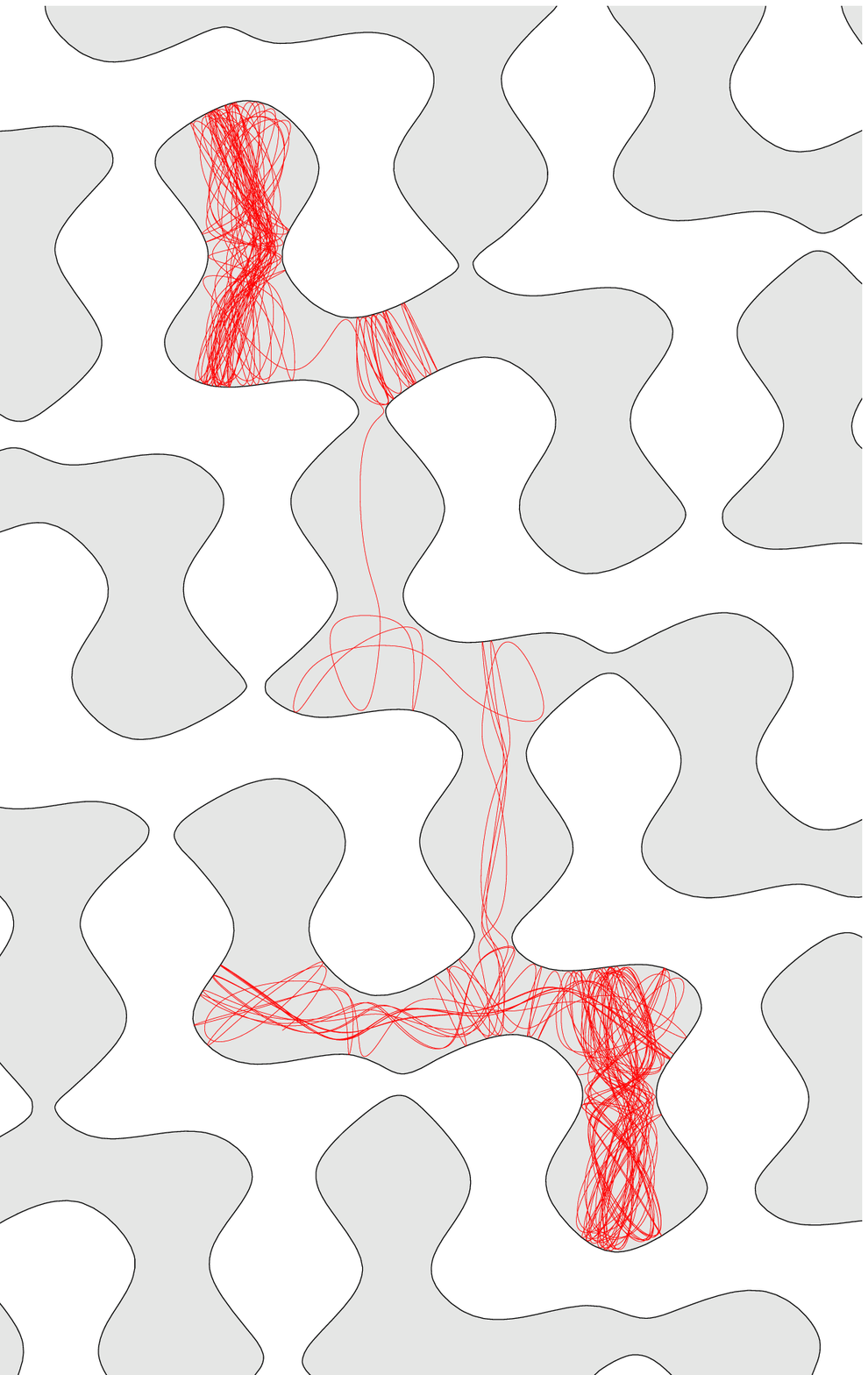}  
\end{center}
\caption{Non-integrable dynamics bounded by invariant tori in the 
phase space in the case of the ``chaotic'' potential (\ref{Potentsial3}) at 
zero total energy.}
\label{Fig29}
\end{figure}

 In addition, in certain regions of initial data, one can observe 
non-integrable dynamics, which is bounded, however, by some invariant 
tori in the manifold $\, \epsilon = 0 \, $. Such dynamics is also 
easily distinguishable from other types when it is projected onto the 
coordinate space (Fig.~\ref{Fig29}).

\begin{figure}[t]
\begin{center}
\includegraphics[width=0.9\linewidth]{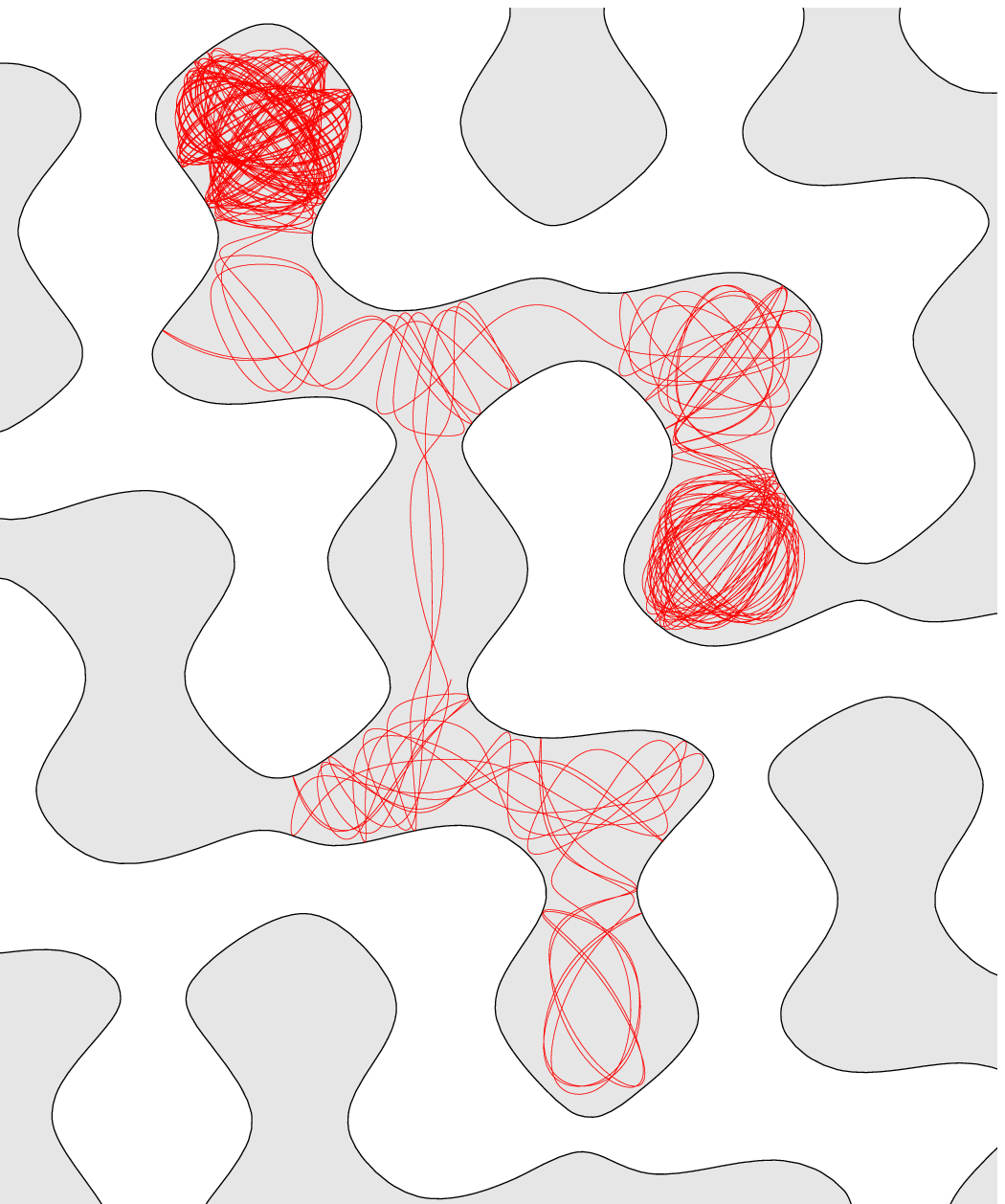}  
\end{center}
\caption{Levy flights in the case of the ``chaotic'' potential 
(\ref{Potentsial3}) at zero total energy.}
\label{Fig30}
\end{figure}

 As in the previous cases, for certain initial conditions 
at the zero energy level, one can also observe the ``sticking'' of 
the particle trajectory to invariant tori for a rather long time, 
interspersed with Levy flights at certain moments 
(Fig.~\ref{Fig30}).

\begin{figure}[t]
\begin{center}
\includegraphics[width=0.9\linewidth]{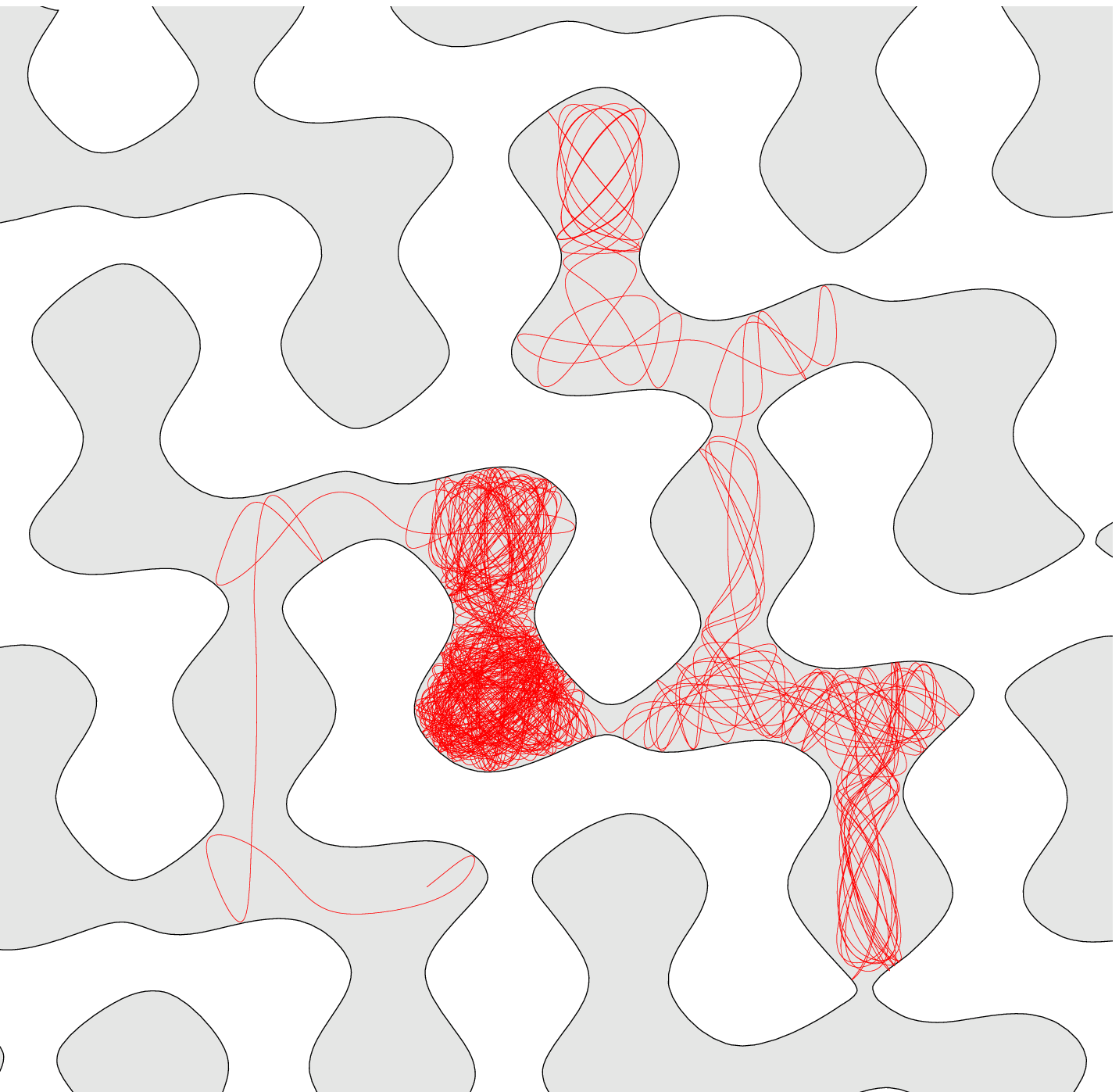}
\end{center}
\vspace{5mm}
\begin{center}
\includegraphics[width=0.9\linewidth]{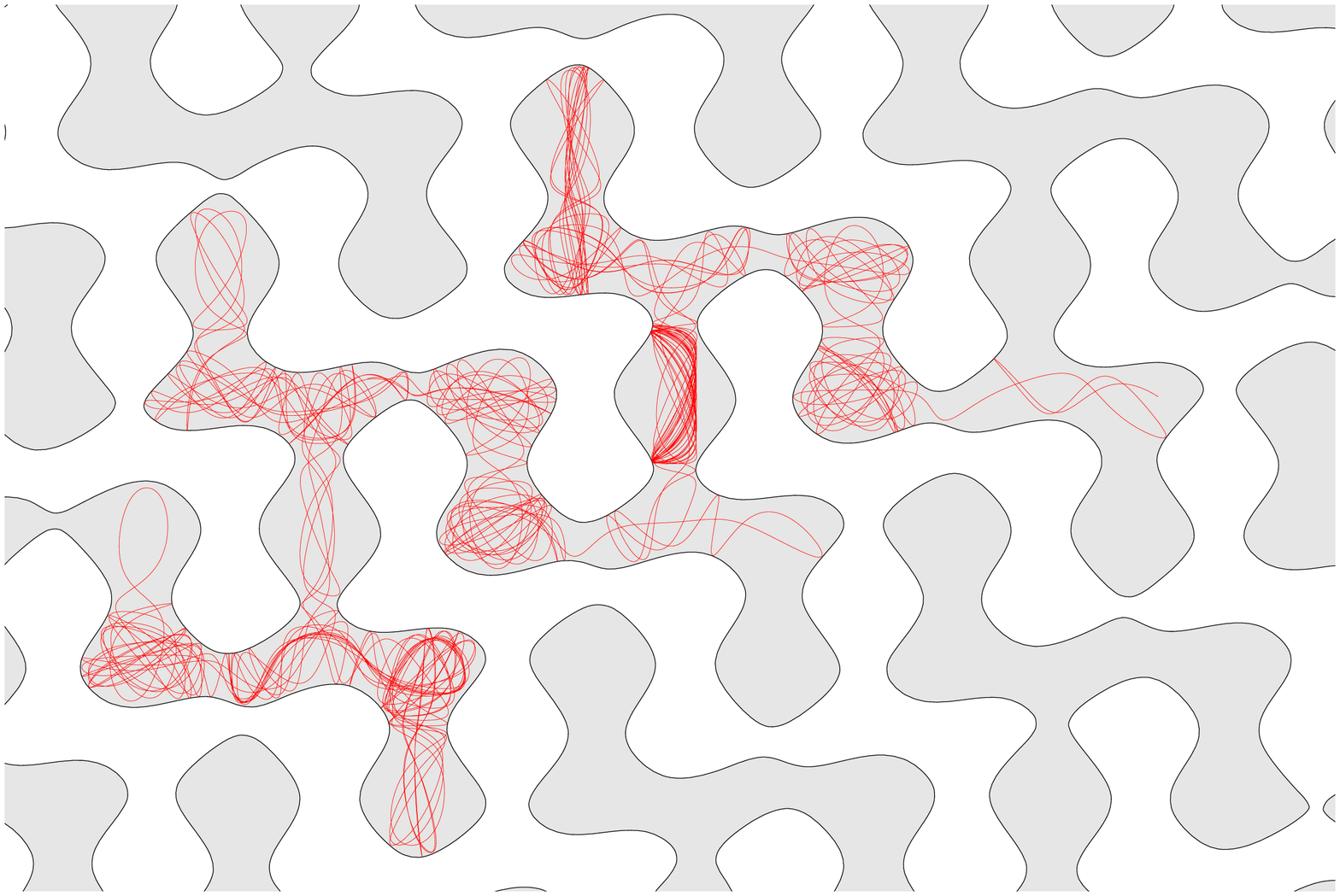}
\end{center}
\caption{Transition to diffusion dynamics under a change of the 
initial data in the case of the ``chaotic'' potential (\ref{Potentsial3}) 
at zero total particle energy.}
\label{Fig31}
\end{figure}

 Also, as in the previous case, by changing the initial data, 
we can achieve the complication of the geometry of the tori 
and the transition to the diffusion dynamics of particles 
(Fig.~\ref{Fig31}). Diffusion dynamics here is also limited to 
the ``region of accessibility'', which now has a completely 
different geometry and itself has, in a sense, 
``diffusion properties''. It must be said that, apparently, 
in the classical limit, the transport properties of particles 
at the zero energy level here are close to the transport 
properties of localized (although in large regions) particles, 
since the probability of distant diffusion in the region under 
consideration is very small. It can be noted, however, that in 
the case of three quasi-periods, such regions always contain 
segments of the boundary that are very close to each other, 
where quantum tunneling should be possible. In this case, it 
is quantum tunneling that, apparently, should play an important 
role for transport phenomena at $\, \epsilon = 0 $.

\begin{figure}[t]
\begin{center}
\includegraphics[width=0.9\linewidth]{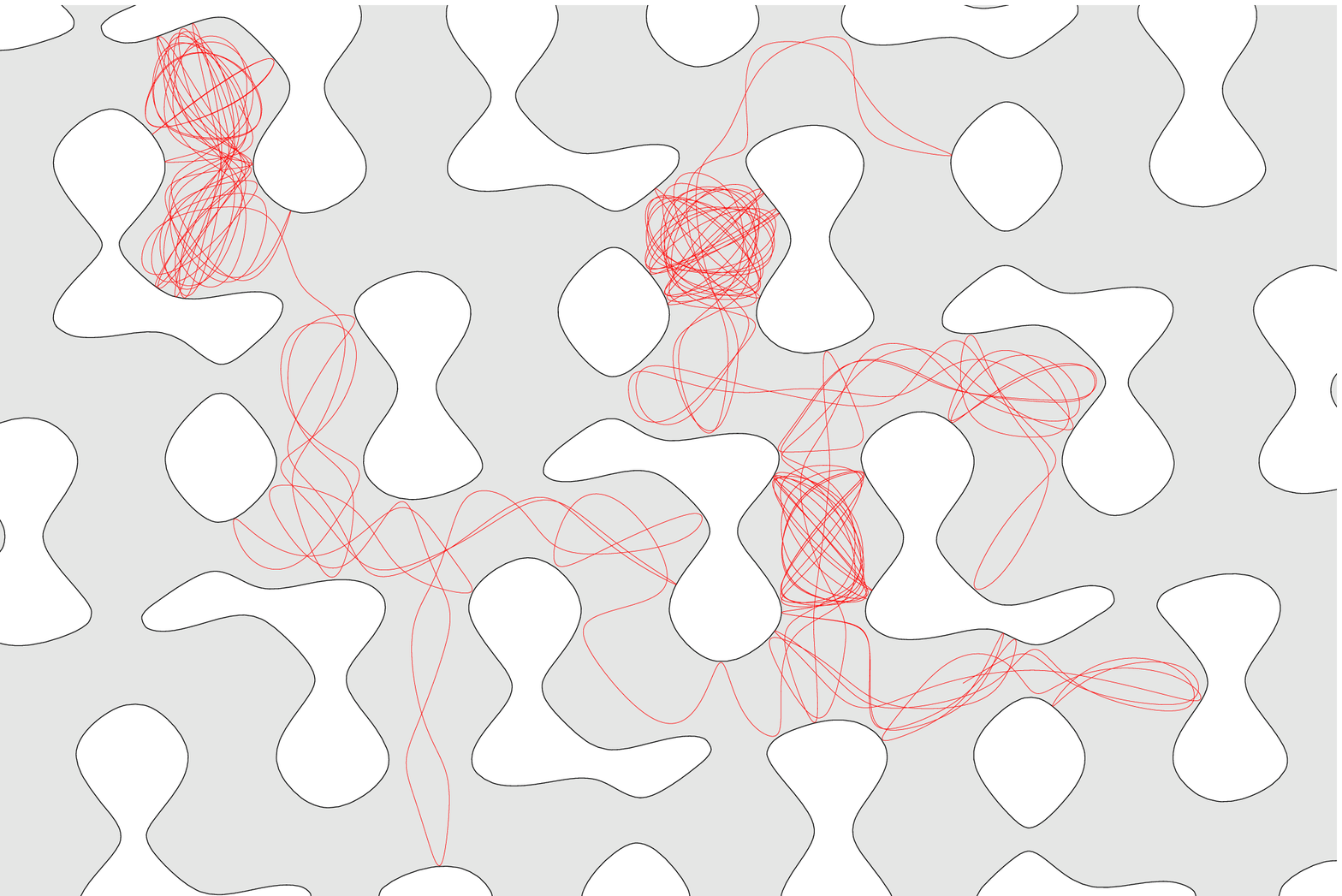}
\end{center}
\vspace{5mm}
\begin{center}
\includegraphics[width=0.9\linewidth]{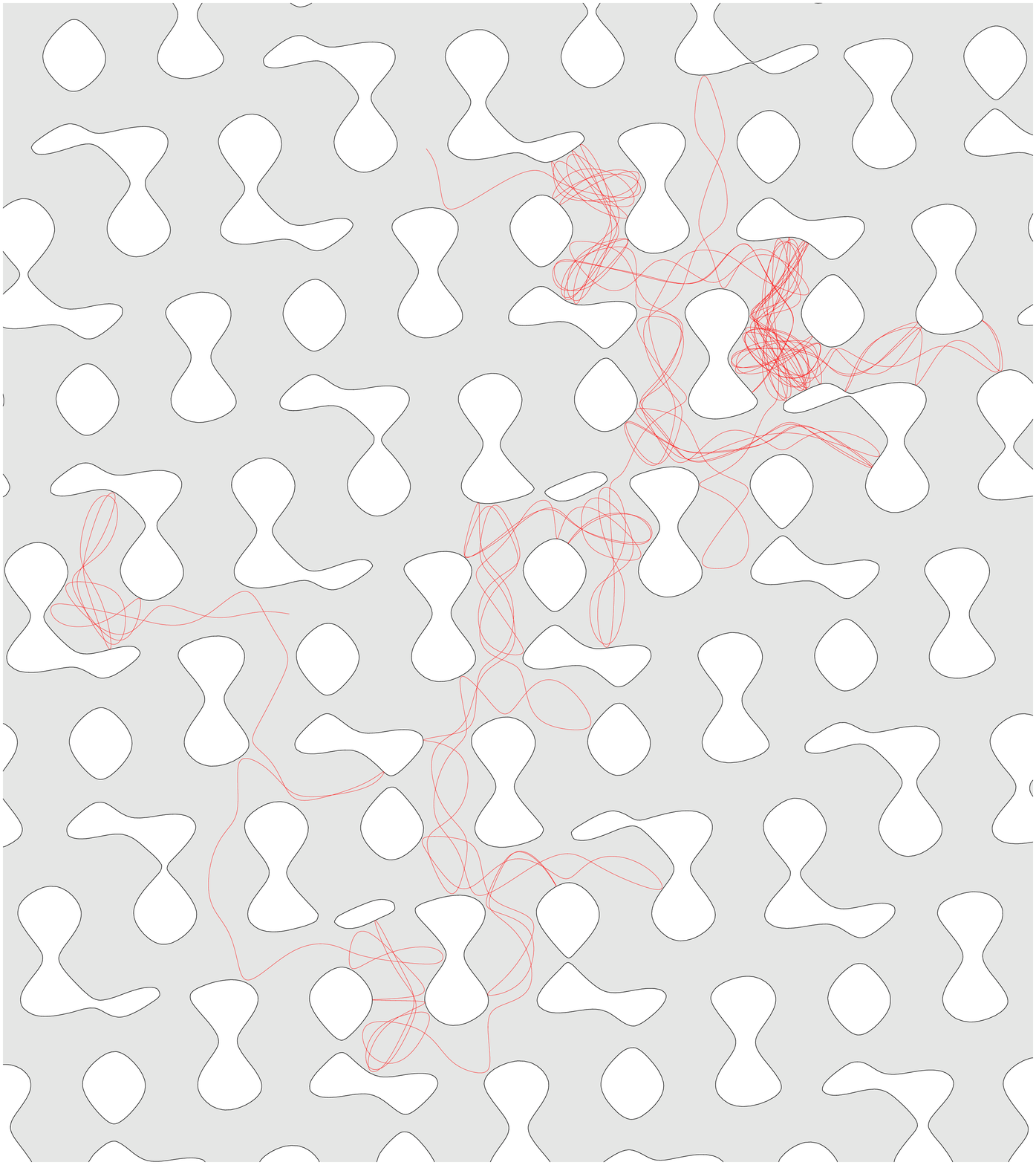}
\end{center}
\caption{Accessibility regions and diffusion dynamics in 
the case of the ``chaotic'' potential (\ref{Potentsial3}) at positive 
particle energies ($\epsilon = 0.64$ and $\epsilon = 0.72$).}
\label{Fig32}
\end{figure}

 In general, with a gradual increase in the energy of 
particles in the ensemble, the transport properties of an atomic gas in the 
described potential should (in the classical limit) change 
significantly when particles with positive energies appear in 
the ensemble. Indeed, with increasing energy, the accessibility 
regions expand and become non-simply connected, in contrast to 
the case $\, \epsilon = 0 \, $. We can say that, in a certain 
sense, such regions have the property of ``percolation''. 
At the same time, they retain for some time also certain 
``diffusion'' form, which should manifest itself in the 
transport properties of an atomic gas. As we can see in
Fig.~\ref{Fig32}, the diffusion properties of the particle 
dynamics in such potentials rapidly increase with an increase 
in the value of $\, \epsilon $. As the energy of the particles 
decreases, the ``accessible regions'' become bounded regions in 
the plane. Thus, one can see here a noticeable difference from
the potentials with ``regular'' level lines, where the picture 
does not change significantly when the particle energy is varied 
near zero.

\begin{figure}[t]
\begin{tabular}{lr}
\includegraphics[width=0.55\linewidth]{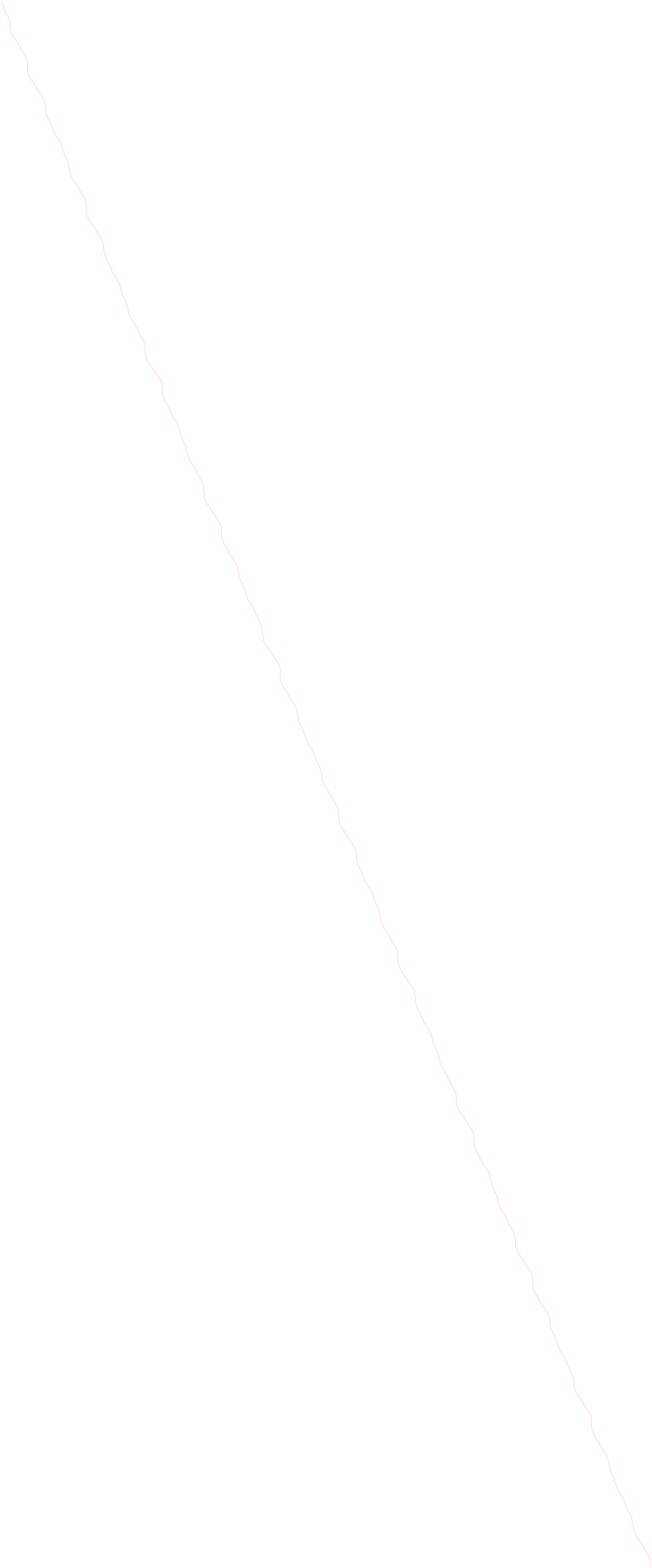}  &
\hspace{5mm}
\includegraphics[width=0.35\linewidth]{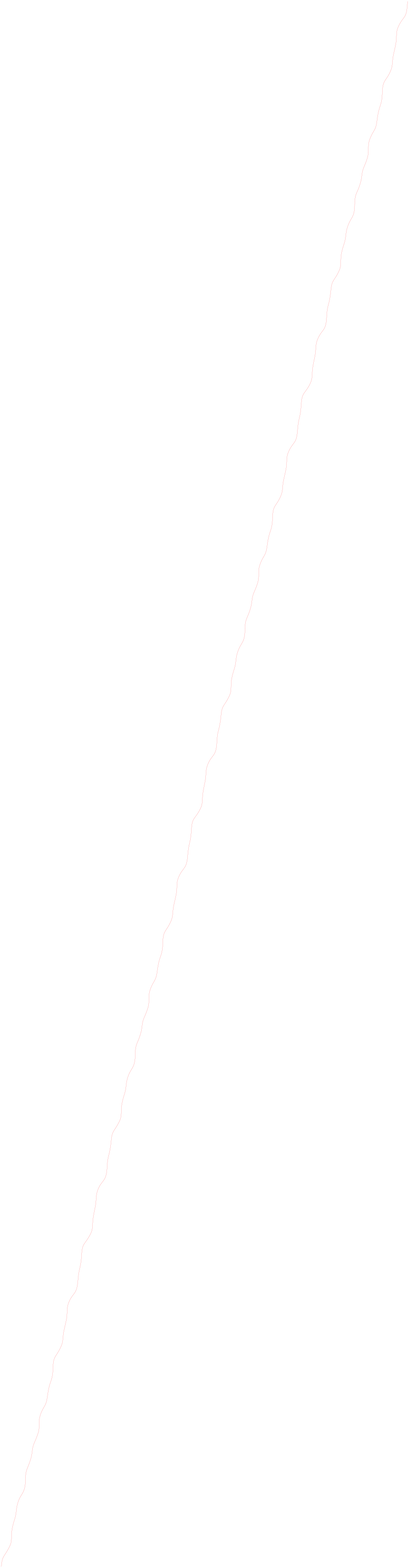}
\end{tabular}
\begin{center}
\includegraphics[width=0.95\linewidth]{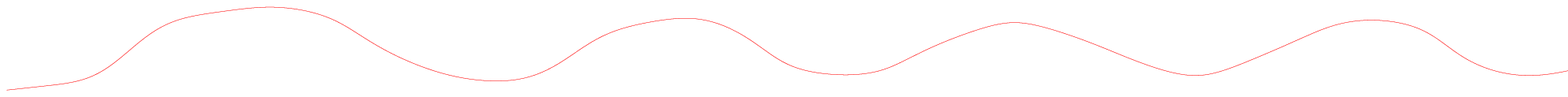}
\end{center}
\caption{The appearance of ballistic trajectories in the 
case of the ``chaotic'' potential (\ref{Potentsial3}) with increasing 
particle energy ($\epsilon = 5$).}
\label{Fig33}
\end{figure}

\begin{figure}[t]
\begin{center}
\includegraphics[width=0.9\linewidth]{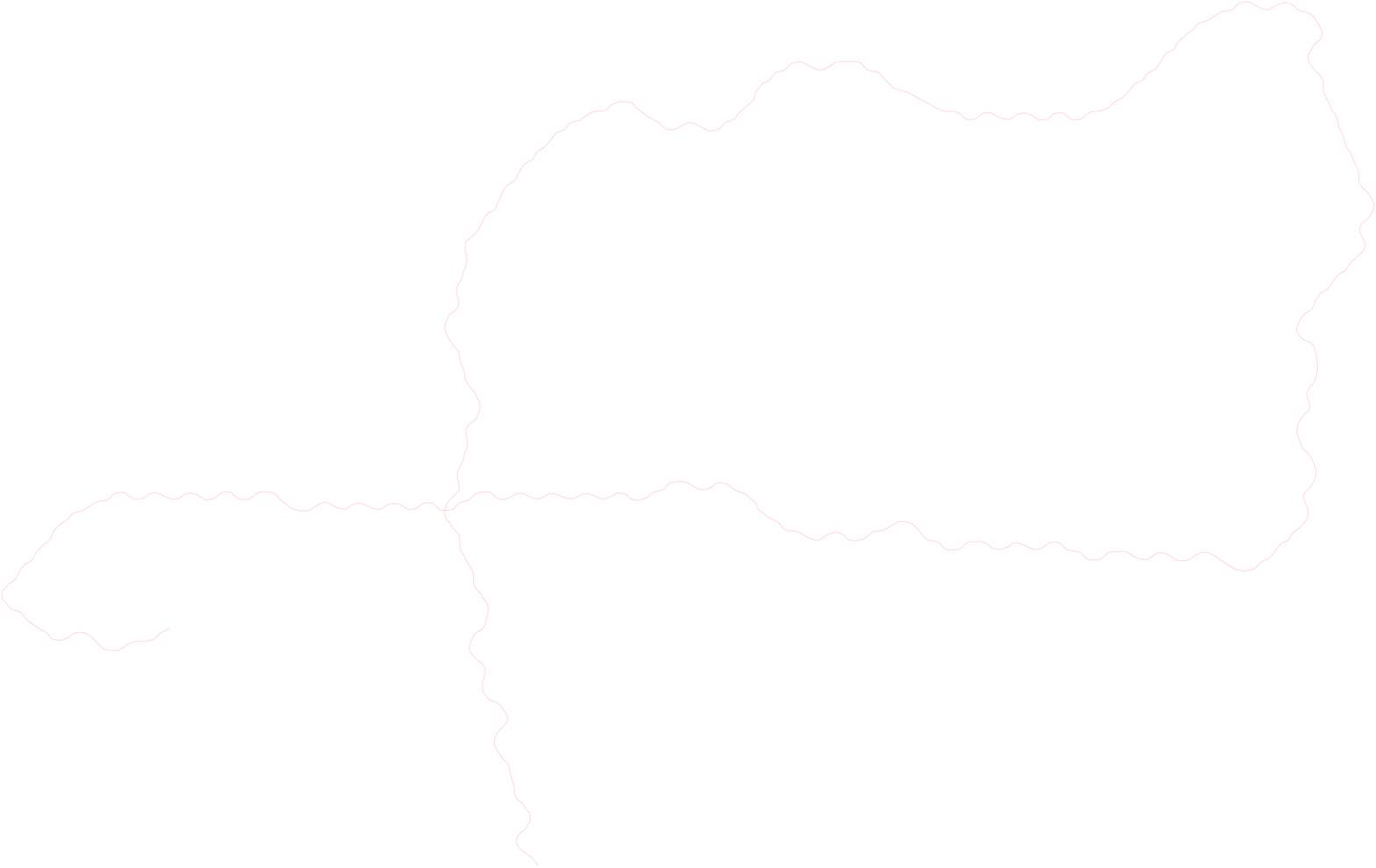}
\end{center}
\vspace{5mm}
\begin{center}
\includegraphics[width=0.9\linewidth]{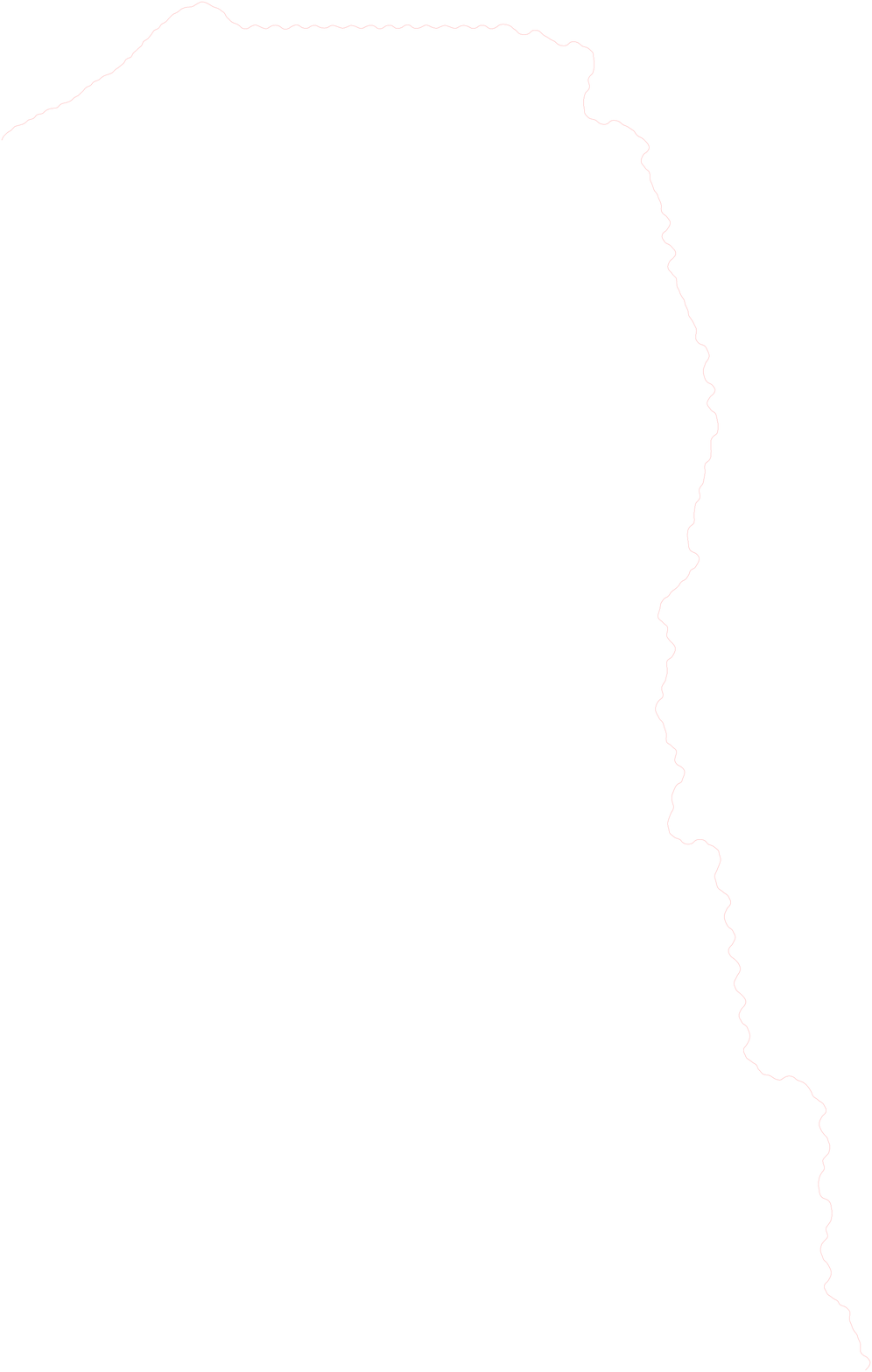}
\end{center}
\vspace{5mm}
\caption{``Almost'' ballistic trajectories in case of the chaotic 
potential (\ref{Potentsial3}) ($\epsilon = 4$).}
\label{Fig34}
\end{figure}

 With a significant increase in energy, ballistic trajectories 
with stable directions appear for the potential (\ref{Potentsial3}) as well
(Fig.~\ref{Fig33}). As in the previous two cases, the main stable 
directions are the directions of the level lines of the cosines 
constituting the potential (\ref{Potentsial3}) according to
formula (\ref{GeneralFormula}). The corresponding trajectories 
appear at the lowest energy levels; with a further increase in 
energy, the number of such directions increases. As in the previous 
two cases, the final phase volume is also occupied by 
``quasi-ballistic'' trajectories, consisting of long segments
of ballistic trajectories joined by short intermediate 
segments (Fig.~\ref{Fig34}). As we have already said above, 
to determine the geometric features of a ``chaotic'' potential, 
as in the ``regular'' case, it seems most appropriate to study 
the contribution of the ``diffusion'' and ``jumping'' trajectories 
appearing at ``intermediate'' energy values.

\clearpage

\section{Conclusion}

  The paper considers questions related to the geometry of 
quasiperiodic potentials on a plane, the methods of their creation, 
their dependence on control parameters, and the dynamics of 
semiclassical particles in such potentials. The main 
considerations are related to the situation of the emergence
of such potentials in systems of ultracold atoms in 
magneto-optical traps, although our results are actually 
valid for the most general types of quasiperiodic potentials. 
It is shown that, in the general case, quasiperiodic potentials 
on the plane can be naturally divided into two main classes, 
namely, potentials with a ``regular'' behavior of open level 
lines and potentials with a ``chaotic'' behavior of open level 
lines. In each family of quasiperiodic potentials, depending 
smoothly on some set of parameters, potentials from these classes 
are parametrized by sets having different structures, complementing 
each other in the full space of parameters. Namely, the first 
set has the form of a union of (countably many) regions with piecewise 
smooth boundaries, while the second has fractal properties.
According to the behavior of open level lines, the former
potentials can be attributed the ``regular'' type 
(approaching periodic potentials), while the latter can be 
considered as a model of random potentials. The non-dissipative 
dynamics of ultracold atoms in the considered potentials is 
integrable at lower energy levels, gradually becoming chaotic 
with increasing the energy of atoms. As a rule, in the interval 
of the existence of open level lines of the potential, both types 
(integrable and chaotic) of dynamics are present, and the 
properties of the chaotic dynamics substantially depend on the 
geometry of the potential level lines. The study of the 
transport properties of an atomic gas in quasiperiodic 
potentials in the presence of particles with the corresponding 
energies in the ensemble can thus allow observing the 
differences between potentials of both types and also provide 
more detailed information on the geometry of their open level 
lines.

\end{document}